\documentclass[]{llncs}

\usepackage{pslatex}
\usepackage{bussproofs}
\usepackage{listings}
\usepackage{fancybox}
\usepackage{mathtools}
\usepackage{tikz}
\usetikzlibrary{arrows.meta, shapes, arrows, automata, fit, matrix,
  positioning, decorations.pathreplacing, calligraphy}
\usepackage{amssymb}
\usepackage{amsmath}
\usepackage{ifthen}
\usepackage{thm-restate}
\usepackage{caption}
\usepackage{subcaption}
\usepackage{paralist}
\usepackage[hidelinks]{hyperref}

\usepackage{algorithm}
\usepackage{algorithmicx}
\usepackage[noend]{algpseudocode}

\usepackage[draft]{commenting}

\declareauthor{ri}{Radu}{blue}
\declareauthor{mb}{Marius}{red}
\declareauthor{lb}{Lucas}{green}

\newif\ifLongVersion\LongVersiontrue

\ifLongVersion
\usepackage[createShortEnv,conf={normal}]{proof-at-the-end}
\else
\usepackage[createShortEnv,conf={end,no link to proof,restate}]{proof-at-the-end}
\fi

\ifLongVersion
\newenvironment{myLemmaE}{\begin{lemmaE}}{\end{lemmaE}}
\newenvironment{myTextE}{}{}
\else
\newenvironment{myLemmaE}{\begin{lemmaE}[][all end]}{\end{lemmaE}}
\newenvironment{myTextE}{\begin{textAtEnd}}{\end{textAtEnd}}
\fi

\newcommand{\nat}{\mathbb{N}}
\newcommand{\arityof}[1]{\#{#1}}
\newcommand{\lenof}[1]{|{#1}|}
\newcommand{\sizeof}[1]{\mathrm{size}({#1})}
\newcommand{\maxarityof}[1]{\mathrm{arity}({#1})}
\newcommand{\maxwidthof}[1]{\mathrm{width}({#1})}
\newcommand{\maxintersize}[1]{\mathrm{intersize}({#1})}
\newcommand{\at}[2]{\tuple{{#1}}_{#2}}

\newcommand{\universe}{\mathbb{C}}
\newcommand{\vars}{\mathbb{V}}

\newcommand{\preds}{\mathbb{A}}
\newcommand{\isdef}{\stackrel{\scriptscriptstyle{\mathsf{def}}}{=}}
\newcommand{\interv}[2]{[{#1},{#2}]}
\newcommand{\tuple}[1]{\langle {#1} \rangle}
\newcommand{\Tuple}[1]{\left\langle {#1} \right\rangle}
\newcommand{\set}[1]{\{ {#1} \}}
\newcommand{\Set}[1]{\left\{ {#1} \right\}}
\newcommand{\pow}[1]{\mathrm{pow}({#1})}
\newcommand{\mpow}[1]{\mathrm{mpow}({#1})}
\newcommand{\dom}[1]{\mathrm{dom}({#1})}

\newcommand{\cardof}[1]{|\!| {#1} |\!|}

\newcommand{\finsubseteq}{\subseteq_{\mathit{fin}}}
\newcommand{\notof}[1]{\mathrm{not}(#1)}

\newcommand{\bigO}{\mathcal{O}}
\newcommand{\twoexptime}{$2\mathsf{EXP}$}
\newcommand{\threeexptime}{$3\mathsf{EXP}$}
\newcommand{\fourexptime}{$4\mathsf{EXP}$}
\newcommand{\np}{$\mathsf{NP}$}
\newcommand{\conp}{$\mathsf{co}$-$\mathsf{NP}$}
\newcommand{\exptime}{$\mathsf{EXP}$}

\newcommand{\prob}[2]{\mathsf{P}[{#1},{#2}]}
\newcommand{\klprob}[4]{\mathsf{P}^{({#3},{#4})}[{#1},{#2}]}
\newcommand{\sat}[2]{\mathsf{Sat}[{#1},{#2}]}
\newcommand{\klsat}[4]{\mathsf{Sat}^{({#3},{#4})}[{#1},{#2}]}
\newcommand{\tight}[2]{\mathsf{Tight}[{#1},{#2}]}
\newcommand{\kltight}[4]{\mathsf{Tight}^{({#3},{#4})}[{#1},{#2}]}
\newcommand{\loose}[2]{\mathsf{Loose}[{#1},{#2}]}
\newcommand{\klloose}[4]{\mathsf{Loose}^{({#3},{#4})}[{#1},{#2}]}
\newcommand{\bound}[2]{\mathsf{Bnd}[{#1},{#2}]}
\newcommand{\klbound}[4]{\mathsf{Bnd}^{({#3},{#4})}[{#1},{#2}]}
\newcommand{\entl}[3]{\mathsf{Entl}[{#1},{#2},{#3}]}
\newcommand{\klentl}[5]{\mathsf{Entl}^{{#4},{#5}}[{#1},{#2},{#3}]}

\newcommand{\comps}{\mathcal{C}}

\newcommand{\interacs}{\mathcal{I}}

\newcommand{\interac}[4]{\tuple{{#1}.{\mathit{#2}}, {#3}.{\mathit{#4}}}}

\newcommand{\intertypes}{\mathsf{Inter}}
\newcommand{\intertype}{\tau}

\newcommand{\beh}{\mathbb{B}}

\newcommand{\states}{\mathcal{Q}}

\newcommand{\ports}{\mathcal{P}}

\newcommand{\arrow}[2]{\xrightarrow{{\scriptstyle #1}}_{{\scriptstyle #2}}}
\newcommand{\Arrow}[2]{\xRightarrow{{\scriptstyle #1}}_{\raisebox{4pt}{\!$\scriptstyle{#2}$}}}

\makeatletter
\newcommand*{\da@rightarrow}{\mathchar"0\hexnumber@\symAMSa 4B }
\newcommand*{\da@leftarrow}{\mathchar"0\hexnumber@\symAMSa 4C }
\newcommand*{\xdashrightarrow}[2][]{%
  \mathrel{%
    \mathpalette{\da@xarrow{#1}{#2}{}\da@rightarrow{\,}{}}{}%
  }%
}
\newcommand{\xdashleftarrow}[2][]{%
  \mathrel{%
    \mathpalette{\da@xarrow{#1}{#2}\da@leftarrow{}{}{\,}}{}%
  }%
}
\newcommand*{\da@xarrow}[7]{%
  \sbox0{$\ifx#7\scriptstyle\scriptscriptstyle\else\scriptstyle\fi#5#1#6\m@th$}%
  \sbox2{$\ifx#7\scriptstyle\scriptscriptstyle\else\scriptstyle\fi#5#2#6\m@th$}%
  \sbox4{$#7\dabar@\m@th$}%
  \dimen@=\wd0 %
  \ifdim\wd2 >\dimen@
    \dimen@=\wd2 %
  \fi
  \count@=2 %
  \def\da@bars{\dabar@\dabar@}%
  \@whiledim\count@\wd4<\dimen@\do{%
    \advance\count@\@ne
    \expandafter\def\expandafter\da@bars\expandafter{%
      \da@bars
      \dabar@ 
    }%
  }%
  \mathrel{#3}%
  \mathrel{%
    \mathop{\da@bars}\limits
    \ifx\\#1\\%
    \else
      _{\copy0}%
    \fi
    \ifx\\#2\\%
    \else
      ^{\copy2}%
    \fi
  }%
  \mathrel{#4}%
}
\makeatother

\newcommand{\store}{\nu}

\newcommand{\statemap}{\varrho}

\newcommand{\aconfig}{\gamma}
\newcommand{\nodesof}[1]{\mathrm{nodes}({#1})}

\newcommand{\nodes}[1]{\mathrm{nodes}({#1})}
\newcommand{\degreenode}[2]{\delta_{#1}({#2})}
\newcommand{\degreeof}[1]{\degreenode{}{#1}}

\newcommand{\comp}{\bullet}

\newcommand{\subconfig}{\sqsubseteq}
\newcommand{\bigcomp}{\scalebox{2}{$\comp$}}

\newcommand{\cl}{\textsf{CL}}
\newcommand{\predname}[1]{\mathsf{#1}}
\newcommand{\apred}{\predname{A}}
\newcommand{\bpred}{\predname{B}}
\newcommand{\appred}{{\apred'}}
\newcommand{\bppred}{{\bpred'}}
\newcommand{\psid}{{\widetilde{\asid}}}
\newcommand{\emp}{\predname{emp}}

\let\Asterisk\undefined
\newcommand{\Asterisk}{\mathop{\scalebox{1.9}{\raisebox{-0.2ex}{$\ast$}}}\hspace*{1pt}}%
\renewcommand{\vec}[1]{\mathbf #1}
\newcommand{\fv}[1]{\mathrm{fv}({#1})}

\newcommand{\compin}[2]{{#1}@{#2}}
\newcommand{\compact}[1]{[{#1}]}
\newcommand{\compactin}[2]{\compin{\compact{#1}}{#2}}
\newcommand{\interacn}[4]{\tuple{{#1}.\mathit{#2}, \ldots, {#3}.\mathit{#4}}}
\newcommand{\interactwo}[4]{\tuple{{#1}.\mathit{#2}, {#3}.\mathit{#4}}}
\newcommand{\formeq}[1]{\approx_{#1}}
\newcommand{\formneq}[1]{{\not\approx}_{#1}}

\newcommand{\seplog}{\textsf{SL}}
\newcommand{\csl}{\textsf{CSL}}
\newcommand{\finmap}{\rightharpoonup_{\scriptscriptstyle\mathit{fin}}}
\newcommand{\rank}{\mathfrak{K}}
\newcommand{\heap}{\mathsf{h}}

\newcommand{\slstore}{\overline{\store}}
\newcommand{\slsid}{{\overline{\asid}}}
\newcommand{\slmodels}{\Vdash}
\newcommand{\slmodelsid}{\slmodels_{\scriptscriptstyle\slsid}}

\newcommand{\profile}[1]{\lambda_{\scriptscriptstyle #1}}

\newcommand{\pureform}{\pi}
\newcommand{\ppureform}{\overline{\pi}}

\newcommand{\pos}[3]{\mathrm{pos}({#1},{#2},{#3})}
\newcommand{\ipos}[2]{{\mathrm{inter}}({#1},{#2})}
\newcommand{\xipos}[2]{\mathcal{Z}_{#1}({#2})}
\newcommand{\spos}[1]{{\mathrm{state}}({#1})}
\newcommand{\gaifman}[1]{\mathbb{G}({#1})}
\newcommand{\gaifimg}[1]{\eta({#1})}
\newcommand{\intermap}{\iota}
\newcommand{\annotate}[2]{\overline{#2}^{#1}_{\intermap^{#1}}}
\newcommand{\xannot}[3]{\overline{#3}^{#1}_{#2}}
\newcommand{\xiatoms}[3]{\mathcal{I}_{#1}^{#2}({#3})}
\newcommand{\xituples}[3]{\mathsf{Tuples}_{#1}^{#2}({#3})}
\newcommand{\compstate}[2]{\mathsf{CompStates}_{#1}({#2})}
\newcommand{\interparam}[2]{\mathsf{InterAtoms}_{#1}({#2})}
\newcommand{\poly}[1]{\mathsf{poly}\left({#1}\right)}

\newcommand{\asid}{\Delta}
\newcommand{\modelsid}{\models_{\scriptscriptstyle\asid}}

\newcommand{\arule}{\mathsf{r}}
\newcommand{\ring}[2]{\predname{ring}_{{#1},{#2}}}
\newcommand{\chain}[2]{\predname{chain}_{{#1},{#2}}}

\newcommand{\size}[1]{\mathrm{size}({#1})}
\newcommand{\width}[1]{\mathrm{width}({#1})}

\newcommand{\defn}[2]{\mathrm{def}_{#1}({#2})}
\newcommand{\defs}[2]{\mathrm{def}^*_{#1}({#2})}
\newcommand{\defnof}[1]{\mathrm{def}({#1})}

\newcommand{\delete}{\predname{delete}}
\newcommand{\connect}{\predname{connect}}
\newcommand{\disconnect}{\predname{disconnect}}

\renewcommand{\mod}{~\mathrm{mod}~}

\newcommand{\hoare}[3]{\{ {#1} \} ~\mathsf{#2}~ \{ {#3} \}}

\newcommand{\lang}[1]{\mathcal{L}({#1})}

\newcommand{\proj}[2]{{#1}\!\!\downarrow_{\scriptscriptstyle{#2}}}

\newcommand{\toktoken}{\mathsf{T}}
\newcommand{\toknotok}{\mathsf{H}}

\newcommand{\tokin}{\textit{in}}
\newcommand{\tokout}{\textit{out}}

\newcommand{\codesize}{\footnotesize}
\newcommand{\boxaround}[1]{\ovalbox{${#1}$}}

\newcommand{\closureof}[1]{\mathrm{cl}({#1})}
\newcommand{\constrof}[1]{\mathrm{dist}({#1})}
\newcommand{\basecomps}{\comps^\sharp}
\newcommand{\pbasecomps}{\overline{\comps}^\sharp}
\newcommand{\baseinteracs}{\interacs^\sharp}
\newcommand{\pbaseinteracs}{\overline{\interacs}^\sharp}
\newcommand{\basetuple}{(\basecomps, \baseinteracs, \pureform)}
\newcommand{\basetuplen}[1]{(\basecomps_{#1}, \baseinteracs_{#1}, \pureform_{#1})}
\newcommand{\pbasetuplen}[1]{(\pbasecomps_{#1}, \pbaseinteracs_{#1}, \ppureform_{#1})}
\newcommand{\satbasetuples}{\mathsf{SatBase}}
\newcommand{\abasetuple}{\mathfrak{t}}
\newcommand{\anbasetuple}{\mathfrak{u}}

\newcommand{\basetupleof}[2]{\mathsf{Base}({#1},{#2})}

\newcommand{\basecomp}{\otimes}

\newcommand{\Basecomp}{\bigotimes}
\newcommand{\basesid}{\asid^\sharp}

\newcommand{\basevartuple}{\overrightarrow{\mathcal{X}}}
\newcommand{\basevarof}[1]{\mathcal{X}({#1})}
\newcommand{\leastbase}{\mu\basevartuple.\basesid}

\newcommand{\leastbaseof}[1]{\leastbase({#1})}

\newcommand{\reprof}[3]{{\{\!\!\{{#1}\}\!\!\}}^{\scriptscriptstyle{#2}}_{#3}}
\newcommand{\nordsubsets}[2]{S_{#1,#2}}
\newcommand{\polynomial}{\mathit{poly}}
\newcommand{\looseof}[1]{\widetilde{#1}}

\newcommand{\squig}{$\scriptsize$\sim$\normalsize$\!}

\newcommand{\rsquigend}{$\scriptsize\rule{.1ex}{0ex}$\rhd$\normalsize$}

\newcounter{index}

\newcommand\squigs[1]{%
  \setcounter{index}{0}%
  \whiledo {\value{index}< #1}
  {\addtocounter{index}{1}\squig}
}

\newcommand\rsquigarrow[2]{$
  \setbox0\hbox{$\squigs{#2}\rsquigend$}%
  \tiny$%
  \!\!\!\!\begin{array}{c}%
  {#1}\\%
  \usebox0%
  \end{array}%
  $\normalsize$\!\!%
}

\newcommand{\depof}[2]{\raisebox{4pt}{$~\rsquigarrow{({#1},{#2})}{4}~$}}
\newcommand{\Depof}[2]{\raisebox{4pt}{$~\rsquigarrow{({#1},{#2})}{8}~$}}

\newcommand{\graphof}[1]{\mathcal{G}({#1})}

\newcommand{\unfold}[2]{\Arrow{#1}{#2}}

\newcommand{\degreebound}{\mathfrak{B}}
\newcommand{\interactionvars}[1]{\mathrm{iv}({#1})}

\lstdefinelanguage{JavaScript}{
  keywords={typeof, true, false, catch, function, return, null, catch, switch, var, if, in, while, do, od, else, case, break, when, with, assume},
  ndkeywords={class, export, boolean, throw, implements, import, this},
  sensitive=false,
  comment=[l]{//},
  morecomment=[s]{/*}{*/},
  morecomment=[s]{$}{$},
  morestring=[b]',
  morestring=[b]"
}
\lstnewenvironment{code}[1][]{%
    \lstset{%
	    basicstyle=\codesize\ttfamily, 
	    breaklines=false,
	    commentstyle=\color[HTML]{073642}\ttfamily\itshape,
	    identifierstyle=\color{black},
	    inputencoding=latin1,
	    keywordstyle=\color[HTML]{228B22}\bfseries,
	    language=JavaScript,
	    ndkeywordstyle=\color[HTML]{228B22}\bfseries,
	    numbers=left,
	    numbersep=5pt,
	    xleftmargin=30pt,
	    xrightmargin=10pt,
	    frame=Tb,
	    framexleftmargin=15pt,
	    numberstyle=\scriptsize, 
	    prebreak = \raisebox{0ex}[0ex][0ex]{\ensuremath{\hookleftarrow}},
	    showstringspaces=false,
	    stringstyle=\color[rgb]{0.639,0.082,0.082}\ttfamily,
	    upquote=true,
	    mathescape=true,
    	escapebegin=\color[HTML]{073642},
	    tabsize=2,
	    #1
    }
}{}

\lstnewenvironment{examplecode}[1][]{%
    \lstset{%
      basicstyle=\codesize\ttfamily, 
      breaklines=false,
      commentstyle=\color[HTML]{073642}\ttfamily\itshape,
      identifierstyle=\color{black},
      inputencoding=latin1,
      keywordstyle=\color[HTML]{228B22}\bfseries,
      language=JavaScript,
      ndkeywordstyle=\color[HTML]{228B22}\bfseries,
      numbers=left,
      numbersep=5pt,
      xleftmargin=75pt,
      xrightmargin=60pt,
      frame=Tb,
      framexleftmargin=15pt,
      numberstyle=\scriptsize, 
      prebreak = \raisebox{0ex}[0ex][0ex]{\ensuremath{\hookleftarrow}},
      showstringspaces=false,
      stringstyle=\color[rgb]{0.639,0.082,0.082}\ttfamily,
      upquote=true,
      mathescape=true,
      escapebegin=\color[HTML]{073642},
      tabsize=2,
      #1
    }
}{}

\begin{document}

\title{Decision Problems in a Logic for Reasoning about Reconfigurable
  Distributed Systems}

\author{Marius Bozga\inst{1} \and Lucas Bueri\inst{1} \and Radu Iosif\inst{1}}
\institute{Univ. Grenoble Alpes, CNRS, Grenoble INP, VERIMAG, 38000, France}

\maketitle

\begin{abstract}
  We consider a logic used to describe sets of configurations of
  distributed systems, whose network topologies can be changed at
  runtime, by reconfiguration programs. The logic uses inductive
  definitions to describe networks with an unbounded number of
  components and interactions, written using a multiplicative
  conjunction, reminiscent of Bunched Implications \cite{PymOHearn99}
  and Separation Logic \cite{Reynolds02}. We study the complexity of
  the satisfiability and entailment problems for the configuration
  logic under consideration. Additionally, we consider \ifLongVersion
  robustness properties, such as tightness (are all interactions
  entirely connected to components?) and \fi degree boundedness (is every component involved in a bounded
  number of interactions?), \ifLongVersion the latter being \fi an
  ingredient for decidability of entailments.
\end{abstract}

\section{Introduction}

Distributed systems are increasingly used as critical parts of the
infrastructure of our digital society, as in e.g., datacenters,
e-banking and social networking. In order to address maintenance
(e.g., replacement of faulty and obsolete network nodes by new ones)
and data traffic issues (e.g., managing the traffic inside a
datacenter \cite{DBLP:journals/comsur/Noormohammadpour18}), the
distributed systems community has recently put massive effort in
designing algorithms for \emph{reconfigurable systems}, whose network
topologies change at runtime
\cite{DBLP:journals/sigact/FoersterS19}. However, dynamic
reconfiguration is an important souce of bugs that may result in e.g.,
denial of services or even data corruption\footnote{
\url{https://status.cloud.google.com/incident/appengine/19007}}.

This paper contributes to a logical framework that addresses the
timely problems of formal \emph{modeling} and \emph{verification} of
reconfigurable distributed systems. The basic building blocks of this
framework are \begin{inparaenum}[(i)]
\item a Hoare-style program proof calculus
  \cite{AhrensBozgaIosifKatoen21} used to write formal proofs of
  correctness of reconfiguration programs, and
\item an invariant synthesis method \cite{BozgaIosifSifakis21} that
  proves the safety (i.e., absence of reachable error configurations)
  of the configurations defined by the assertions that annotate a
  reconfiguration program.
\end{inparaenum}
These methods are combined to prove that an initially correct
distributed system cannot reach an error state, following the
execution of a given reconfiguration sequence.

The assertions of the proof calculus are written in a logic that
defines infinite sets of configurations, consisting of
\emph{components} (i.e., processes running on different nodes of the
network) connected by \emph{interactions} (i.e., multi-party channels
alongside which messages between components are transfered). Systems
that share the same architectural style (e.g., pipeline, ring, star,
tree, etc.) and differ by the number of components and interactions
are described using inductively defined predicates. Such
configurations can be modified either by \begin{inparaenum}[(a)]
\item\label{it:reconfiguration} adding or removing components and
  interactions (reconfiguration), or
\item\label{it:havoc} changing the local states of components, by
  firing interactions.
\end{inparaenum}

The assertion logic views components and interactions as
\emph{resources}, that can be created or deleted, in the spirit of
resource logics \emph{\`{a} la} Bunched Implications
\cite{PymOHearn99}, or Separation Logic \cite{Reynolds02}. The main
advantage of using resource logics is their support for \emph{local
reasoning} \cite{CalcagnoOHearnYan07}: reconfiguration actions are
specified by pre- and postconditions mentioning only the resources
involved, while framing out the rest of the configuration.

The price to pay for this expressive power is the difficulty of
automating the reasoning in these logics. This paper makes several
contributions in the direction of proof automation, by studying the
complexity of the \emph{satisfiability} and \emph{entailment}
problems, for the configuration logic under
consideration. Additionally, we study the complexity of \ifLongVersion
two robustness properties \cite{JansenKatelaanMathejaNollZuleger17},
namely \emph{tightness} (are all interactions entirely connected to
components?) and \fi \emph{degree
  boundedness} (is every component involved in a bounded number of
interactions?). In particular, the latter problem is used as a
prerequisite for defining a fragment with a decidable entailment
problem.

\subsection{Motivating Example}
\label{sec:running-example}

\begin{figure}[t!]
  \begin{center}
    \begin{minipage}{.55\textwidth}
      \hspace*{-5mm}\centerline{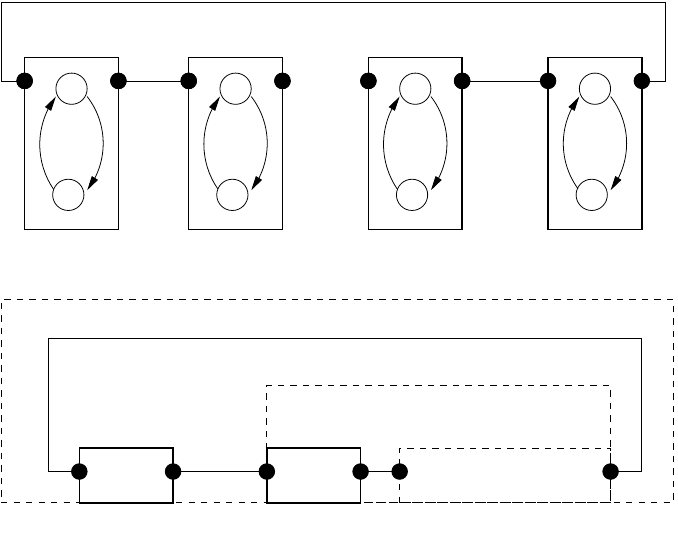}
    \end{minipage}
    \begin{minipage}{.44\textwidth}
      {\begin{lstlisting}
$\set{\ring{h}{t}(y)}$ /*assuming $h\geq2,~t\geq1$*/
$\set{\boxaround{\compactin{y}{\toknotok} * \interac{y}{out}{z}{in} * \chain{h-1}{t}(z,x)} * \interac{x}{out}{y}{in}}$ 
disconnect(x.$\mathit{out}$, y.$\mathit{in}$);
$\set{\boxaround{\compactin{y}{\toknotok}} * \interac{y}{out}{z}{in} * \boxaround{\chain{h-t}{t}(z,x)}}$ 
disconnect(y.$\mathit{out}$, z.$\mathit{in}$); 
$\set{\compactin{y}{\toknotok} * \boxaround{\chain{h-1}{t}(z,x)}}$ 
delete(y); 
$\set{\boxaround{\chain{h-1}{t}(z,x)}}$ 
connect(x.$\mathit{out}$, z.$\mathit{in}$)
$\set{\chain{h-1}{t}(z,x) * \interac{x}{out}{z}{in}}$
$\set{\ring{h-1}{t}(z)}$
      \end{lstlisting}}
      \vspace*{-.7\baselineskip}
      \centerline{\footnotesize(c)}
    \end{minipage}
  \end{center}
  \vspace*{-\baselineskip}
  \caption{Inductive Specification and Reconfiguration of a Token Ring}
  \vspace*{-2\baselineskip}
  \label{fig:ring}
\end{figure}

The logic studied in this paper is motivated by the need for an
assertion language that supports reasoning about dynamic
reconfigurations in a distributed system. For instance, consider a
distributed system consisting of a finite (but unknown) number of
\emph{components} (processes) placed in a ring, executing the same
finite-state program and communicating via \emph{interactions} that
connect the \emph{out} port of a component to the \emph{in} port of
its right neighbour, in a round-robin fashion, as in
Fig. \ref{fig:ring} (a). The behavior of a component is a machine with
two states, $\toktoken$ and $\toknotok$, denoting whether the
component has a token ($\toktoken$) or not ($\toknotok$). A component
$c_i$ without a token may receive one, by executing a transition
$\toknotok \arrow{\mathit{in}}{} \toktoken$, simultaneously with its
left neighbour $c_j$, that executes the transition transition
$\toktoken \arrow{\mathit{out}}{} \toknotok$. Then, we say that the
interaction $(c_j, \mathit{out}, c_i, \mathit{in})$ has fired, moving
a token one position to the right in the ring. Note that there can be
more than one token, moving independently in the system, as long as no
token overtakes another token.

The token ring system is formally specified by the following inductive
rules:
\begin{align*}
\ring{h}{t}(x) & \leftarrow \exists y \exists z ~.~ \compactin{x}{q} * \interactwo{x}{out}{z}{in} * \chain{h'}{t'}(z,y) * \interactwo{y}{out}{x}{in} \\
\chain{h}{t}(x,y) & \leftarrow \exists z.~\compactin{x}{q} * \interac{x}{out}{z}{in} * \chain{h'}{t'}(z,y) \\
\chain{0}{1}(x,x) & \leftarrow \compactin{x}{\toktoken} \hspace*{8mm}
\chain{1}{0}(x,x) \leftarrow \compactin{x}{\toknotok} \hspace*{8mm}
\chain{0}{0}(x,x) \leftarrow \compact{x} \\
\text{where } h' & \isdef \left\{\begin{array}{ll} \max(h-1,0) & \text{, if } q = \toknotok \\
h & \text{, if } q = \toktoken \end{array}\right. \text{ and }
t' \isdef \left\{\begin{array}{ll} \max(t-1,0) & \text{, if } q = \toktoken \\
t & \text{, if } q = \toknotok \end{array}\right. 
\end{align*}
The predicate $\ring{h}{t}(x)$ describes a ring with at least two
components, such that at least $h$ (resp. $t$) components are in state
$\toknotok$ (resp. $\toktoken$). The ring consists of a component $x$
in state $q$, described by the formula $\compactin{x}{q}$, an
interaction from the $\mathit{out}$ port of $x$ to the $\mathit{in}$
port of another component $z$, described as
$\interactwo{x}{out}{z}{in}$, a separate chain of components
stretching from $z$ to $y$ ($\chain{h'}{t'}(z,y)$), and an interaction
connecting the $\mathit{out}$ port of component $y$ to the
$\mathit{in}$ port of component $x$
($\interactwo{y}{out}{x}{in}$). Inductively, a chain consists of a
component $\compactin{x}{q}$, an interaction
$\interactwo{x}{out}{z}{in}$ and a separate
$\chain{h'}{t'}(z,y)$. Fig. \ref{fig:ring} (b) depicts the unfolding
of the inductive definition of the token ring, with the existentially
quantified variables $z$ from the above rules $\alpha$-renamed to
$z^1, z^2, \ldots$ to avoid confusion. 

A \emph{reconfiguration program} takes as input a mapping of program
variables to components and executes a sequence of \emph{basic
  operations} i.e., component/interaction creation/deletion, involving
the components and interactions denoted by these variables. For
instance, the reconfiguration program in Fig. \ref{fig:ring} (c) takes
as input three adjacent components, mapped to the variables
$\mathtt{x}$, $\mathtt{y}$ and $\mathtt{z}$, respectively, removes the
component $\mathtt{y}$ together with its left and right interactions
and reconnects $\mathtt{x}$ directly with $\mathtt{z}$.  Programming
reconfigurations is error-prone, because the interleaving between
reconfiguration actions and interactions in a distributed system may
lead to bugs that are hard to trace. For instance, if a
reconfiguration program removes the last component in state
$\toktoken$ (resp. $\toknotok$) from the system, no token transfer
interaction may fire and the system deadlocks.

We prove absence of such errors using a Hoare-style proof system
\cite{AhrensBozgaIosifKatoen21}, based on the above logic as assertion
language. For instance, the proof from Fig.  \ref{fig:ring} (c) shows
that the reconfiguration sequence applied to a component $\mathtt{y}$
in state $\toknotok$ (i.e., $\compactin{y}{\toknotok}$) in a ring with
at least $h\geq2$ components in state $\toknotok $ and at least
$t\geq1$ components in state $\toktoken$ leads to a ring with at least
$h-1$ components in state $\toknotok$ and at least $t$ in state
$\toktoken$; note that the states of the components may change during
the execution of the reconfiguration program, as tokens are moved by
interactions.

\ifLongVersion
For reasons of proof scalability, a basic operation is specified only
with regard to the components and interactions required to avoid
faulting. For instance $\hoare{\compactin{x}{q}}{\delete(x)}{\emp}$
(resp. $\hoare{\interactwo{x}{out}{y}{in}}{\disconnect(\mathtt{x}.\mathit{out},
  \mathtt{y}.\mathit{in})}{\emp}$) means that $\delete$
(resp. $\disconnect$) requires a component (resp. interaction)
and returns an empty configuration, whereas
$\hoare{\emp}{\connect(\mathtt{x}.\mathit{out},\mathtt{y}.\mathit{in})}{\interactwo{x}{out}{y}{in}}$
means that $\connect$ requires nothing and creates an interaction
between the given ports of the given components. These local
specifications are plugged into a context described by a frame formula
$F$, using the \emph{frame rule} $\hoare{\phi}{\predname{P}}{\psi}
\Rightarrow \hoare{\phi*\boxaround{F}}{\predname{P}}{\psi*F}$; for
readability, the frame formul{\ae} (from the preconditions of the
conclusion of the frame rule applications) are enclosed in boxes.
\fi

The proof also uses the \emph{consequence rule} $\hoare{\phi}{P}{\psi}
\Rightarrow \hoare{\phi'}{P}{\psi'}$ that applies if $\phi'$ is
stronger than $\phi$ and $\psi'$ is weaker than $\psi$. The side
conditions of the consequence rule require checking the validity of
the entailments $\ring{h}{t}(y) \models \exists x \exists z ~.~
\interac{x}{out}{y}{in} * \compactin{y}{\toknotok} *
\interac{y}{out}{z}{in} * \chain{h-1}{t}(z,x)$ and
$\chain{h-1}{t}(z,x) * \interac{x}{out}{z}{in} \models
\ring{h-1}{t}(z)$, for all $h\geq2$ and $t\geq1$. These side
conditions can be automatically discharged using the results on the
decidability of entailments given in this paper. Additionally,
checking the satisfiability of a precondition is used to detect
trivially valid Hoare triples.

\subsection{Related Work}
\label{sec:related-work}

Formal modeling coordinating architectures of component-based systems
has received lots of attention, with the development of architecture
description languages (ADL), such as BIP \cite{basu2006modeling} or
REO \cite{Arbab04}. Many such ADLs have extensions that describe
programmed reconfiguration, e.g.,
\cite{DR-BIP-STTT,KrauseMaraikarLazovikArbab11}, classified according
to the underlying formalism used to define their operational
semantics: \emph{process algebras}
\cite{CavalcanteBO15,magee1996dynamic}, \emph{graph rewriting}
\cite{taentzer1998dynamic,DBLP:journals/scp/WermelingerF02,LeMetayer},
\emph{chemical reactions} \cite{wermelinger1998towards} (see the
surveys
\cite{bradbury2004survey,rumpe2017classification}). Unfortunately,
only few ADLs support formal verification, mainly in the flavour of
runtime verification
\cite{BucchiaroneG08,DormoyKL10,LanoixDK11,DBLP:conf/sac/El-HokayemBS21}
or finite-state model checking \cite{Clarke08}.

Parameterized verification of unbounded networks of distributed
processes uses mostly hard-coded coordinating architectures (see
\cite{BloemJacobsKhalimovKonnovRubinVeithWidder15} for a survey). A
first attempt at specifying architectures by logic is the
\emph{interaction logic} of Konnov et al. \cite{KonnovKWVBS16}, a
combination of Presburger arithmetic with monadic uninterpreted
function symbols, that can describe cliques, stars and rings. More
structured architectures (pipelines and trees) can be described using
a second-order extension \cite{MavridouBBS17}. However, these
interaction logics are undecidable and lack support for automated
reasoning.

Specifying parameterized component-based systems by inductive
definitions is not new. \emph{Network grammars}
\cite{ShtadlerGrumberg89,LeMetayer,Hirsch} use context-free grammar
rules to describe systems with linear (pipeline, token-ring)
architectures obtained by composition of an unbounded number of
processes. In contrast, we use predicates of unrestricted arities to
describe architectural styles that are, in general, more complex than
trees. Moreover, we write inductive definitions using a resource
logic, suitable also for writing Hoare logic proofs of reconfiguration
programs, based on local reasoning \cite{CalcagnoOHearnYan07}.

Local reasoning about concurrent programs has been traditionally the
focus of Concurrent Separation Logic (\csl), based on a parallel
composition rule \cite{DBLP:journals/tcs/OHearn07}, initially with a
non-interfering (race-free) semantics \cite{Brookes:2016} and later
combining ideas of assume- and rely-guarantee
\cite{Owicki1978,DBLP:phd/ethos/Jones81} with local reasoning
\cite{FengFerreiraShao07,Vafeiadis07} and abstract notions of framing
\cite{Dinsdale-Young10,Dinsdale-Young13,Farka21}. However, the body of
work on CSL deals almost entirely with shared-memory multithreading
programs, instead of distributed systems, which is the aim of our
work. In contrast, we develop a resource logic in which the processes
do not just share and own resources, but become mutable resources
themselves.

The techniques developed in this paper are inspired by existing
techniques for similar problems in the context of Separation Logic
(\seplog) \cite{Reynolds02}. For instance, we use an abstract domain
similar to the one defined by Brotherston et
al. \cite{DBLP:conf/csl/BrotherstonFPG14} for checking satisfiability
of symbolic heaps in \seplog\ and reduce a fragment of the entailment
problem in our logic to \seplog\ entailment
\cite{EchenimIosifPeltier21}. In particular, the use of existing
automated reasoning techniques for \seplog\ has pointed out several
differences between the expressiveness of our logic and that of
\seplog. First, the configuration logic describes hypergraph
structures, in which edges are $\ell$-tuples for $\ell\geq2$, instead
of directed graphs as in \seplog, where $\ell$ is a parameter of the
problem: considering $\ell$ to be a constant strictly decreases the
complexity of the problem. Second, the degree (number of hyperedges
containing a given vertex) is unbounded, unlike in \seplog, where the
degree of heaps is constant. Therefore, we dedicate an entire section
(\S\ref{sec:boundedness}) to the problem of deciding the existence of
a bound (and computing a cut-off) on the degree of the models of a
formula, used as a prerequisite for the encoding of the entailment
problems from the configuration logic as \seplog\ entailments.

\section{Definitions}
\label{sec:definitions}

We denote by $\nat$ the set of positive integers. For a set $A$, we
define $A^1 \isdef A$, $A^{i+1} \isdef A^i \times A$, for all $i \geq
0$, and $A^+ = \bigcup_{i\geq1} A^i$, where $\times$ denotes the
Cartesian product. We denote by $\pow{A}$ the powerset of $A$ and by
$\mpow{A}$ the power-multiset (set of multisets) of $A$. The
cardinality of a finite set $A$ is denoted as $\cardof{A}$. By writing
$A \finsubseteq B$ we mean that $A$ is a finite subset of $B$. Given
integers $i$ and $j$, we write $\interv{i}{j}$ for the set
$\set{i,i+1,\ldots,j}$, assumed to be empty if $i>j$. For a tuple
$\vec{t} = \tuple{t_1, \ldots, t_n}$, we define $\lenof{\vec{t}}
\isdef n$, $\at{\vec{t}}{i} \isdef t_i$ and
$\at{\vec{t}}{\interv{i}{j}} \isdef \tuple{t_i, \ldots, t_j}$. By
writing $x=\polynomial(y)$, for given $x,y\in\nat$, we mean that there
exists a polynomial function $f : \nat \rightarrow \nat$, such that $x
\leq f(y)$.

\subsection{Configurations}

We model distributed systems as hypergraphs, whose vertices are
\emph{components} (i.e., the nodes of the network) and hyperedges are
\emph{interactions} (i.e., describing the way the components
communicate with each other). The components are taken from a
countably infinite set $\universe$, called the \emph{universe}. We
consider that each component executes its own copy of the same
\emph{behavior}, represented as a finite-state machine
$\beh=(\ports,\states,\arrow{}{})$, where $\ports$ is a finite set of
\emph{ports}, $\states$ is a finite set of \emph{states} and
$\arrow{}{} \subseteq \states \times \ports \times \states$ is a
transition relation. Intuitively, each transition $q \arrow{p}{} q'$
of the behavior is triggerred by a visible event, represented by the
port $p$. For instance, the behavior of the components of the token
ring system from Fig. \ref{fig:ring} (a) is
$\beh=(\set{\mathit{in},\mathit{out}}, \set{\toknotok,\toktoken},
\set{\toknotok\arrow{\mathit{in}}{}\toktoken,
  \toktoken\arrow{\mathit{out}}{}\toknotok})$. The universe
$\universe$ and the behavior $\beh=(\ports,\states,\arrow{}{})$ are
considered fixed in the rest of this paper.

We introduce a logic for describing infinite sets of
\emph{configurations} of distributed systems with unboundedly many
components and interactions. A configuration is a snapshot of the
system, describing the topology of the network (i.e., the set of
present components and interactions) together with the local state of
each component:

\begin{definition}\label{def:configuration}
  A \emph{configuration} is a tuple $\aconfig = (\comps,\interacs,
  \statemap)$, where: \begin{compactitem}
  \item $\comps \finsubseteq \universe$ is a finite set of \emph{components},
    that are present in the configuration,
  \item $\interacs \finsubseteq (\universe\times\ports)^+$ is a finite
    set of \emph{interactions}, where each interaction is a sequence
    $(c_1, p_1, \ldots, c_n, p_n) \in (\universe \times \ports)^n$
    that binds together the ports $p_1, \ldots, p_n$ of the pairwise
    distinct components $c_1, \ldots, c_n$, respectively.
  \item $\statemap : \universe \rightarrow \states$ is a \emph{state
    map} associating each (possibly absent) component, a state of the
    behavior $\beh$, such that the set $\set{c \in \universe \mid
      \statemap(c) = q}$ is infinite, for each $q \in \states$.
  \end{compactitem}
\end{definition}
The last condition requires that there is an infinite pool of
components in each state $q \in \states$; since $\universe$ is
infinite and $\states$ is finite, this condition is feasible. For
example, the configurations of the token ring from Fig. \ref{fig:ring}
(a) are $(\{c_1, \ldots, c_n\}, \{(c_i,\mathit{out},c_{(i \mod n) +
  1},\mathit{in}) \mid i \in \interv{1}{n}\}, \statemap)$, where
$\statemap:\universe\rightarrow\set{\toknotok,\toktoken}$ is a state
map. The ring topology is described by the set of components $\{c_1,
\ldots, c_n\}$ and interactions $\{(c_i,\mathit{out}, c_{(i \mod n) +
  1},\mathit{in}) \mid i \in \interv{1}{n}\}$.

Intuitively, an interaction $(c_1, p_1, \ldots, c_n, p_n)$
synchronizes transitions labeled by the ports $p_1, \ldots, p_n$ from
the behaviors (i.e., replicas of the state machine $\beh$) of $c_1,
\ldots, c_n$, respectively.  The interactions are classified according
to their sequence of ports, called the \emph{interaction type} and let
$\intertypes \isdef \ports^+$ be the set of interaction types; an
interaction type models, for instance, the passing of a certain kind
of message (e.g., request, acknowledgement, etc.).
From an operational point of view, two interactions that differ by a
permutation of indices e.g., $(c_1, p_1, \ldots, c_n, p_n)$ and
$(c_{i_1}, p_{i_1}, \ldots, c_{i_n}, p_{i_n})$ such that $\set{i_1,
  \ldots, i_n} = \interv{1}{n}$, are equivalent, since the set of
transitions is the same; nevertheless, we chose to distinguish them in
the following, exclusively for reasons of simplicity.

\ifLongVersion
Note that Def. \ref{def:configuration} allows configurations with
interactions that involve absent components (i.e., not from the set
$\comps$ of present components in the given configuration). The
following definition distinguishes such configurations:

\begin{definition}\label{def:tightness}
  Let $\aconfig = (\comps,\interacs,\statemap)$ be a configuration. An
  interaction $(c_1, p_1, \ldots, c_n, p_n)$ is \emph{loose} in
  $\aconfig$ if and only if $c_i \not\in \comps$, for some $i \in
  \interv{1}{n}$. If $\interacs$ contains at least one interaction
  that is loose in $\aconfig$, we say that $\aconfig$ is
  \emph{loose}. An interaction (resp. configuration) that is not loose
  is said to be \emph{tight}.
\end{definition}
For instance, every configuration of the system from
Fig. \ref{fig:ring} (a) is tight and becomes loose if a component is
deleted. Moreover, the reconfiguration program from
Fig. \ref{fig:ring} (c) manipulates tight configurations only. In
particular, loose configurations are useful for the definition of a
composition operation, as the union of disjoint sets of components and
interactions: \fi

\begin{definition}\label{def:composition}
  The composition of two configurations $\aconfig_i = (\comps_i,
  \interacs_i, \statemap)$, for $i = 1,2$, such that $\comps_1 \cap
  \comps_2 = \emptyset$ and $\interacs_1 \cap \interacs_2 =
  \emptyset$, is defined as $\aconfig_1 \comp \aconfig_2 \isdef
  (\comps_1 \cup \comps_2, \interacs_1 \cup \interacs_2,
  \statemap)$. The composition $\aconfig_1 \comp \aconfig_2$ is
  undefined if $\comps_1 \cap \comps_2 \neq \emptyset$ or $\interacs_1
  \cap \interacs_2 \neq \emptyset$.
\end{definition}
\ifLongVersion
Note that a tight configuration may be the result of composing two
loose configurations, whereas the composition of tight configurations
is always tight. The example below shows that, in most cases, a
non-trivial decomposition of a tight configuration necessarily
involves loose configurations: 

\begin{example}\label{ex:composition}
  Let $\aconfig_i = (\comps_i, \interacs_i, \statemap)$ be loose
  configurations, where $\comps_i = \set{c_i}$, $\interacs_i =
  \set{(c_i, \mathit{out}, c_{3-i}, \mathit{in})}$, for all $i =
  1,2$. Then $\aconfig \isdef \aconfig_1 \comp \aconfig_2$ is the
  tight configuration $\aconfig = (\set{c_1, c_2}, \set{(c_1,
    \mathit{out}, c_2, \mathit{in}), (c_2, \mathit{out}, c_1,
    \mathit{in})}, \statemap)$. The only way of decomposing $\aconfig$
  into two tight subconfigurations $\aconfig'_1$ and $\aconfig'_2$ is
  taking $\aconfig'_1 \isdef \aconfig$ and $\aconfig'_2 \isdef
  (\emptyset, \emptyset, \statemap)$, or
  viceversa. \hfill$\blacksquare$
\end{example}
\fi In analogy with graphs, the \emph{degree} of a configuration is
the maximum number of interactions from the configuration that involve
a (possibly absent) component:

\begin{definition}\label{def:degree}
  The \emph{degree} of a configuration $\aconfig = (\comps, \interacs,
  \statemap)$ is defined as $\degreeof{\aconfig} \isdef
  \max_{c\in\universe} \degreenode{c}{\aconfig}$, where
  $\degreenode{c}{\aconfig} \isdef \cardof{\{(c_1, p_1, \ldots, c_n,
    p_n) \in \interacs \mid c = c_i,~ i \in \interv{1}{n}\}}$.
\end{definition}
For instance, the configuration of the system from Fig. \ref{fig:ring}
(a) has degree two.

\subsection{Configuration Logic}

Let $\vars$ and $\preds$ be countably infinite sets of
\emph{variables} and \emph{predicates}, respectively. For each
predicate $\apred \in \preds$, we denote its arity by
$\arityof{\apred}$. The formul{\ae} of the \emph{Configuration Logic}
(\cl) are described inductively by the following syntax:
\[\phi := \emp \mid \compact{x} \mid \interacn{x_1}{p_1}{x_n}{p_n} \mid \compin{x}{q}
\mid x=y \mid x\neq y \mid \apred(x_1, \ldots, x_{\arityof{\apred}})
\mid \phi * \phi \mid \exists x ~.~ \phi\] where $x, y, x_1, \ldots
\in \vars$, $q \in \states$ and $\apred \in \preds$. A formula
$\compact{x}$, $\interacn{x_1}{p_1}{x_n}{p_n}$, $\compin{x}{q}$ and
$\apred(x_1, \ldots, x_{\arityof{\apred}})$ is called a
\emph{component}, \emph{interaction}, \emph{state} and
\emph{predicate} atom, respectively. Sometimes, we use the shorthand
$\compactin{x}{q} \isdef \compact{x} * \compin{x}{q}$. Intuitively,
the formula $\compactin{x}{q} * \compactin{y}{q'} *
\interactwo{x}{out}{y}{in} * \interactwo{x}{in}{y}{out}$ describes a
configuration consisting of two distinct components, denoted by the
values of $x$ and $y$, in states $q$ and $q'$, respectively, and two
interactions binding the $\mathit{out}$ port of one to the
$\mathit{in}$ port of the other component. \ifLongVersion For
instance, $\aconfig = \aconfig_1 \comp \aconfig_2$ from Example
\ref{ex:composition} is such a configuration. \fi

A formula is said to be \emph{pure} if and only if it consists of
state atoms, equalities and disequalities. A formula with no
occurrences of predicate atoms (resp. existential quantifiers) is
called \emph{predicate-free} (resp. \emph{quantifier-free}). A
variable is \emph{free} if it does not occur within the scope of an
existential quantifier and let $\fv{\phi}$ be the set of free
variables of $\phi$. A \emph{sentence} is a formula with no free
variables. A \emph{substitution} $\phi[x_1/y_1 \ldots x_n/y_n]$
replaces simultaneously every free occurrence of $x_i$ by $y_i$ in
$\phi$, for all $i \in \interv{1}{n}$. Before defining the semantics
of \cl\ formul{\ae}, we introduce the set of inductive definitions
that assigns meaning to predicates:

\begin{definition}\label{def:sid}
  A \emph{set of inductive definitions (SID)} $\asid$ consists of
  \emph{rules} $\apred(x_1, \ldots, x_{\arityof{\apred}}) \leftarrow
  \phi$, where $x_1, \ldots, x_{\arityof{\apred}}$ are pairwise
  distinct variables, called \emph{parameters}, such that $\fv{\phi}
  \subseteq \set{x_1, \ldots, x_{\arityof{\apred}}}$. The rule
  $\apred(x_1, \ldots, x_{\arityof{\apred}}) \leftarrow \phi$
  \emph{defines} $\apred$ and we denote by $\defn{\asid}{\apred}$ the
  set of rules from $\asid$ that define $\apred$.
\end{definition}
Note that having distinct parameters in a rule is without loss of
generality, as e.g., a rule $\apred(x_1, x_1) \leftarrow \phi$ can be
equivalently written as $\apred(x_1, x_2) \leftarrow x_1 = x_2 *
\phi$. As a convention, we shall always use the names $x_1, \ldots,
x_{\arityof{\apred}}$ for the parameters of a rule that defines
$\apred$.

The semantics of \cl\ formul{\ae} is defined by a satisfaction
relation $\aconfig \models^\store_\asid \phi$ between configurations
and formul{\ae}. This relation is parameterized by a \emph{store}
$\store : \vars \rightarrow \universe$ mapping the free variables of a
formula into components from the universe (possibly absent from
$\aconfig$) and an SID $\asid$. We write $\store[x \leftarrow c]$ for
the store that maps $x$ into $c$ and agrees with $\store$ on all
variables other than $x$. The definition of the satisfaction relation
is by induction on the structure of formul{\ae}, where $\aconfig =
(\comps, \interacs, \statemap)$ is a configuration
(Def. \ref{def:configuration}):
\[\begin{array}{rclcl}
\aconfig & \models^\store_\asid & \emp & \iff & \comps = \emptyset \text{ and } \interacs = \emptyset \\
\aconfig & \models^\store_\asid & \compact{x} & \iff & \comps = \set{\store(x)} \text{ and } \interacs = \emptyset \\
\aconfig & \models^\store_\asid & \interacn{x_1}{p_1}{x_n}{p_n} & \iff & \comps = \emptyset \text{ and } \interacs = \set{(\nu(x_1), p_1, \ldots, \nu(x_n), p_n)} \\
\aconfig & \models^\store_\asid & \compin{x}{q} & \iff & \aconfig \models^\store_\asid \emp \text{ and } \statemap(\store(x)) = q \\
\aconfig & \models^\store_\asid & x \sim y & \iff & \aconfig \models^\store_\asid \emp \text{ and } \store(x)\sim\store(y) \text{, for all } \sim \in \set{=,\neq} \\
\aconfig & \models^\store_\asid & \apred(y_1, \ldots, y_{\arityof{\apred}}) & \iff & \aconfig \models^\store_\asid \phi[x_1/y_1, \ldots, x_{\arityof{\apred}}/y_{\arityof{\apred}}] \text{, for some rule } \\
&&&& \apred(x_1, \ldots, x_{\arityof{\apred}}) \leftarrow \phi \text{ from } \asid \\
\aconfig & \models^\store_\asid & \phi_1 * \phi_2 & \iff & \text{exist } \aconfig_1, \aconfig_2 \text{, such that } \aconfig = \aconfig_1 \comp \aconfig_2 \text{ and }
\aconfig_i \models^\store_\asid \phi_i \text{, for } i = 1,2 \\
\aconfig & \models^\store_\asid & \exists x ~.~ \phi & \iff & \aconfig \models^{\store[x \leftarrow c]}_\asid \phi \text{, for some } c \in \universe
\end{array}\]
If $\phi$ is a sentence, the satisfaction relation $\aconfig
\models^\store_\asid \phi$ does not depend on the store, written
$\aconfig \models_\asid \phi$, in which case we say that $\aconfig$ is
a \emph{model} of $\phi$. If $\phi$ is a predicate-free formula, the
satisfaction relation does not depend on the SID, written $\aconfig
\models^\store \phi$. A formula $\phi$ is \emph{satisfiable} if and
only if the sentence $\exists x_1 \ldots \exists x_n ~.~ \phi$ has a
model, where $\fv{\phi} = \set{x_1, \ldots, x_n}$. A formula $\phi$
\emph{entails} a formula $\psi$, written $\phi \models_\asid \psi$ if
and only if, for any configuration $\aconfig$ and store $\store$, we
have $\aconfig \models^\store_\asid \phi$ only if $\aconfig
\models^\store_\asid \psi$.

\subsection{Separation Logic}
\label{sec:sl}

Separation Logic (\seplog) \cite{Reynolds02} will be used in the
following to prove several technical results concerning the
decidability and complexity of certain decision problems for \cl. For
self-containment reasons, we define \seplog\ below. The syntax of
\seplog\ formul{\ae} is described by the following grammar:
\[\phi := \emp \mid x_0 \mapsto (x_1, \ldots, x_\rank) \mid x = y \mid x \neq y \mid \apred(x_1, \ldots, x_{\arityof{\apred}})
\mid \phi * \phi \mid \exists x ~.~ \phi\] where $x, y, x_0, x_1,
\ldots \in \vars$, $\apred \in \preds$ and $\rank \geq 1$ is an
integer constant. Formul{\ae} of \seplog\ are interpreted over finite
partial functions $\heap : \universe \finmap \universe^\rank$, called
\emph{heaps}\footnote{We use the universe $\universe$ here for
simplicity, the definition works with any countably infinite set.}, by
a satisfaction relation $\heap \slmodels^\store \phi$, defined
inductively as follows:
\[\begin{array}{rclcl}
\heap & \slmodels^\store_\asid & \emp & \iff & \heap = \emptyset \\
\heap & \slmodels^\store_\asid & x_0 \mapsto (x_1,\ldots,x_\rank) & \iff & \dom{\heap} = \set{\store(x_0)}
\text{ and } \heap(\store(x_0)) = \tuple{\store(x_1), \ldots, \store(x_\rank)} \\
\heap & \slmodels^\store & \phi_1 * \phi_2 & \iff & \text{there exist } \heap_1, \heap_2 \text{ such that }
\dom{\heap_1} \cap \dom{\heap_2} = \emptyset, \\
&&&& \heap = \heap_1 \cup \heap_2 \text{ and } \heap_i \slmodels^\store_\asid \phi_i \text{, for both } i = 1,2 \\
\end{array}\]
where $\dom{\heap} \isdef \set{c \in \universe \mid \heap(c) \text{ is
    defined}}$ is the domain of the heap and (dis-)equalities,
predicate atoms and existential quantifiers are defined same as for
\cl.

\subsection{Decision Problems}

We define the decision problems that are the focus of the upcoming
sections. As usual, a decision problem is a class of yes/no queries
that differ only in their input. In our case, the input consists of an
SID and one or two predicates, written between square brackets.

\begin{definition}\label{def:decision}
  We consider the following problems, for a SID $\asid$ and predicates
  $\apred, \bpred \in \preds$: \begin{enumerate}
  \item\label{decision:sat} $\sat{\asid}{\apred}$: is the sentence
    $\exists x_1 \ldots \exists x_{\arityof{\apred}} ~.~ \apred(x_1,
    \ldots, x_{\arityof{\apred}})$ satisfiable for $\asid$?
    \ifLongVersion
  \item\label{decision:tight} $\tight{\asid}{\apred}$: is every model
    $\aconfig$ of the sentence $\exists x_1 \ldots \exists
    x_{\arityof{\apred}} ~.~ \apred(x_1, \ldots,
    x_{\arityof{\apred}})$ a tight configuration?
    \fi
  \item\label{decision:bound} $\bound{\asid}{\apred}$: is the set
    $\set{\degreeof{\aconfig} \mid \aconfig \models_\asid \exists x_1
    \ldots \exists x_{\arityof{\apred}} ~.~ \apred(x_1, \ldots,
    x_{\arityof{\apred}})}$ finite?
    %
  \item\label{decision:entl} $\entl{\asid}{\apred}{\bpred}$: does
    $\apred(x_1, \ldots, x_{\arityof{\apred}}) \models_\asid \exists
    x_{\arityof{\bpred}+1} \ldots \exists x_{\arityof{\apred}} ~.~
    \bpred(x_1, \ldots, x_{\arityof{\bpred}})$ hold?
  \end{enumerate}
\end{definition}
We define the size of a formula $\phi$ as the total number of
occurrences of symbols needed to write it down, denoted by
$\sizeof{\phi}$. The size of a SID $\asid$ is $\sizeof{\asid} \isdef
\sum_{\apred(x_1, \ldots, x_{\arityof{\apred}}) \leftarrow \phi \in
  \asid} \sizeof{\phi} + \arityof{\apred} + 1$. Other parameters of a
SID $\asid$ are its: \begin{compactitem}
\item \emph{maximal arity}, denoted as $\maxarityof{\asid} \isdef
\max\set{\arityof{\apred} \mid \apred(x_1, \ldots,
  x_{\arityof{\apred}}) \leftarrow \phi \in \asid}$,
\item \emph{width}, denoted as $\maxwidthof{\asid} \isdef
  \max\set{\sizeof{\phi} \mid \apred(x_1, \ldots,
    x_{\arityof{\apred}}) \leftarrow \phi \in \asid}$,
\item \emph{maximal interaction size}, denoted as
  $\maxintersize{\asid} \isdef \max\{n \mid
  \interacn{x_1}{p_1}{x_n}{p_n} \text{ occurs in } \phi,
  \\ \apred(x_1, \ldots, x_{\arityof{\apred}}) \leftarrow \phi \in
  \asid\}$.
\end{compactitem}
For each decision problem $\prob{\asid}{\apred,\bpred}$, we consider
its $(k,\ell)$-bounded versions
$\klprob{\asid}{\apred,\bpred}{k}{\ell}$, obtained by restricting the
predicates and interaction atoms occurring $\asid$ to
$\maxarityof{\asid} \leq k$ and $\maxintersize{\asid} \leq \ell$,
respectively, where $k$ and $\ell$ are either positive integers or
infinity. We consider, for each $\prob{\asid}{\apred,\bpred}$, the
subproblems $\klprob{\asid}{\apred,\bpred}{k}{\ell}$ corresponding to
the three cases \begin{inparaenum}[(1)] \item $k<\infty$ and
  $\ell=\infty$, \item $k=\infty$ and $\ell<\infty$, and \item
  $k=\infty$ and $\ell=\infty$. \end{inparaenum} As we explain next,
this is because, for the decision problems considered
(Def. \ref{def:decision}), the complexity for the case $k<\infty,
\ell<\infty$ matches the one for the case $k<\infty, \ell=\infty$.

\ifLongVersion
Moreover, for each problem $\prob{\asid}{\apred}$
(resp. $\prob{\asid}{\apred,\bpred}$), we consider its general version
$\prob{\asid}{\phi}$ (resp. $\prob{\asid}{\phi,\psi}$), where $\phi$
and $\psi$ are \cl\ formul{\ae}, whose predicates are interpreted by
the rules in $\asid$. The generalized problems $\prob{\asid}{\phi}$
involving one predicate atom (points \ref{decision:sat} and
\ref{decision:bound} of Def. \ref{def:decision}) can be reduced to
their restricted versions $\prob{\asid}{\apred}$, by introducing a
fresh predicate $\apred_\phi$ (not occurring in $\asid$), of arity
$n\geq0$ and a rule $\apred_\phi(x_1, \ldots, x_n) \leftarrow \phi$,
where $\fv{\phi} = \set{x_1, \ldots, x_n}$. This reduction is linear
in the size of the input and changes none of the following complexity
results. Concerning the generalized entailment problem
$\entl{\asid}{\phi}{\psi}$, the reduction to the problem
$\entl{\asid}{\apred}{\bpred}$ (Def. \ref{def:decision}
\ref{decision:entl}) might affect its decidability status, which is
subject to syntactic restrictions on the rules in $\asid$ (details
will be given in \S\ref{sec:entailment}).

Satisfiability (\ref{decision:sat}) and entailment
(\ref{decision:entl}) arise naturally during verification of
reconfiguration programs. For instance, $\sat{\asid}{\phi}$ asks
whether a specification $\phi$ of a set configurations (e.g., a pre-,
post-condition, or a loop invariant) is empty or not (e.g., an empty
precondition typically denotes a vacuous verification condition),
whereas $\entl{\asid}{\phi}{\psi}$ is used as a side condition for the
Hoare rule of consequence, as in e.g., the proof from
Fig. \ref{fig:ring} (c). Moreover, entailments must be proved when
checking inductiveness of a user-provided loop invariant.

In contrast, the applications of the tightness (\ref{decision:tight})
and boundedness (\ref{decision:bound}) problems are less obvious and
require a few explanations. The $\tight{\asid}{\phi}$ problem is
relevant in the context of compositional verification of distributed
systems. Suppose we have a distributed system consisting of two
interacting subsystems, whose sets of initial configurations are
described by $\phi_1$ and $\phi_2$, respectively i.e., the initial
configurations of the system are described by $\phi_1 * \phi_2$. The
compositional verification of a reconfiguration program $\mathsf{P}$
reduces checking the validity of a Hoare triple $\hoare{\phi_1 *
  \phi_2}{P}{\psi_1 * \psi_2}$ to checking the validity of the simpler
$\hoare{\phi_i}{P}{\psi_i}$, for $i = 1,2$. Unfortunately, this
appealing method faces the problem of \emph{interference} between the
subsystems described by $\phi_1$ and $\phi_2$, namely the loose
interactions of $\phi_i$ might connect to present components of
$\phi_{3-i}$ and change their states during the execution. In this
case, it is sufficient to infer the sets of cross-boundary
interactions $\mathcal{F}_{i,3-i}$, describing those interactions from
$\phi_i$ that connect to components from $\phi_{3-i}$, and check the
validity of the triples $\hoare{\phi_i *
  \mathcal{F}_{3-i,i}}{P}{\psi_i * \mathcal{F}_{3-i,i}}$, under a
relaxed semantics which considers that the interactions in
$\mathcal{F}_{3-i,i}$ can fire anytime, or according to the order
described by some regular language. However, if
$\tight{\asid}{\phi_1}$ (resp. $\tight{\asid}{\phi_2}$) has a negative
answer, the set of cross-boundary interactions may be unbounded, hence
not representable by a finite separating conjunction of interaction
atoms $\mathcal{F}_{1,2}$ (resp. $\mathcal{F}_{2,1}$). Thus, the
tightness problem is important in establishing necessary conditions
under which a compositional proof rule can be applied to checking
correctness of reconfigurations in a distributed system.

The $\bound{\asid}{\phi}$ problem is used to check a necessary
condition for the decidability of entailments i.e.,
$\entl{\asid}{\phi}{\psi}$. If $\bound{\asid}{\phi}$ has a positive
answer, we can reduce the problem $\entl{\asid}{\phi}{\psi}$ to an
entailment problem for \seplog, which is always interpreted over heaps
of bounded degree \cite{EchenimIosifPeltier21}. Otherwise, the
decidability status of the entailment problem is open, for
configurations of unbounded degree, such as the one described by the
example below.

\begin{example}\label{ex:star}
  The following SID describes star topologies with a central
  controller connected to an unbounded number of workers stations:
\[\mathit{Star}(x) \leftarrow \compact{x} * \mathit{Worker}(x),~ 
\mathit{Worker}(x) \leftarrow \emp \mid \exists y ~.~ \interactwo{x}{out}{y}{in} * \compact{y} * \mathit{Worker}(x)~
\blacksquare\]
\end{example}
\fi

    
    



\section{Satisfiability}
\label{sec:satisfiability}

We show that the satisfiability problem (Def. \ref{def:decision},
point \ref{decision:sat}) is decidable, using a method similar to the
one pioneered by Brotherston et
al.~\cite{DBLP:conf/csl/BrotherstonFPG14}, for checking satisfiability
of inductively defined symbolic heaps in \seplog. We recall that a
formula $\pureform$ is \emph{pure} if and only if it is a separating
conjunction of equalities, disequalities and state atoms.

\begin{definition}\label{def:closure}
  The \emph{closure} $\closureof{\pureform}$ of a pure formula
  $\pureform$ is the limit of the sequence $\pureform^0, \pureform^1,
  \pureform^2, \ldots$ such that $\pureform^0 = \pureform$ and, for
  each $i \geq 0$, $\pureform^{i+1}$ is obtained by joining (with $*$)
  all of the following formul{\ae} to
  $\pureform^i$: \begin{compactitem}
  \item $x = z$, where $x$ and $z$ are the same variable, or $x = y$
    and $y = z$ both occur in $\pureform^i$,
  \item $x \neq z$, where $x = y$ and $y \neq z$ both occur in
    $\pureform^i$, or 
  \item $\compin{y}{q}$, where $\compin{x}{q}$ and $x = y$ both occur
    in $\pureform^i$.
  \end{compactitem}
\end{definition}
Because only finitely many such formul{\ae} can be added, the sequence
of pure formul{\ae} from Def. \ref{def:closure} is bound to stabilize
after polynomially many steps. A pure formula is satisfiable if and
only if its closure does not contain contradictory literals i.e., $x =
y$ and $x \neq y$, or $\compin{x}{q}$ and $\compin{x}{q'}$, for $q
\neq q' \in \states$. We write $x \formeq{\pureform} y$ (resp. $x
~\formneq{\pureform} y$) if and only if $x = y$ (resp. $x \neq y$)
occurs in $\closureof{\pureform}$ and $\notof{x \formeq{\pureform} y}$
(resp. $\notof{x ~\formneq{\pureform} y}$) whenever $x
\formeq{\pureform} y$ (resp. $x \formneq{\pureform} y$) does not hold.
Note that e.g., $\notof{x \formeq{\pureform} y}$ is not the same as $x
~\formneq{\pureform} y$.

\begin{myLemmaE}\label{lemma:pureform-sat}
  A pure formula $\pureform$ is satisfiable if and only if the
  following hold: \begin{enumerate}
  \item for all $x, y \in \fv{\pureform}$, $x=y$ and $x\neq y$ do not
    occur both in $\closureof{\pureform}$,
  \item for all $x \in \fv{\pureform}$ and $q \neq r \in \states$,
    $\compin{x}{q}$ and $\compin{x}{r}$ do not occur both in
    $\closureof{\pureform}$.
  \end{enumerate}
\end{myLemmaE}
\begin{proofE}
  A pure formula $\pureform$ is satisfiable if and only if there
  exists a store $\store$ and a configuration $(\emptyset, \emptyset,
  \statemap)$, such that $(\emptyset,\emptyset,\statemap)
  \models^\store \pureform$. ``$\Leftarrow$'' It is easy to see that
  $\formeq{\pureform}$ is an equivalence relation, for each pure
  formula $\pureform$. Given any state map $\statemap$, we define
  $\store$ by assigning each equivalence class of $\formeq{\pureform}$
  a distinct component $c$, such that $\statemap(c)=q$ if
  $\compin{y}{q}$ occurs in $\pureform$, for a variable $y$ in the
  class. By the conditions of the Lemma, $\statemap$ and $\store$ are
  well defined and we have $(\emptyset,\emptyset,\statemap)
  \models^\store \pureform$, by definition. ``$\Rightarrow$'' If
  $(\emptyset,\emptyset,\statemap) \models^\store \pureform$ then
  $(\emptyset,\emptyset,\statemap) \models^\store
  \closureof{\pureform}$, because each additional formula in
  $\closureof{\pureform}$ is a logical consequence of
  $\pureform$. Since $\closureof{\pureform}$ is satisfiable, the two
  conditions of the Lemma must hold. \qed
\end{proofE}
  
\emph{Base tuples} constitute the abstract domain used by the
algorithms for checking satisfiability (point \ref{decision:sat} of
Def. \ref{def:decision}) and boundedness (point \ref{decision:bound}
of Def. \ref{def:decision}), defined as follows:

\begin{definition}\label{def:base-tuple}
  A \emph{base tuple} is a triple $\abasetuple = (\basecomps, \baseinteracs,
  \pureform)$, where: \begin{compactitem}
  \item $\basecomps \in \mpow{\vars}$ is a multiset of variables
    denoting present components,
  \item $\baseinteracs : \intertypes \rightarrow \mpow{\vars^+}$ maps
    each interaction type $\intertype \in \intertypes$ into a multiset
    of tuples of variables of length $\lenof{\intertype}$ each, and
  \item $\pureform$ is a pure formula. 
  \end{compactitem}
  A base tuple is called \emph{satisfiable} if and only if $\pureform$ is
  satisfiable and the following hold: \begin{compactenum}
  \item\label{it1:base-tuple} for all $x,y \in \basecomps$, $\notof{x \formeq{\pureform}
    y}$,
  \item\label{it2:base-tuple} for all $\intertype \in \intertypes$, $\tuple{x_1, \ldots,
    x_{\lenof{\intertype}}}, \tuple{y_1, \ldots,
    y_{\lenof{\intertype}}} \in \baseinteracs(\intertype)$, there
    exists $i \in \interv{1}{\lenof{\intertype}}$ such that
    $\notof{x_i \formeq{\pureform} y_i}$,
  \item\label{it3:base-tuple} for all $\intertype \in \intertypes$,
    $\tuple{x_1, \ldots, x_{\arityof{\intertype}}} \in
    \baseinteracs(\intertype)$ and $1 \leq i < j \leq
    \lenof{\intertype}$, we have $\notof{x_i \formeq{\pureform} x_j}$.
  \end{compactenum}
  We denote by $\satbasetuples$ the set of satisfiable base tuples.
\end{definition}
Note that a base tuple $\basetuple$ is unsatisfiable if $\basecomps$
($\baseinteracs$) contains the same variable (tuple of variables)
twice (for the same interaction type), hence the use of multisets in
the definition of base tuples. It is easy to see that checking the
satisfiability of a given base tuple $\basetuple$ can be done in time
$\polynomial(\cardof{\basecomps}+\sum_{\intertype\in\intertypes}
\cardof{\baseinteracs(\intertype)}+\sizeof{\pureform})$.

We define a partial \emph{composition} operation on satisfiable base tuples, as follows:
\[
(\basecomps_1, \baseinteracs_1, \pureform_1) \basecomp (\basecomps_2,
\baseinteracs_2, \pureform_2) \isdef (\basecomps_1 \cup \basecomps_2,
\baseinteracs_1 \cup \baseinteracs_2, \pureform_1 * \pureform_2)
\]
where the union of multisets is lifted to functions $\intertypes
\rightarrow \mpow{\vars^+}$ in the usual way. The composition
operation $\basecomp$ is undefined if $(\basecomps_1, \baseinteracs_1,
\pureform_1) \basecomp (\basecomps_2, \baseinteracs_2, \pureform_2)$
is not satisfiable e.g., if $\basecomps_1 \cap \basecomps_2 \neq
\emptyset$, $\baseinteracs_1(\intertype) \cap
\baseinteracs_2(\intertype) \neq \emptyset$, for some
$\intertype\in\intertypes$, or $\pureform_1 * \pureform_2$ is not
satisfiable.

Given a pure formula $\pureform$ and a set of variables $X$, the
projection $\proj{\pureform}{X}$ removes from $\pureform$ all atomic
propositions $\alpha$, such that $\fv{\alpha} \not\subseteq X$. The
\emph{projection} of a base tuple
$(\basecomps,\baseinteracs,\pureform)$ on a variable set $X$ is
formally defined below:
\[\begin{array}{rcl}
\proj{(\basecomps,\baseinteracs,\pureform)}{X} & \isdef &
\left(\basecomps \cap X, \lambda \intertype ~.~ \set{\tuple{x_1, \ldots, x_{\lenof{\intertype}}} \in
  \baseinteracs(\intertype) \mid x_1, \ldots, x_{\lenof{\intertype}} \in X},
\proj{\closureof{\constrof{\baseinteracs} * \pureform}}{X}\right) \\
\text{where } \constrof{\baseinteracs} & \isdef &
\Asterisk_{\intertype\in\intertypes}\Asterisk_{\tuple{x_1, \ldots,
    x_{\lenof{\intertype}}} \in \baseinteracs(\intertype)} \Asterisk_{1 \leq i < j \leq
  \lenof{\intertype}} ~x_i \neq x_j
\end{array}\]

The \emph{substitution} operation $(\basecomps, \baseinteracs,
\pureform)[x_1/y_1, \ldots, x_n/y_n]$ replaces simultaneously each
$x_i$ with $y_i$ in $\basecomps$, $\baseinteracs$ and $\pureform$,
respectively.
\begin{myTextE}
For a store $\store$, we denote by $\nu[x_1/y_1, \ldots, x_n/y_n]$ the
store such that $\nu[x_1/y_1, \ldots, x_n/y_n](x_i) = \nu(y_i)$ and
agrees with $\nu$ over $\vars\setminus\set{x_1, \ldots, x_n}$.
\end{myTextE}
We lift the composition, projection and substitution operations to
sets of satisfiable base tuples, as usual.

\begin{myLemmaE}\label{lemma:pure-subst}
  Given a formula $\phi$ and a substitution $\sigma = [x_1/y_1,
    \ldots, x_n/y_n]$, for any configuration $(\comps, \interacs,
  \statemap)$ and store $\store$, $(\comps,\interacs,\statemap)
  \models^\store_\asid \phi\sigma$ only if
  $(\comps,\interacs,\statemap) \models^{\store\sigma} \phi$.
\end{myLemmaE}
\begin{proofE}
  By induction on the definition of $\models^\store_\asid$. \qed
\end{proofE}

Next, we define the base tuple corresponding to a quantifier- and
predicate-free formula $\phi = \psi * \pureform$, where $\psi$
consists of component and interaction atoms and $\pureform$ is pure.
Since, moreover, we are interested in those components and
interactions that are visible through a given indexed set of
parameters $X = \set{x_1, \ldots, x_n}$, for a variable $y$, we denote
by $\reprof{y}{X}{\pureform}$ the parameter $x_i$ with the least
index, such that $y \formeq{\pureform} x_i$, or $y$ itself, if no such
parameter exists. We define the following sets of formul{\ae}:
\[\begin{array}{rcl}
\basetupleof{\phi}{X} & \isdef & \left\{\begin{array}{ll} \set{(\basecomps, \baseinteracs, \pureform)} &
\text{, if } (\basecomps, \baseinteracs, \pureform) \text{ is satisfiable} \\
\emptyset & \text{, otherwise}
\end{array}\right. \\
\text{where } \basecomps & \isdef & \set{\reprof{x}{X}{\pureform} \mid \compact{x} \text{ occurs in } \psi} \\
\baseinteracs & \isdef & \lambda \tuple{p_1, \ldots, p_s} .
\Set{\Tuple{\reprof{y_1}{X}{\pureform}, \ldots, \reprof{y_s}{X}{\pureform}} \mid \interacn{y_1}{p_1}{y_s}{p_s} \text{ occurs in } \psi}
\end{array}\]
We consider a tuple of variables $\basevartuple$, having a variable
$\basevarof{\apred}$ ranging over $\pow{\satbasetuples}$, for each
predicate $\apred$ that occurs in $\asid$. With these definitions,
each rule of $\asid$:
\begin{equation*}
\apred(x_1, \ldots, x_{\arityof{\apred}}) \leftarrow \exists y_1 \ldots \exists y_m ~.~
\phi * \bpred_1(z^1_1, \ldots, z^1_{\arityof{\bpred_1}}) * \ldots * \bpred_h(z^h_1, \ldots, z^h_{\arityof{\bpred_h}})
\end{equation*}
where $\phi$ is a quantifier- and predicate-free formula, induces the
constraint:
\vspace*{-.5\baselineskip}
\begin{equation}\label{eq:basesid}
  \hspace*{-3.5mm}\basevarof{\apred} \supseteq \proj{\big(\basetupleof{\phi}{\set{x_1, \ldots, x_{\arityof{\apred}}}}
  \basecomp \Basecomp_{\ell=1}^h \basevarof{\bpred_\ell}[x_1/z^\ell_1, \ldots,
    x_{\arityof{\bpred_\ell}}/z^\ell_{\arityof{\bpred_\ell}}]\big)}{x_1,
    \ldots, x_{\arityof{\apred}}}
\end{equation} Let $\basesid$ be the set of such
constraints, corresponding to the rules in $\asid$ and let
$\leastbase$ be the tuple of least solutions of the constraint system
generated from $\asid$, indexed by the tuple of predicates that occur
in $\asid$, such that $\leastbaseof{\apred}$ denotes the entry of
$\leastbase$ correponding to $\apred$. Since the composition and
projection are monotonic operations, such a least solution exists and
is unique. Moreover, since $\satbasetuples$ is finite, the least
solution can be attained in a finite number of steps, using a standard
Kleene iteration\ifLongVersion(see Fig. \ref{fig:base-sets})\fi.

\ifLongVersion
\begin{figure}[t!]
  {\small\begin{algorithmic}[0]
    \State \textbf{input}: a SID $\asid$
    \State \textbf{output}: $\leastbase$
  \end{algorithmic}}
  {\small\begin{algorithmic}[1]
    \State initially $\leastbase := \lambda \apred ~.~ \emptyset$
    
    \For{$\apred(x_1,\ldots,x_{\arityof{\apred}}) \leftarrow \exists y_1 \ldots \exists y_m ~.~ \phi \in \asid$, with $\phi$ quantifier- and predicate-free}

    \State $\leastbaseof{\apred} := \leastbaseof{\apred} \cup
    \proj{\basetupleof{\phi}{\set{x_1,\ldots,x_{\arityof{\apred}}}}}{x_1,\ldots,x_{\arityof{\apred}}}$

    \EndFor

    \While{$\leastbase$ still change}

    \For{$\arule : \apred(x_1,\ldots,x_{\arityof{\apred}}) \leftarrow \exists y_1 \ldots \exists y_m ~.~
      \phi * \Asterisk_{\ell=1}^h \bpred_\ell(z^\ell_1,\ldots,z^\ell_{\arityof{\bpred_\ell}}) \in \asid$}

    \If{there exist $\abasetuple_1 \in \leastbaseof{\bpred_1}, \ldots, \abasetuple_h \in \leastbaseof{\bpred_h}$}

    \State $\leastbaseof{\apred} := \leastbaseof{\apred} ~\cup~
    \proj{\left(\basetupleof{\phi}{\set{x_1,\ldots,x_{\arityof{\apred}}}}
      \basecomp \Basecomp_{\ell=1}^h \abasetuple_\ell[x_1/z^\ell_1,
          \ldots,
          x_{\arityof{\bpred_\ell}}/z^\ell_{\arityof{\bpred_\ell}}]\right)}{x_1,\ldots,x_{\arityof{\apred}}}$    
    
    \EndIf \EndFor \EndWhile
  \end{algorithmic}}
  \caption{Algorithm for the Computation of the Least Solution}
  \label{fig:base-sets}
\end{figure}
\fi

\begin{myTextE}
Given a base tuple $(\basecomps, \baseinteracs, \pureform)$ and a
store $\store$, we define the following sets of components and
interactions, respectively:
\begin{align*}
  \store(\basecomps) & \isdef \set{\nu(x) \mid x \in \basecomps} \\
  \store(\baseinteracs) & \isdef \bigcup_{\tuple{p_1, \ldots,
    p_n}\in\intertypes} \set{(\nu(x_1), p_1, \ldots, \nu(x_n), p_n)
  \mid (x_1, \ldots, x_n) \in \baseinteracs(\tuple{p_1, \ldots, p_n})}
\end{align*}
\end{myTextE}
We state below the main result leading to an elementary recursive
algorithm for the satisfiability problem (Thm. \ref{thm:sat}).

\begin{myLemmaE}\label{lemma:sat-soundness}
  Given a base tuple $\basetuple \in \leastbaseof{\apred}[x_1/y_1,
    \ldots, x_{\arityof{\apred}}/y_{\arityof{\apred}}]$, a state map
  $\statemap$ and a store $\store$ such that $(\emptyset, \emptyset,
  \statemap) \models^\store \pureform$, a set of components
  $\mathcal{D}$ disjoint from $\store(\basecomps)$ and a set of
  interactions $\mathcal{J}$ disjoint from $\store(\baseinteracs)$,
  there exists a configuration $(\comps,\interacs,\statemap)$, such
  that $(\comps,\interacs,\statemap) \models^\store_\asid \apred(y_1,
  \ldots, y_{\arityof{\apred}})$, $\comps \cap \mathcal{D} =
  \emptyset$ and $\interacs \cap \mathcal{J} = \emptyset$.
\end{myLemmaE}
\begin{proofE}
  Let $\sigma \isdef [x_1/y_1, \ldots,
    x_{\arityof{\apred}}/y_{\arityof{\apred}}]$ be a substitution and
  $\basetuplen{0} \in \leastbaseof{\apred}$ be a base pair, such that
  $\basetuple = \basetuplen{0}\sigma$. Since $(\emptyset, \emptyset,
  \statemap) \models^\store \pureform$, we obtain $(\emptyset,
  \emptyset, \statemap) \models^{\store_0} \pureform_0$, by Lemma
  \ref{lemma:pure-subst}, where we define $\store_0 \isdef
  \store\sigma$.  Let \[\mathcal{K} \isdef \mathcal{D} \cup \set{c_i
    \mid (c_1, p_1, \ldots, c_n, p_n) \in \mathcal{J},~ i \in
    \interv{1}{n}}\] The proof is by fixpoint induction on the
  definition of $\basetuplen{0}$. Assume that:
  \[\basetuplen{0} \in \proj{\big(\basetupleof{\psi*\pureform'}{\set{x_1, \ldots,
        x_{\arityof{\apred}}}} \basecomp \Basecomp_{\ell=2}^h
    \leastbaseof{\bpred_\ell}[x_1/z^\ell_1, \ldots,
      x_{\arityof{\bpred_\ell}}/z^\ell_{\arityof{\bpred_\ell}}]\big)}{x_1,
    \ldots, x_{\arityof{\apred}}}\] for a rule \(\apred(x_1, \ldots,
  x_{\arityof{\apred}}) \leftarrow \exists y_1 \ldots \exists y_m ~.~
  \psi * \pureform' * \Asterisk_{\ell=2}^h\bpred_\ell(z^\ell_1,
  \ldots, z^\ell_{\arityof{\bpred_\ell}})\) of $\asid$, such that
  $\psi * \pureform'$ is quantifier-free, $\psi$ consists of component
  and interaction atoms and $\pureform'$ is the largest pure
  subformula of $\psi * \pureform'$. Then there exist base tuples
  $\basetuplen{1}, \ldots, \basetuplen{h}$, such
  that: \begin{compactitem}
  \item $\basetuplen{1} \in \basetupleof{\psi*\pureform'}{\set{x_1,
      \ldots, x_{\arityof{\apred}}}}$,
  \item $\basetuplen{\ell} \in \leastbaseof{\bpred_{\ell}}[x_1/z^{\ell}_1,
    \ldots,
    x_{\arityof{\bpred_{\ell}}}/z^{\ell}_{\arityof{\bpred_{\ell}}}]$, for
    all $\ell \in \interv{2}{h}$,
  \item $\basetuplen{0} = \proj{\big(\basetuplen{1} \otimes \ldots
    \otimes \basetuplen{h}\big)}{x_1, \ldots,
    x_{\arityof{\apred}}}$.
  \end{compactitem}
  From the first and last points, we deduce $\pureform_0 = \proj{
    \closureof{ \pureform' * \Asterisk_{\ell=2}^{h}
      (\constrof{\baseinteracs_\ell} * \pureform_{\ell}) }}{x_1,
    \ldots, x_{\arityof{\apred}}}$. Let $\pureform'' \isdef \pureform'
  * \Asterisk_{\ell=2}^{h} (\constrof{\baseinteracs_\ell} *
  \pureform_{\ell})$ and define a store $\store'_0$, by assigning each
  $\formeq{\pureform''}$-equivalence class the following
  component: \begin{compactitem}
  \item $\store_0(x_i)$, if $x_i$ belongs to the class, for some $i
    \in \interv{1}{\arityof{\apred}}$, 
  \item else, if the class is disjoint from $\set{x_1, \ldots,
    x_{\arityof{\apred}}}$ and $\compin{y}{q}$ occurs in $\pureform'$,
    for a variable $y$ in the class, we assign $c \in
    \universe\setminus\mathcal{K}$, such that $\statemap(c) = q$;
    since $\pureform'$ is satisfiable, there are no two state atoms
    $\compin{y}{q}$ and $\compin{z}{r}$, such that $y
    \formeq{\pureform'} z$ and $q \neq r$ in $\pureform'$ and,
    moreover, chosing $c$ is always possible, by the last point of
    Def. \ref{def:configuration}, 
  \item otherwise, the class is assigned an arbitrary component $c \in
    \universe \setminus \mathcal{K}$.
  \end{compactitem}
  Such a store exists, because $\pureform''$ is satisfiable and,
  moreover, $(\emptyset,\emptyset,\statemap) \models^{\store'_0}
  \pureform''$, hence also $(\emptyset,\emptyset,\statemap)
  \models^{\store'_0} \pureform_{\ell}$, for all
  $\ell\in\interv{2}{h}$. We define two sequences of sets of
  components $\comps_1, \ldots, \comps_{h}$ and interactions
  $\interacs_1, \ldots, \interacs_{h}$, as
  follows: \begin{compactitem}
  \item $\comps_1 \isdef \set{\store'_0(y) \mid \compact{y} \text{
      occurs in } \psi}$,
  \item $\interacs_1 \isdef \set{(\store'_0(z_1), p_1, \ldots,
    \store'_0(z_t), p_t) \mid \interacn{z_1}{p_1}{z_t}{p_t} \text{
      occurs in } \psi}$,
  \item for all $\ell \in \interv{2}{h}$, assume that $\comps_1,
    \ldots, \comps_{\ell-1}$ and $\interacs_1, \ldots,
    \interacs_{\ell-1}$ have been defined and let us define:
    \[\begin{array}{rcl}
    \mathcal{D}_\ell & \isdef & \mathcal{D} ~\cup~ \bigcup_{j=1}^{\ell-1} \comps_j ~\cup~ \bigcup_{j=\ell+1}^h \store'_0(\basecomps_j) \\
    \mathcal{J}_\ell & \isdef & \mathcal{J} ~\cup~ \bigcup_{j=1}^{\ell-1} \interacs_j ~\cup~ \bigcup_{j=\ell+1}^h \store'_0(\baseinteracs_j)
    \end{array}\]
    First, we prove that $\mathcal{D}_\ell \cap
    \store'_0(\basecomps_\ell) = \emptyset$ and $\mathcal{J}_\ell \cap
    \store'_0(\baseinteracs_\ell) = \emptyset$ (we prove the first
    point, the second using a similar argument).  Suppose, for a
    contradiction, that $c \in \mathcal{D}_\ell \cap
    \store'_0(\basecomps_\ell)$. We distinguish the following
    cases: \begin{compactitem}
    \item if $c \in \mathcal{D} \cap \store'_0(\basecomps_\ell)$, then
      $c \in \mathcal{D} \cap \store'_0(\basecomps_0)$, because
      $\basecomps_\ell \subseteq \basecomps_0$, contradiction with
      $\mathcal{D} \cap \store'_0(\basecomps_0) = \mathcal{D} \cap
      \store(\basecomps) = \emptyset$. 
    \item else, if $c \in \comps_j \cap \store'_0(\basecomps_\ell)$,
      for some $j \in \interv{1}{\ell-1}$, then $c \in \comps_j \cap
      \mathcal{D}_j$, because $\store'_0(\basecomps_\ell) \subseteq
      \mathcal{D}_j$, contradiction with the inductive hypothesis
      $\comps_j \cap \mathcal{D}_j = \emptyset$.
    \item otherwise, $c \in \store'_0(\basecomps_j) \cap
      \store'_0(\basecomps_\ell)$, hence there exist variables $y_j
      \in \basecomps_j$ and $y_\ell \in \basecomps_\ell$, such that
      $y_j \formeq{\pureform''} y_\ell$, contradiction with the fact
      that $\basetuplen{j} \basecomp \basetuplen{\ell}$ is
      satisfiable. 
    \end{compactitem}
    Second, we apply the inductive hypothesis to obtain configurations
    $(\comps_\ell, \interacs_\ell, \statemap)$, such that
    $(\comps_\ell, \interacs_\ell, \statemap)
    \models^{\store'_0}_\asid \bpred_\ell(t^\ell_1, \ldots,
    t^\ell_{\arityof{\bpred_\ell}})$, $\comps_\ell \cap
    \mathcal{D}_\ell = \emptyset$ and $\interacs_\ell \cap
    \mathcal{J}_\ell = \emptyset$, for all $\ell \in
    \interv{2}{h}$. By the definitions of $\mathcal{D}_\ell$ and
    $\mathcal{J}_\ell$, the sets $\comps_\ell$ and $\interacs_\ell$
    are pairwise disjoint, respectively, hence the composition
    $(\comps, \interacs, \statemap) \isdef \bigcomp_{\ell=1}^h
    (\comps_\ell, \interacs_\ell, \statemap)$ is defined. Moreover,
    $(\comps, \interacs, \statemap) \models^{\store'_0}_\asid \psi *
    \pureform' * \Asterisk_{\ell=2}^h \bpred_\ell(t^\ell_1, \ldots,
    t^\ell_{\arityof{\bpred_\ell}})$, hence $(\comps, \interacs,
    \statemap) \models^{\store'_0}_\asid \apred(x_1, \ldots,
    x_{\arityof{\apred}})$, leading to $(\comps, \interacs, \statemap)
    \models^{\store} \apred(y_1, \ldots, y_{\arityof{\apred}})$.

    Finally, we are left with proving that $\comps \cap \mathcal{D} =
    \emptyset$ and $\interacs \cap \mathcal{J} = \emptyset$ (we prove
    the first point only, the second uses a similar reasoning). Since
    $\comps = \bigcup_{\ell=1}^h \comps_\ell$, this is equivalent to
    proving the following: \begin{compactitem}
    \item $\comps_1 \cap \mathcal{D} = \emptyset$: suppose, for a
      contradiction, that $\store'_0(y) \in \mathcal{D}$, for a
      variable $y$, such that $\compact{y}$ occurs in $\psi$. By the
      definition of $\store'_0$, we have either $\store'_0(y) \in
      \store'_0(\basecomps_0)$ or $\store'_0(y) \in
      \mathcal{K}$. Since $\store'_0(\basecomps_0) \cap \mathcal{D} =
      \mathcal{K} \cap \mathcal{D} = \emptyset$, both cases lead to a
      contradiction.
    \item $\comps_\ell \cap \mathcal{D} = \emptyset$, for all $\ell
      \in \interv{2}{h}$: because $\comps_\ell \cap \mathcal{D}_\ell =
      \emptyset$ and $\mathcal{D} \subseteq \mathcal{D}_\ell$, by
      definition of $\mathcal{D}_\ell$, for all $\ell \in
      \interv{2}{h}$. \qed
    \end{compactitem}
  \end{compactitem}
\end{proofE}

\begin{myLemmaE}\label{lemma:sat-completeness}
  Given a predicate atom $\apred(y_1, \ldots, y_{\arityof{\apred}})$,
  a store $\store$ and a configuration $(\comps,\interacs,\statemap)$,
  such that $(\comps,\interacs,\statemap) \models^\store_\asid
  \apred(y_1, \ldots, y_{\arityof{\apred}})$, there exists a base
  tuple $(\basecomps, \baseinteracs, \pureform) \in
  \leastbaseof{\apred}[x_1/y_1, \ldots,
    x_{\arityof{\apred}}/y_{\arityof{\apred}}]$, such that
  $\store(\basecomps) \subseteq \comps$, $\store(\baseinteracs)
  \subseteq \interacs$ and $(\emptyset, \emptyset, \statemap)
  \models^\store \pureform$.
\end{myLemmaE}
\begin{proofE}
  By fixpoint induction on the definition of the satisfaction relation
  $\models^\store_\asid$. Since $(\comps,\interacs,\statemap)
  \models^\store_\asid \apred(y_1, \ldots, y_{\arityof{\apred}})$, by
  Lemma \ref{lemma:pure-subst}, we have $(\comps,\interacs,\statemap)
  \models^{\store_0}_\asid \apred(x_1, \ldots, x_{\arityof{\apred}})$,
  where $\store_0 \isdef \store[x_1/y_1, \ldots,
    x_{\arityof{\apred}}/y_{\arityof{\apred}}]$. Hence, $\asid$ has a
  rule \(\apred(x_1, \ldots, x_{\arityof{\apred}}) \leftarrow \exists
  y_1 \ldots \exists y_m ~.~ \psi * \pureform' *
  \Asterisk_{\ell=2}^h\bpred_\ell(z^\ell_1, \ldots,
  z^\ell_{\arityof{\bpred_\ell}})\), such that $\psi * \pureform'$ is
  quantifier-free, $\psi$ consists of component and interaction atoms
  and $\pureform'$ is pure and there exists a store $\store'_0$, that
  agrees with $\store_0$ over $x_1, \ldots, x_{\arityof{\apred}}$ and
  configurations $(\comps_1, \interacs_1, \statemap)$, $\ldots,
  (\comps_\ell, \interacs_\ell, \statemap)$, such
  that: \begin{compactitem}
  \item $(\comps_1, \interacs_1, \statemap) \models^{\store'_0} \psi *
    \pureform'$, 
  \item $(\comps_\ell, \interacs_\ell, \statemap)
    \models^{\store'_0}_\asid \bpred_\ell(z^\ell_1, \ldots,
    z^\ell_{\arityof{\bpred_\ell}})$, for all $\ell \in \interv{2}{h}$,
    and
  \item $(\comps,\interacs,\statemap) = (\comps_1, \interacs_1,
    \statemap) \comp \ldots \comp (\comps_h, \interacs_h, \statemap)$.
  \end{compactitem}
  We consider the following base tuples: \begin{compactitem}
  \item $\basetuplen{1} \isdef \basetupleof{\psi *
    \pureform'}{\set{x_1, \ldots, x_{\arityof{\apred}}}}$,
  \item $\basetuplen{\ell} \in \leastbaseof{\bpred_\ell}[x_1/z^\ell_1,
    \ldots,
    x_{\arityof{\bpred_\ell}}/z^\ell_{\arityof{\bpred_\ell}}]$, such
    that $\store'_0(\basecomps_\ell) \subseteq \comps_\ell$,
    $\store'_0(\baseinteracs_\ell) \subseteq \interacs_\ell$ and
    $(\emptyset, \emptyset, \statemap) \models^{\store'_0}
    \pureform_\ell$, whose existence is guaranteed by the inductive
    hypothesis, for all $\ell \in \interv{2}{h}$.
  \end{compactitem}
  By the definition of $\basetupleof{\psi * \pureform'}{\set{x_1,
      \ldots, x_{\arityof{\apred}}}}$ and the fact that $(\comps_1,
  \interacs_1, \statemap) \models^{\store'_0} \psi * \pureform'$, we
  obtain $\store'_0(\basecomps_1) = \comps_1$ and
  $\store'_0(\baseinteracs_1) = \interacs_1$. Since the composition
  $\bigcomp_{\ell=1}^h (\comps_\ell, \interacs_\ell, \statemap)$ is
  defined, the sets $\comps_1, \ldots, \comps_h$ and $\interacs_1,
  \ldots, \interacs_h$ are pairwise disjoint, respectively. Since
  $\store'_0(\basecomps_\ell) \subseteq \comps_\ell$ and
  $\store'_0(\baseinteracs) \subseteq \interacs_\ell$, for all $\ell
  \in \interv{1}{h}$, we deduce that $\Basecomp_{\ell=1}^h
  \basetuplen{\ell}$ is satisfiable, because: \begin{compactitem}
  \item for all $1 \leq i < j \leq h$, for any two variables $y \in
    \basecomps_i$ and $z \in \basecomps_j$ we have $\notof{y
      \formeq{\pureform'} z}$, because $\store'_0(\basecomps_i) \cap
    \store'_0(\basecomps_j) = \emptyset$,
  \item for all $1 \leq i < j \leq h$, all $\intertype \in
    \intertypes$, for any two tuples $\tuple{y_1, \ldots,
      y_{\lenof{\intertype}}} \in \baseinteracs_i(\intertype)$ and
    $\tuple{z_1, \ldots, z_{\lenof{\intertype}}} \in
    \baseinteracs_j(\intertype)$, we have $\notof{y_k
      \formeq{\pureform'} z_k}$, for at least some $k \in
    \interv{1}{\lenof{\intertype}}$, because
    $\store'_0(\baseinteracs_i) \cap \store'_0(\baseinteracs_j) =
    \emptyset$,
  \item for each tuple $\tuple{y_1, \ldots, y_{\lenof{\intertype}}}
    \in \baseinteracs_\ell(\tuple{p_1, \ldots, p_n})$, for $\ell \in
    \interv{1}{h}$, we have $\notof{y_i \formeq{\pureform'} y_j}$, for
    all $1 \leq i < j \leq n$, because $(\store'_0(y_1), p_1, \ldots,
    \store'_0(y_n), p_n) \in \interacs_\ell$, hence $\store'_0(y_1),
    \ldots,\store'_0(y_n)$ are pairwise distinct,
  \item $(\emptyset, \emptyset, \statemap) \models^{\store'_0}
    \pureform' * \Asterisk_{\ell=2}^h \pureform_\ell$, hence
    $(\emptyset, \emptyset, \statemap) \models^{\store'_0} \pureform'
    * \Asterisk_{\ell=2}^h \constrof{\baseinteracs_\ell} *
    \pureform_\ell$, by the previous point.
  \end{compactitem}
  Then we define $\basetuplen{0} \isdef
  \proj{\left(\Basecomp_{\ell=1}^h \basetuplen{\ell} \right)}{x_1,
    \ldots, x_{\arityof{\apred}}}$ and $\basetuple \isdef
  \basetuplen{0}[x_1/y_1, \ldots,
    x_{\arityof{\apred}}/y_{\arityof{\apred}}]$. By the definition of
  $\basesid$, we have:
  \[\leastbaseof{\apred} \supseteq \proj{\big(\basetupleof{\psi*\pureform'}{\set{x_1, \ldots,
        x_{\arityof{\apred}}}} \basecomp \Basecomp_{\ell=2}^h
    \leastbaseof{\bpred_\ell}[x_1/z^\ell_1, \ldots,
      x_{\arityof{\bpred_\ell}}/z^\ell_{\arityof{\bpred_\ell}}]\big)}{x_1,
    \ldots, x_{\arityof{\apred}}}\]
  and, since, by the construction of $\basetuplen{0}$, 
  \[\basetuplen{0} \in \proj{\big(\basetupleof{\psi*\pureform'}{\set{x_1, \ldots,
        x_{\arityof{\apred}}}} \basecomp \Basecomp_{\ell=2}^h
    \leastbaseof{\bpred_\ell}[x_1/z^\ell_1, \ldots,
      x_{\arityof{\bpred_\ell}}/z^\ell_{\arityof{\bpred_\ell}}]\big)}{x_1,
    \ldots, x_{\arityof{\apred}}}\] we obtain $\basetuplen{0} \in
  \leastbaseof{\apred}$, leading to $\basetuple \in
  \leastbaseof{\apred}[x_1/y_1, \ldots,
    x_{\arityof{\apred}}/y_{\arityof{\apred}}]$. Next, we check that
  $\store(\basecomps) \subseteq \bigcup_{\ell=1}^h
  \store'_0(\basecomps_\ell) \subseteq \bigcup_{\ell=1}^h \comps_\ell
  = \comps$ and $\store(\baseinteracs) \subseteq \bigcup_{\ell=1}^h
  \store'_0(\baseinteracs_\ell) \subseteq \bigcup_{\ell=1}^h
  \interacs_\ell = \interacs$. Finally, the requirement
  $(\emptyset,\emptyset,\statemap) \models^\store \pureform$ follows
  from the following: \begin{compactitem}
  \item $\pureform = \pureform_0[x_1/y_1, \ldots,
    x_{\arityof{\apred}}/y_{\arityof{\apred}}]$, by the definition of
    $\basetuple$,
  \item $(\emptyset,\emptyset,\statemap) \models^{\store'_0}
    \pureform'$ and $(\emptyset,\emptyset,\statemap)
    \models^{\store'_0} \pureform_\ell$, for all $\ell
    \in\interv{2}{h}$,
  \item $\pureform_0 = \proj{\closureof{\pureform' *
      \Asterisk_{\ell=2}^h \constrof{\baseinteracs_\ell} *
      \pureform_\ell}}{x_1, \ldots, x_{\arityof{\apred}}}$, where
    $(\emptyset,\emptyset,\statemap) \models^{\store'_0}
    \constrof{\baseinteracs_\ell}$ follows from the satisfiability of
    $\basetuplen{\ell}$, for all $\ell \in \interv{2}{h}$. \qed
  \end{compactitem}
\end{proofE}
  
\begin{lemmaE}\label{lemma:sat}
  $\sat{\asid}{\apred}$ has a positive answer if and only if
  $\leastbaseof{\apred} \neq \emptyset$.
\end{lemmaE}
\begin{proofE} ``$\Leftarrow$'' follows from Lemma \ref{lemma:sat-soundness}
  and ``$\Rightarrow$'' follows from Lemma
  \ref{lemma:sat-completeness}.  \qed \end{proofE}

If the maximal arity of the predicates occurring in $\asid$ is bound
by a constant $k$, no satisfiable base tuple $\basetuple$ can have a
tuple $\tuple{y_1, \ldots, y_{\lenof{\intertype}}} \in
\baseinteracs(\intertype)$, for some $\intertype\in\intertypes$, such
that $\lenof{\intertype} > k$, since all variables $y_1, \ldots,
y_{\lenof{\intertype}}$ are parameters denoting distinct components
(point \ref{it3:base-tuple} of Def. \ref{def:base-tuple}). Hence, the
upper bound on the size of a satisfiable base tuple is constant, in
both the $k<\infty, \ell<\infty$ and $k<\infty, \ell=\infty$ cases,
which are, moreover indistinguishable complexity-wise (i.e., both are
\np-complete). In contrast, in the cases $k=\infty, \ell<\infty$ and
$k=\infty, \ell=\infty$, the upper bound on the size of satisfiable
base tuples is polynomial and simply exponential in $\sizeof{\asid}$,
incurring a complexity gap of one and two exponentials, respectively.
The theorem below states the main result of this section:

\begin{theoremE}\label{thm:sat}
  $\klsat{\asid}{\apred}{k}{\infty}$ is \np-complete for $k\ge 4$,
  $\klsat{\asid}{\apred}{\infty}{\ell}$ is \exptime-complete and
  $\sat{\asid}{\apred}$ is in \twoexptime.
\end{theoremE}
\begin{proofE}
  \emph{Membership (upper bounds).}
  For non-negative integers $m \le n$ denote by $\nordsubsets{n}{m} \isdef
  \frac{n!}{(n-m)!}$ the number of ordered $m$-element subsets of a
  $n$-element set.

  Let $\alpha=\maxarityof{\asid}$, $\beta=\maxintersize{\asid}$ and $p = \cardof{\ports}$.  The
  maximum length of a satisfiable base tuple is $B \isdef \alpha +
  (\sum_{j=1}^{\min(\alpha,\beta)} p^j \cdot \nordsubsets{\alpha}{j}) +
  (2\alpha^2 + \alpha)$, that is, size of the set of components
  $\basecomps$ plus the size of the set of interactions
  $\baseinteracs$ plus the length of the longest pure formula
  $\pureform$.  In general, for any non-negative integer $j$ there
  exists at most $p^j$ interaction types of arity $j$ with ports from
  $\ports$; moreover, for any such interaction type there exists at
  most $\nordsubsets{\alpha}{j}$ interactions relating distinct
  components from an $\alpha$-element set.  Moreover, no such
  interaction exists neither if $j > \alpha$ nor $j > \beta$.  

  For any $u \le \alpha$ it holds that $\sum_{j=1}^u p^j \cdot
  \nordsubsets{\alpha}{j} \le p^u \alpha^u$ (an easy check by
  induction on $u$).  We use the inequality above with
  $u=\min(\alpha,\beta)$ and obtain that $B \le 2\alpha + 2\alpha^2 +
  p^{\min(\alpha,\beta)} \alpha^{\min(\alpha,\beta)} \isdef B^*$.  We
  distinguish the three cases:\begin{compactenum}
  \item $k<\infty$, $\ell=\infty$: since $\alpha \le k$ then $\alpha$ is constant and
    $B^* = \bigO(1)$,
  \item $k=\infty$, $\ell < \infty$: since $\beta\le \ell$ and
    $\alpha=\bigO(\size{\asid})$ then
    $B^*=\polynomial(\size{\asid})$,
  \item $k=\infty$, $\ell=\infty$: since $\alpha=\bigO(\size{\asid})$
    then $B^*=2^{\polynomial(\size{\asid)}}$.
  \end{compactenum}
  Let $N \isdef 2^{B^*}$, that is, (an over-approximation of) the
  total number of base tuples.  Clearly, $N$ is constant in case (1)
  and respectively $2^{\polynomial(\size{\asid)}}$ and
  $2^{2^{\polynomial(\size{\asid)}}}$ in cases (2), (3).  Let $L$ be
  the number of predicates occuring in $\asid$ and $H$ be the maximum
  number of predicates used in a term in $\asid$.  Let observe that
  both $L$ and $H$ are in general $\bigO(\size{\asid})$. Then the
  least solution $\leastbase$ has at most $N$ base tuples for each
  predicate, hence at most $L \cdot N$ base tuples. Furthermore, for
  each rule of $\asid$ the time to check and/or produce the base tuple
  $(\basecomps_0,\baseinteracs_0,\pureform_0)$ with respect to the
  rule constraint (\ref{eq:basesid}) and given arguments
  $(\basecomps_j,\baseinteracs_j,\pureform_j)_{j=1,h}$ is polynomial
  $\polynomial(B^*,\size{\asid}))$.  That is, both composition
  and projection take at most $(H+1) B^* + \size{\asid}^3$ time as
  they need to process (union or scan) at most $H+1$ base tuples of
  length $B^*$ each plus the closure of pure formula with at most
  $\size{\asid}$ variables. \begin{compactenum}
  \item $k < \infty$, $\ell=\infty$: We define a non-determinstic algorithm as
    follows.  Let $(\asid,\apred)$ be the input instance. We guess a
    witness $\tuple{W_1,\ldots,W_K}$ for a least solution, where $1
    \le K \le L\cdot N$ and each $W_i$ entry is of the form
    $(T_i,r_i,e_{i,1},\ldots,e_{i,h_i})$, where $T_i$ is a base tuple,
    $r_i$ an index of a rule of $\asid$ and $e_{i,1},\ldots,e_{i,h}$
    are index values from $\set{i+1,\ldots,K}$ for $0 \le h_i \le H$.
    The length of every witness entry is therefore at most $B^* +
    \lceil \log_2(\size{\asid}) \rceil + H \lceil \log_2(L\cdot N)
    \rceil$.  As $N$ is constant when $k < \infty$, and $L$ and $H$
    are $\bigO(\size{\asid})$ it follows that the number of guesses is
    polynomial for building $\tuple{W_1,\ldots,W_K}$. We now check
    that $\tuple{W_1,\ldots,W_K}$ represents indeed a valid
    computation of a base tuples from the least solution i.e.,
    $\basetuple \in \leastbaseof{\apred}$. For this, we need to check:
    (a) every entry is well-formed, that is, the rule indexed by $r_i$
    instantiates precisely $h_i$ predicates; moreover, for every $1
    \le j \le h_i$ the index $e_{i,j}$ designates an entry
    $W_{e_{i,j}}$ whose rule defines the $j$-th predicate instantiated
    by the rule $r_i$; (b) the base tuple of every entry is
    satisfiable and correctly computed, that is, $T_i$ is the result
    of applying the constraint (\ref{eq:basesid}) for rule $r_i$ with
    actual arguments $T_{e_{i,1}},\ldots, T_{e_{i,h_i}}$ from the
    referred entries; (c) the rule $r_1$ of the first entry $W_1$
    defines the predicate $\apred$.  Again, as $B^*$ and $N$ are
    constant in this case, all these checks are done in polynomial
    time. Since both the generation and the checking of the witness
    are polynomial time, this ensures membership in \np.
        
  \item $k = \infty$, $\ell < \infty$: Consider the computation of the
    least solution $\leastbase$ using standard Kleene iteration.  At
    every step, a rule of $\asid$ and a tuple of at most $H$ base
    tuples arguments are selected to produce a new base tuple. Thus,
    in the worst case, at most $\size{\asid}$ rules in combination
    with at most $N^H$ base tuples need to be selected and evaluated.
    If no new base tuple is generated the fixpoint is reached and the
    algorithms stops.  Since there are at most $L \cdot N$ base tuples
    in the least solution, the total time will be therefore $L\cdot N
    \cdot \size{\asid} \cdot N^H \cdot t(B^*,\size{\asid})$ where
    $t(B^*,\size{\asid})$ is the (polynomial) time to process one
    selection. It is an easy check that the above is
    $2^{\polynomial(\size{\asid})}$ since
    $N=2^{\polynomial(\size{\asid})}$ in this case.
       
  \item $k=\infty$, $\ell=\infty$: Following the same reasoning as in
    the previous case the complexity is
    $2^{2^{\polynomial(\size{\asid)}}}$ as
    $N=2^{2^{\polynomial(\size{\asid})}}$ in this case.
  \end{compactenum}
  
  \emph{Hardness (lower bounds)}.  The restricted fragment of \cl\ to
  $*$, $=$, $\not=$ is equisatisfiable to the restricted fragment of
  \seplog\ restricted to $*$, $=$, $\not=$.  The satisfiability of the
  above \seplog\ fragment has been proven respectively \np-hard, if
  the arities of predicates are bounded by a constant $k \ge 3$
  \cite[Theorem 4.9]{DBLP:conf/csl/BrotherstonFPG14} and
  \exptime-hard, in general \cite[Theorem
    4.15]{DBLP:conf/csl/BrotherstonFPG14}. Yet, the reductions
  considered in these proofs rely on the use of a predefined
  \emph{nil} constant symbol in the \seplog\ logic; this constant can
  be nevertheless replaced by a variable consistently propagated along
  the SID, that is, at the price of increasing the arities of all
  predicates by one.  Therefore, it follows immediately that
  $\klsat{\asid}{\phi}{k}{\infty}$ is \np-hard for $k\ge 4$ and
  $\klsat{\asid}{\phi}{\infty}{\ell}$, $\sat{\asid}{\phi}$ are both
  \exptime-hard. \qed
\end{proofE}
  
\begin{example}\label{ex:sat-worst-case}
  The doubly-exponential upper bound for the algorithm computing the
  least solution of a system of constraints of the form
  (\ref{eq:basesid}) is necessary, in general, as illustrated by the following
  worst-case example. Let $n$ be a fixed parameter and consider the
  $n$-arity predicates $A_1,\ldots,A_n$ defined by the following SID:
    \[\begin{array}{rcll}
    A_i(x_1,\ldots,x_n) & \rightarrow & \Asterisk_{j=0}^{n-i} ~A_{i+1}(x_1,\ldots,x_{i-1}, [x_i,\ldots,x_n]^j)
    & \text{, for all } i \in \interv{1}{n-1} \\
    A_n(x_1,\ldots,x_n) & \rightarrow & \interacn{x_1}{p}{x_n}{p} \\
    A_n(x_1,\ldots,x_n) & \rightarrow & \emp
    \end{array}\]
    where, for a list of variables $x_i,\ldots,x_n$ and an integer
    $j\geq0$, we write $[x_i,\ldots,x_n]^j$ for the list rotated to
    the left $j$ times (e.g.,
    $[x_1,x_2,x_3,x_4,x_5]^2=x_3,x_4,x_5,x_1,x_2$). In this example,
    when starting with $A_1(x_1,\ldots,x_n)$ one eventually obtains
    predicate atoms $A_n(x_{i_1},\ldots,x_{i_n})$, for any permutation
    $x_{i_1},\ldots,x_{i_n}$ of $x_1,\ldots,x_n$.  Since $A_n$ may
    choose to create or not an interaction with that permutation of
    variables, the total number of base tuples generated for $A_1$ is
    $2^{n!}$. That is, the fixpoint iteration generates $2^{2^{\bigO(n
        \log n)}}$ base tuples, whereas the size of the input of $\sat{\asid}{\apred}$ is
    $\polynomial(n)$. \hfill$\blacksquare$
\end{example}

\ifLongVersion
\section{Tightness}
\label{sec:tightness}

The tightness problem (Def. \ref{def:decision}, point
\ref{decision:tight}) is the complement of a problem slightly stronger
than satisfiability (\ref{decision:sat}): given a SID $\asid$ and a
formula $\phi$, such that $\fv{\phi} = \set{x_1, \ldots, x_n}$, the
\emph{looseness problem} $\loose{\asid}{\apred}$ asks for the
existence of a loose configuration $\aconfig$
(Def. \ref{def:tightness}), such that $\aconfig \models_\asid \exists
x_1 \ldots \exists x_n ~.~ \phi$. We establish upper and lower bounds
for the complexity of the looseness problem by a reduction to and from
the satisfiability problem. The bounds for the tightness problem
follow by standard complementation of the complexity classes for the
looseness problem.

\paragraph{From Looseness to Satisfiability.}
Let $\asid$ be a given SID and $\apred$ be a predicate. For each
predicate $\bpred$ that occurs in $\asid$, we consider a fresh
predicate $\bppred$, not occurring in $\asid$, such that
$\arityof{\bppred}=\arityof{\bpred}+1$. The SID $\psid$ consists of
$\asid$ and, for each rule of $\asid$ of the form:
\[\bpred_0(x_1, \ldots, x_{\arityof{\bpred_0}}) \leftarrow \exists y_1 \ldots \exists y_m ~.~ \phi *
\bpred_1(t^1_1, \ldots, t^1_{\arityof{\bpred_1}}) * \ldots *
\bpred_h(t^h_1, \ldots, t^h_{\arityof{\bpred_h}})\] where $\phi$ is a
quantifier- and predicate-free formula, $\psid$ has the following
rules:
\[\bppred_0(x_1, \ldots, x_{\arityof{\bpred_0}+1}) \leftarrow \exists y_1 \ldots \exists y_m ~.~ \phi ~*~ x_{\arityof{\bpred_0}+1} = z ~*~
\bpred_1(t^1_1, \ldots, t^1_{\arityof{\bpred_1}}) ~* \ldots *~
\bpred_h(t^h_1, \ldots, t^h_{\arityof{\bpred_h}})\]
if $z$ occurs in an interaction atom from $\phi$, and: 
\[\begin{array}{rcl}
\bppred_0(x_1, \ldots, x_{\arityof{\bpred_0}+1}) & \leftarrow & \exists y_1 \ldots \exists y_m ~.~ \phi * 
\bpred_1(t^1_1, \ldots, t^1_{\arityof{\bpred_1}}) * \ldots * \bppred_i(t^i_1, \ldots, t^i_{\arityof{\bpred_i}}, x_{\arityof{\bpred_0}+1}) \\
&& \hspace*{2cm} * \ldots * \bpred_h(t^h_1, \ldots, t^h_{\arityof{\bpred_h}})
\end{array}\]
for some $i \in \interv{1}{h}$, if $h \geq 1$. Moreover, for each rule of
$\asid$ of the form above, with no predicate atoms (i.e., $h=0$),
$\psid$ contains the rule:
\[\begin{array}{rcll}
\bppred(x_1, \ldots, x_{\arityof{\bpred}+1}) & \leftarrow & \exists y_1 \ldots \exists y_m ~.~ \phi * \compact{x_{\arityof{\bpred_0}+1}}
\end{array}\]
if and only if $\phi$ contains no predicate atoms. Finally, there is a
fresh predicate $\looseof{\apred}$, of arity $\arityof{\apred}$, with
a rule:
\[\looseof{\apred}(x_1, \ldots, x_{\arityof{\apred}}) \leftarrow \exists y ~.~ \appred(x_1, \ldots, x_{\arityof{\apred}}, y) * \compact{y}\]
Intuitively, the last parameter of a $\bppred$ predicate binds to an
arbitrary variable of an interaction atom. A configuration $\aconfig$
is loose if and only if the value (component) of some variable
occurring in an interaction atom is absent, in which case the
component can be added to $\aconfig$ (without clashing with a present
component of $\aconfig$) by the last rule. The reduction is
polynomial, since the number of rules in $\psid$ is linear in the
number of rules in $\asid$ and the size of each newly added rule is
increased by a constant. The following lemma states the correctness of
the reduction:

\begin{lemma}\label{lemma:loose-sat}
  Given a SID $\asid$ and a predicate $\apred$, the problem
  $\loose{\asid}{\apred}$ has a positive answer if and only if the
  problem $\sat{\psid}{\looseof{\apred}}$ has a positive answer.
\end{lemma}
\proof{
  ``$\Rightarrow$'' Let $\aconfig \isdef (\comps, \interacs, \statemap)$ be
a loose configuration, such that $\aconfig \models^{\store}_\asid
\apred(x_1, \ldots, x_{\arityof{\apred}})$, for some store
$\store$. Since $\aconfig$ is loose, there exists an interaction
$(c_1, p_1, \ldots, c_n, p_n) \in \interacs$, such that $c_i \not\in
\comps$, for some $i \in \interv{1}{n}$. We prove that $\aconfig
\models_{\psid}^{\store[y \leftarrow c_i]} \appred(x_1, \ldots,
x_{\arityof{\apred}},y)$, by fixpoint induction on the definition of
$\aconfig \models^{\store}_\asid \apred(x_1, \ldots,
x_{\arityof{\apred}})$. This is sufficient, because then we obtain
$(\comps\cup\set{c_i},\interacs,\statemap) \models_{\psid}^{\store}
\looseof{\apred}(x_1, \ldots, x_{\arityof{\apred}})$, thus
$\sat{\psid}{\looseof{\apred}}$ has a positive answer. Consider the
rule of $\asid$:
\[\apred(x_1, \ldots, x_{\arityof{\apred}}) \leftarrow \exists y_1 \ldots \exists y_m ~.~ \phi *
\bpred_1(t^1_1, \ldots, t^1_{\arityof{\bpred_1}}) * \ldots *
\bpred_h(t^h_1, \ldots, t^h_{\arityof{\bpred_h}})\] where $\phi$ is
quantifier- and predicate-free and $c'_1, \ldots, c'_m \in \universe$
are components, such that $(\comps,\interacs,\statemap)
\models_{\asid}^{\store'} \phi * \bpred_1(t^1_1, \ldots,
t^1_{\arityof{\bpred_1}}) * \ldots * \bpred_h(t^h_1, \ldots,
t^h_{\arityof{\bpred_h}})$, where $\store' \isdef \store[y_1
  \leftarrow c'_1, \ldots, y_m \leftarrow c'_m]$. We distinguish the
following cases: \begin{compactitem}
\item if $(c_1, p_1, \ldots, c_n, p_n) \in \interacs$ because of an
  interaction atom $\interacn{z_1}{p_1}{z_n}{p_n}$ from $\phi$, such
  that $\store'(z_i) = c_i$, for all $i \in \interv{1}{n}$, then
  $(\comps,\interacs,\statemap) \models_{\psid}^{\store'[y \leftarrow
      c_i]} \phi * y = z_i * \bpred_1(t^1_1, \ldots,
  t^1_{\arityof{\bpred_1}}) * \ldots * \bpred_h(t^h_1, \ldots,
  t^h_{\arityof{\bpred_h}})$, hence $(\comps,\interacs,\statemap)
  \models_{\psid}^{\store[y \leftarrow c_i]} \appred(x_1, \ldots,
  x_{\arityof{\apred}}, y)$, by the definition of $\psid$.
\item else $(c_1, p_1, \ldots, c_n, p_n) \in \interacs$ because of a
  configuration $\aconfig'$, such that $\aconfig = \aconfig' \comp
  \aconfig''$, for some configuration $\aconfig''$, and $\aconfig'
  \models^{\store'}_\asid \bpred_i(t^i_1, \ldots,
  t^i_{\arityof{\bpred_i}})$. By the inductive hypothesis, we obtain
  $\aconfig' \models^{\store'[y \leftarrow c_i]}_\asid
  \bppred_i(t^i_1, \ldots, t^i_{\arityof{\bpred_i}},y)$, hence
  $\aconfig \models^{\store[y \leftarrow c_i]}_\asid \appred(x_1,
  \ldots, x_{\arityof{\apred}},y)$, because $\psid$ contains the rule:
  \[\begin{array}{rcl}
  \appred(x_1, \ldots, x_{\arityof{\apred}+1}) & \leftarrow & \exists y_1 \ldots \exists y_m ~.~ \phi * 
  \bpred_1(t^1_1, \ldots, t^1_{\arityof{\bpred_1}}) * \ldots * \bppred_i(t^i_1, \ldots, t^i_{\arityof{\bpred_i}}, x_{\arityof{\apred}+1}) \\
  && \hspace*{2cm} * \ldots * \bpred_h(t^h_1, \ldots, t^h_{\arityof{\bpred_h}})
  \end{array}\]
  and $\aconfig'' \models^{\store'} \phi * \bpred_1(t^1_1, \ldots,
  t^1_{\arityof{\bpred_1}}) * \ldots * \bppred_{i-1}(t^{i-1}_1,
  \ldots, t^{i-1}_{\arityof{\bpred_{i-1}}}) * \bppred_{i+1}(t^{i+1}_1,
  \ldots, t^{i+1}_{\arityof{\bpred_{i+1}}}) * \ldots * \bpred_h(t^h_1,
  \ldots, t^h_{\arityof{\bpred_h}})$ follows from $\aconfig =
  \aconfig' \comp \aconfig''$. 
\end{compactitem}

\noindent''$\Leftarrow$'' Let $\aconfig \isdef
(\comps,\interacs,\statemap)$ be a configuration and $\store$ be a
store, such that $\aconfig \models^{\store}_{\psid}
\looseof{\apred}(x_1, \ldots, x_{\arityof{\apred}})$. Since the only
rule of $\psid$ that defines $\looseof{\apred}$ is:
\[\looseof{\apred}(x_1, \ldots, x_{\arityof{\apred}}) \leftarrow \exists y ~.~ \appred(x_1, \ldots, x_{\arityof{\apred}}, y) * \compact{y}\]
there exists a component $c \in \comps$, such that
$(\comps\setminus\set{c},\interacs,\statemap) \models^{\store[y
    \leftarrow c]}_\psid \appred(x_1, \ldots, x_{\arityof{\apred}},
y)$. We prove the following: \begin{compactitem}
\item there exists an interaction $(c_1, p_1, \ldots, c_n, p_n) \in
  \interacs$, such that $c_i = c$, and
\item $(\comps\setminus\set{c},\interacs,\statemap)
  \models^{\store}_\asid \apred(x_1, \ldots, x_{\arityof{\apred}})$,
\end{compactitem}
by fixpoint induction on the definition of
$(\comps\setminus\set{c},\interacs,\statemap) \models^{\store[y
    \leftarrow c]}_\psid \appred(x_1, \ldots, x_{\arityof{\apred}},
y)$. Based on the definition of $\psid$, we distinguish the following
cases, where $\phi$ is quantifier- and predicate-free, $c'_1, \ldots,
c'_m \in \universe$ are components and $\store' \isdef \store[y_1
  \leftarrow c'_1, \ldots, y_m \leftarrow c'_m]$: \begin{itemize}
\item $(\comps\setminus\set{c},\interacs,\statemap) \models^{\store'[y
    \leftarrow c]}_\psid \phi ~*~ y = z ~*~ \bpred_1(t^1_1, \ldots,
  t^1_{\arityof{\bpred_1}}) ~* \ldots *~ \bpred_h(t^h_1, \ldots,
  t^h_{\arityof{\bpred_h}})$, where $z$ occurs in an interaction atom
  from $\phi$. In this case, there exists an interaction $(c_1, p_1,
  \ldots, c_n, p_n) \in \interacs$, such that $c=c_i$, for some $i \in
  \interv{1}{n}$. Moreover,
  $(\comps\setminus\set{c},\interacs,\statemap) \models^{\store}_\asid
  \apred(x_1, \ldots, x_{\arityof{\apred}})$, because $\asid$ has a
  rule:
  \[\apred(x_1, \ldots, x_{\arityof{\apred}}) \leftarrow \exists y_1 \ldots \exists y_m ~.~ \phi *
  \bpred_1(t^1_1, \ldots, t^1_{\arityof{\bpred_1}}) * \ldots *
  \bpred_h(t^h_1, \ldots, t^h_{\arityof{\bpred_h}})\] such that
  $(\comps\setminus\set{c},\interacs,\statemap) \models^{\store}_\asid
  \phi * \bpred_1(t^1_1, \ldots, t^1_{\arityof{\bpred_1}}) ~* \ldots *~
  \bpred_h(t^h_1, \ldots, t^h_{\arityof{\bpred_h}})$.
\item $(\comps\setminus\set{c},\interacs,\statemap) \models^{\store'[y
    \leftarrow c]}_\psid \phi * \bpred_1(t^1_1, \ldots,
  t^1_{\arityof{\bpred_1}}) * \ldots * \bppred_i(t^i_1, \ldots,
  t^i_{\arityof{\bpred_i}}, y) * \ldots * \bpred_h(t^h_1, \ldots,
  t^h_{\arityof{\bpred_h}})$. In this case, there exists
  configurations $\aconfig'$ and $\aconfig''$, such that
  $(\comps\setminus\set{c},\interacs,\statemap) = \aconfig' \comp
  \aconfig''$, $\aconfig' \models^{\store'[y \leftarrow c]}_\psid
  \bppred_i(t^i_1, \ldots, t^i_{\arityof{\bpred_i}}, y)$ and
  $\aconfig'' \models^\store_\asid \phi * \bpred_1(t^1_1, \ldots,
  t^1_{\arityof{\bpred_1}}) * \ldots * \bpred_h(t^h_1, \ldots,
  t^h_{\arityof{\bpred_h}})$. By the inductive hypothesis, there
  exists an interaction $(c_1,p_1,\ldots,c_n,p_n)$ in $\aconfig'$,
  such that $c=c_i$, for some $i\in\interv{1}{n}$ and $\aconfig'
  \models^\store_\asid \bpred_i(t^i_1, \ldots,
  t^i_{\arityof{\bpred_i}})$. Then $(c_1,p_1,\ldots,c_n,p_n) \in
  \interacs$ and $(\comps\setminus\set{c},\interacs,\statemap) \models^\store_\asid
  \apred(x_1, \ldots, x_{\arityof{\apred}})$, since $\asid$ has a rule:
  \[\apred(x_1, \ldots, x_{\arityof{\apred}}) \leftarrow \exists y_1 \ldots \exists y_m ~.~ \phi *
  \bpred_1(t^1_1, \ldots, t^1_{\arityof{\bpred_1}}) * \ldots *
  \bpred_h(t^h_1, \ldots, t^h_{\arityof{\bpred_h}})\] such that
  $\aconfig' \comp \aconfig'' \models^{\store'}_\asid \phi *
  \bpred_1(t^1_1, \ldots, t^1_{\arityof{\bpred_1}}) ~* \ldots *~
  \bpred_h(t^h_1, \ldots, t^h_{\arityof{\bpred_h}})$.
\item $(\comps\setminus\set{c},\interacs,\statemap) \models^{\store'[y
    \leftarrow c]}_\psid \phi * \compact{y}$ this case contradicts the
  semantics of \cl. \qed
\end{itemize}}

\paragraph{From Satisfiability to Looseness.}
Given a SID $\asid$ and a predicate $\apred$, we build a SID $\psid$
that defines a predicate $\looseof{\apred}$, of equal arity, not
occurring in $\asid$, such that $\sat{\asid}{\apred}$ has a positive
answer if and only if there exists a loose configuration $\aconfig$
and a store $\store$, such that $\aconfig \models^\store_\psid
\looseof{\apred}(x_1, \ldots, x_{\arityof{\apred}})$. The rules of
$\psid$ are the rules of $\asid$, to which the following rule is
added, for some ports $p_1, p_2 \in \ports$: 
\[\looseof{\apred}(x_1, \ldots, x_{\arityof{\apred}}) \leftarrow
\exists y_1 \exists y_2 ~.~ \apred(x_1, \ldots, x_{\arityof{\apred}})
* \interactwo{y_1}{p_1}{y_2}{p_2}\] This reduction is polynomial,
because we add one rule, of size linear on $\arityof{\apred}$.
The following lemma states the correctness of the reduction:

\begin{lemma}\label{lemma:sat-loose}
  Given a SID $\asid$ and a predicate $\apred$, the problem
  $\sat{\asid}{\apred}$ has a positive answer if and only if the
  problem $\loose{\psid}{\looseof{\apred}}$ has a positive answer.
\end{lemma}
\proof{ ``$\Rightarrow$'' If $\sat{\asid}{\apred}$ has a positive
  answer, there exists a configuration $\aconfig \isdef
  (\comps,\interacs,\statemap)$ and a store $\store$, such that
  $\aconfig \models^{\store}_\asid \apred(x_1, \ldots,
  x_{\arityof{\apred}})$. Consider the configuration $\aconfig' \isdef
  (\emptyset, \set{(c_1,p_1,c_2,p_2)}, \statemap)$, for some
  components $c_1,c_2 \not\in \comps$. Then the composition $\aconfig
  \comp \aconfig'$ is defined and we have $\aconfig \comp \aconfig'
  \models^{\store[y_1\leftarrow c_1,y_2 \leftarrow c_2]}_\asid
  \apred(x_1, \ldots, x_{\arityof{\apred}}) *
  \interactwo{y_1}{p_1}{y_2}{p_2}$, leading to $\aconfig \comp
  \aconfig' \models^\store_\psid \looseof{\apred}(x_1, \ldots,
  x_{\arityof{\apred}})$. Moreover, $\aconfig\comp\aconfig'$ is loose,
  because $c_1, c_2 \not\in\comps$. ``$\Leftarrow$'' If $\aconfig
  \models^\store_\psid\looseof{\apred}(x_1, \ldots,
  x_{\arityof{\apred}})$, we necessarily have $\aconfig
  \models^{\store[y_1 \leftarrow c_1, y_2 \leftarrow c_2]}_\asid
  \apred(x_1, \ldots, x_{\arityof{\apred}}) *
  \interactwo{y_1}{p_1}{y_2}{p_2}$, for some components $c_1, c_2 \in
  \universe$, hence there exists a configuration $\aconfig'$, such
  that $\aconfig' \models^\store_\asid \apred(x_1, \ldots,
  x_{\arityof{\apred}})$. \qed}

The polynomial reductions from Lemmas \ref{lemma:loose-sat} and
\ref{lemma:sat-loose} establish the following complexity bounds for
the tightness problem:

\begin{theorem}\label{thm:tightness}
  $\kltight{\asid}{\apred}{k}{\infty}$ is \conp-complete,
  $\kltight{\asid}{\apred}{\infty}{\ell}$ is \exptime-complete and
  $\tight{\asid}{\apred}$ is \twoexptime.
\end{theorem}
\proof{ Since $\klloose{\asid}{\apred}{k}{\infty}$ is
  polynomially-reducible to $\klsat{\asid}{\apred}{k+1}{\infty}$, by
  Theorem \ref{thm:sat}, we obtain that
  $\klloose{\asid}{\apred}{k}{\infty}$ is in \np. Moreover, since
  $\klsat{\asid}{\apred}{k}{\infty}$ is polynomially-reducible to
  $\klloose{\asid}{\apred}{k}{\infty}$, by Theorem \ref{thm:sat}, we
  obtain that $\klloose{\asid}{\apred}{k}{\infty}$ is
  \np-complete. Because $\kltight{\asid}{\apred}{k}{\infty}$ is the
  complement of $\klloose{\asid}{\apred}{k}{\infty}$, we obtain that
  $\kltight{\asid}{\apred}{k}{\infty}$ is \conp-complete. The rest of
  the bounds are obtained by the same polynomial reductions and the
  fact that $\kltight{\asid}{\apred}{k}{\infty}$ is the complement of
  $\klloose{\asid}{\apred}{k}{\infty}$, for any $k$ and $\ell$, either
  integer constants, or infinity. \qed}

\fi

\section{Degree Boundedness}
\label{sec:boundedness}

The boundedness problem (Def. \ref{def:decision}, point
\ref{decision:bound}) asks for the existence of a bound on the degree
(Def. \ref{def:degree}) of the models of a sentence $\exists x_1
\ldots \exists x_{\arityof{\apred}} ~.~
\apred(x_1,\ldots,x_{\arityof{\apred}})$. Intuitively, the
$\bound{\asid}{\apred}$ problem has a negative answer if and only if
there are increasingly large unfoldings (i.e., expansions of a formula
by replacement of a predicate atom with one of its definitions) of
$\apred(x_1, \ldots, x_{\arityof{\apred}})$ repeating a rule that
contains an interaction atom involving a parameter of the rule, which
is always bound to the same component. \ifLongVersion For instance,
the rule $\mathit{Worker}(x) \leftarrow \exists y ~.~
\interactwo{x}{out}{y}{in} * \compact{y} * \mathit{Worker}(x)$
(Example \ref{ex:star}) declares an unbounded number of interactions
$\interactwo{x}{out}{y}{in}$ involving the component to which $x$ is
bound.\fi We formalize the notion of unfolding below:

\begin{definition}\label{def:unfolding}
  Given a predicate $\apred$ and a sequence $(\arule_1, i_1), \ldots,
  (\arule_n, i_n) \in \left(\asid\times\nat\right)^+$, where
  $\arule_1$ is the rule $\apred(x_1, \ldots, x_{\arityof{\apred}})
  \leftarrow \phi \in \asid$, the \emph{unfolding} $\apred(x_1,
  \ldots, x_{\arityof{\apred}}) \unfold{(\arule_1, i_1) \ldots
    (\arule_n, i_n)}{\asid} \psi$ is inductively defined
  as \begin{inparaenum}[(1)]
  \item $\psi = \phi$ if $n=1$, and
  \item$\psi$ is obtained from $\phi$ by replacing its $i_1$-th
    predicate atom $\bpred(y_1, \ldots, y_{\arityof{\bpred}})$ with
    $\psi_1[x_1/y_1, \ldots,
      x_{\arityof{\bpred}}/y_{\arityof{\bpred}}]$, where $\bpred(x_1,
    \ldots, x_{\arityof{\bpred}}) \unfold{(\arule_2,i_2) \ldots
      (\arule_n,i_n)}{\asid} \psi_1$ is an unfolding, if $n>1$.
  \end{inparaenum}
\end{definition}

We show that the $\bound{\asid}{\apred}$ problem can be reduced to the
existence of increasingly large unfoldings or, equivalently, a cycle
in a finite directed graph, built by a variant of the least fixpoint
iteration algorithm used to solve the satisfiability problem
(Fig. \ref{fig:base-graph}).

\begin{definition}\label{def:dependency}
Given satisfiable base pairs $\abasetuple,\anbasetuple \in
\satbasetuples$ and a rule from $\asid$:
\vspace*{-.25\baselineskip}
\[\arule ~:~ \apred(x_1, \ldots, x_{\arityof{\apred}}) \leftarrow
\exists y_1 \ldots \exists y_m ~.~ \phi * \bpred_1(z^1_1, \ldots,
z^1_{\arityof{\bpred_1}}) * \ldots * \bpred_h(z^h_1, \ldots,
z^h_{\arityof{\bpred_h}})\] where $\phi$ is a quantifier- and
predicate-free formula, we write $(\apred,\abasetuple)
\depof{\arule}{i} (\bpred,\anbasetuple)$ if and only if $\bpred =
\bpred_i$ and there exist satisfiable base tuples $\abasetuple_1,
\ldots, \anbasetuple = \abasetuple_i, \ldots, \abasetuple_h \in
\satbasetuples$, such that \(\abasetuple \in
\proj{\big(\basetupleof{\phi}{\set{x_1, \ldots, x_{\arityof{\apred}}}}
  \basecomp \Basecomp_{\ell=1}^h \abasetuple_\ell[x_1/z^\ell_1,
    \ldots,
    x_{\arityof{\bpred_\ell}}/z^\ell_{\arityof{\bpred_\ell}}]\big)}{x_1,
  \ldots, x_{\arityof{\apred}}}\). We define the directed graph with
edges labeled by pairs $(\arule,i) \in \asid \times \nat$:
\vspace*{-.5\baselineskip}
\begin{align*}
\graphof{\asid} & \isdef
\big(\set{\defnof{\asid} \times \satbasetuples},
\set{\tuple{(\apred,\abasetuple),(\arule,i),(\bpred,\anbasetuple)} \mid (\apred,\abasetuple) \depof{\arule}{i} (\bpred,\anbasetuple)}\big) 
\end{align*}
\end{definition}

\begin{figure}[t!]
  {\small\begin{algorithmic}[0]
    \State \textbf{input}: a SID $\asid$
    \State \textbf{output}: $\graphof{\asid} = (V,E)$
  \end{algorithmic}}
  {\small\begin{algorithmic}[1]
    \State initially $V := \emptyset$, $E := \emptyset$
    
    \For{$\apred(x_1,\ldots,x_{\arityof{\apred}}) \leftarrow \exists y_1 \ldots \exists y_m ~.~ \phi \in \asid$, with $\phi$ quantifier- and predicate-free}

    \State $V := V \cup \left(\set{\apred} \times \proj{\basetupleof{\phi}{\set{x_1,\ldots,x_{\arityof{\apred}}}}}{x_1,\ldots,x_{\arityof{\apred}}}\right)$

    \EndFor

    \While{$V$ or $E$ still change}

    \For{$\arule : \apred(x_1,\ldots,x_{\arityof{\apred}}) \leftarrow \exists y_1 \ldots \exists y_m ~.~
      \phi * \Asterisk_{\ell=1}^h \bpred_\ell(z^\ell_1,\ldots,z^\ell_{\arityof{\bpred_\ell}}) \in \asid$}

    \If{there exist $(\bpred_1,\abasetuple_1), \ldots, (\bpred_h,\abasetuple_h) \in V$}

    \State $\mathcal{X} := 
    \proj{\left(\basetupleof{\phi}{\set{x_1,\ldots,x_{\arityof{\apred}}}}
      \basecomp \Basecomp_{\ell=1}^h \abasetuple_\ell[x_1/z^\ell_1,
          \ldots,
          x_{\arityof{\bpred_\ell}}/z^\ell_{\arityof{\bpred_\ell}}]\right)}{x_1,\ldots,x_{\arityof{\apred}}}$    

    \State $V := V \cup (\set{\apred} \times \mathcal{X})$

    \State $E := E \cup \set{\tuple{(\apred,\abasetuple),(\arule,\ell),(\bpred_\ell,\abasetuple_\ell)} \mid \abasetuple \in \mathcal{X},\ell\in\interv{1}{h}}$
    
    \EndIf \EndFor \EndWhile
  \end{algorithmic}}
  \caption{Algorithm for the Construction of $\graphof{\asid}$}
  \label{fig:base-graph}
  \vspace*{-\baselineskip}
\end{figure}

The graph $\graphof{\asid}$ is built by the algorithm in
Fig. \ref{fig:base-graph}, a slight variation of the classical Kleene
iteration algorithm for the computation of the least solution of the
constraints of the form (\ref{eq:basesid})\ifLongVersion (see
Fig. \ref{fig:base-sets})\fi. A path $(\apred_1,\abasetuple_1)
\depof{\arule_1}{i_1} (\apred_2,\abasetuple_2) \depof{\arule_2}{i_2}
\ldots \depof{\arule_n}{i_n} (\apred_n,\abasetuple_n)$ in
$\graphof{\asid}$ induces a unique unfolding
$\apred_1(x_1,\ldots,x_{\arityof{\apred_1}}) \unfold{(\arule_1,i_1)
  \ldots (\arule_n,i_n)}{\asid} \phi$
(Def. \ref{def:unfolding}). Since the vertices of $\graphof{\asid}$
are pairs $(\apred,\abasetuple)$, where $\abasetuple$ is a satisfiable
base tuple and the edges of $\graphof{\asid}$ reflect the construction
of the base tuples from the least solution of the constraints
(\ref{eq:basesid}), the outcome $\phi$ of this unfolding is always a
satisfiable formula.

\begin{myLemmaE}\label{lemma:unfold-soundness}
  Given a path $(\apred_0,\abasetuple_0) \depof{\arule_1}{i_1} \ldots
  \depof{\arule_n}{i_n} (\apred_n,\abasetuple_n)$ in
  $\graphof{\asid}$, where $\abasetuple_0 = \basetuple$, a state map
  $\statemap$ and a store $\store$, such that
  $(\emptyset,\emptyset,\statemap) \models^\store \pureform$, there
  exists a configuration $(\comps,\interacs,\statemap)$, such that
  $(\comps,\interacs,\statemap) \models^\store_\asid \phi$, where
  $\apred_0(x_1, \ldots, x_{\arityof{\apred_0}})
  \unfold{(\arule_1,i_1) \ldots (\arule_n,i_n)}{\asid} \phi$ is the
  unique unfolding corresponding to the path.
\end{myLemmaE}
\begin{proofE}
  Let $\arule_1$ be the following rule:
  \[\apred_0(x_1, \ldots, x_{\arityof{\apred_0}}) \leftarrow \phi \text{, where }
  \phi = \exists y_1 \ldots \exists y_m ~.~ \psi * \pureform *
  \Asterisk_{\ell=2}^h \bpred_\ell(z^\ell_1, \ldots,
  z^\ell_{\arityof{\bpred_\ell}})\] and $\psi*\pureform$ is a
  quantifier- and predicate-free formula and $\pureform$ is, moreover,
  pure. The proof goes by induction on the length $n \geq 1$ of the
  path. For the base case $n=1$, by Def. \ref{def:dependency}, the
  edge $(\apred_0,\abasetuple_0) \depof{\arule_1}{i_1}
  (\apred_1,\abasetuple_1)$ implies the existence of base tuples
  $\anbasetuple_\ell \in \leastbaseof{\bpred_\ell}$, for all $\ell \in
  \interv{2}{h}$, such that $\bpred_{i_1} = \apred_1$,
  $\anbasetuple_{i_1-1} = \abasetuple_1$ and:
  \[\abasetuple_0 \in \proj{\left(\basetupleof{\psi*\pureform}{\set{x_1,\ldots,x_{\arityof{\apred}}}}
      \basecomp \Basecomp_{\ell=2}^h \anbasetuple_\ell[x_1/z^\ell_1,
        \ldots,
        x_{\arityof{\bpred_\ell}}/z^\ell_{\arityof{\bpred_\ell}}]\right)}{x_1,\ldots,x_{\arityof{\apred}}}\]
  Let $\anbasetuple_\ell \isdef \basetuplen{\ell}$, for all $\ell \in
  \interv{2}{h}$ and $\pureform' \isdef \pureform *
  \Asterisk_{\ell=2}^h \pureform_\ell$. Since $\abasetuple_0$ is
  satisfiable, there exists a store $\store'$, that agrees with
  $\store$ over $x_1, \ldots, x_{\arityof{\apred_0}}$, such that,
  moreover:
  \[\begin{array}{rl}
  \store'(x) = \store'(y) \text{ only if } x \formeq{\pureform'} y,
  & \text{for all } x, y \in \fv{\psi * \pureform'} \cup \hspace*{3.5cm} (\dagger) \\
  & \bigcup_{\ell=2}^h \big(\basecomps_\ell \cup  \set{z_i \mid \tuple{z_1, \ldots, z_n} \in \baseinteracs_\ell(\intertype),
    \intertype\in\intertypes}\big)
    \end{array}\] 
  We define the configurations $(\comps_1,\interacs_1,\statemap),
  \ldots, (\comps_h,\interacs_h,\statemap)$ inductively, as
  follows: \begin{compactitem}
  \item $\comps_1 \isdef \set{\store'(y) \mid \compact{y} \text{ occurs
      in } \psi}$,
  \item $\interacs_1 \isdef \set{(\store'(z_1), p_1, \ldots,
    \store'(z_s), p_s) \mid \tuple{z_1.p_1, \ldots, z_t.p_t} \text{
      occurs in } \psi}$,
  \item for all $\ell \in \interv{2}{h}$, assuming $\comps_1, \ldots,
    \comps_{\ell-1}$ and $\interacs_1, \ldots, \interacs_{\ell-1}$ are
    defined, let:
    \[\begin{array}{rcl}
    \mathcal{D}_\ell & \isdef & \bigcup_{i=1}^{\ell-1} \comps_i \cup \bigcup_{i=\ell+1}^h \store'(\basecomps_i) \\
    \mathcal{J}_\ell & \isdef & \bigcup_{i=1}^{\ell-1} \interacs_i \cup \bigcup_{i=\ell+1}^h \store'(\baseinteracs_i)
    \end{array}\]
  \end{compactitem}
  We prove first that $\mathcal{D}_\ell \cap \store'(\basecomps_\ell) =
  \emptyset$ and $\mathcal{J}_\ell \cap \store'(\baseinteracs_\ell) =
  \emptyset$ (we prove only the first point, the second uses a similar
  reasoning), by induction on $\ell\in\interv{2}{h}$. For the base case
  $\mathcal{D}_2 \cap \store'(\basecomps_2) \neq \emptyset$, we
  prove the points below: \begin{compactitem}
    \item $\comps_1 \cap \store'(\basecomps_2) = \emptyset$: suppose,
      for a contradiction, that there exists $c \in \comps_1 \cap
      \store'(\basecomps_2)$, then $c = \store'(y)$, for a component
      atom $\compact{y}$ from $\psi$ and $c = \store'(x)$, for some $x
      \in \basecomps_2$. By ($\dagger$), we obtain $x
      \formeq{\pureform'} y$, contradicting the existence of
      $\abasetuple_0$.
    \item $\store'(\basecomps_i) \cap \store'(\basecomps_2) =
      \emptyset$, for some $i \in \interv{3}{h}$: suppose, for a
      contradiction, that there exists $c \in \store'(\basecomps_i)
      \cap \store'(\basecomps_2)$, then $c = \store'(x) = \store'(y)$,
      for some $x \in \basecomps_i$ and $y \in \basecomps_2$. By
      ($\dagger$), we obtain $x \formeq{\pureform'} y$, contradicting
      the existence of $\abasetuple_0$.
  \end{compactitem}
  We assume that $\mathcal{D}_j \cap \store'(\basecomps_j) =
  \emptyset$, for all $j \in \interv{2}{\ell-1}$. By Lemma
  \ref{lemma:sat-soundness}, there exist configurations $(\comps_j,
  \interacs_j, \statemap)$, such that $\comps_j \cap \mathcal{D}_j =
  \emptyset$ ($\ddagger$) and $(\comps_j, \interacs_j, \statemap)
  \models^{\store'} \bpred_j(x_1, \ldots, x_{\arityof{\bpred_j}})$,
  for all $j \in \interv{2}{\ell-1}$. We prove $\mathcal{D}_\ell \cap
  \store'(\basecomps_\ell) = \emptyset$, by showing the following
  points: \begin{compactitem}
  \item $\comps_j \cap \store'(\basecomps_\ell) = \emptyset$, for all
    $j \in \interv{1}{\ell-1}$: suppose, for a contradiction, that
    there exists $c \in \comps_j \cap \store'(\basecomps_\ell)$, for
    some $j \in \interv{1}{\ell-1}$, then $c \in \comps_j \cap
    \mathcal{D}_j$, because $\mathcal{D}_j \subseteq
    \store'(\basecomps_\ell)$, in contradiction with $\comps_j \cap
    \mathcal{D}_j = \emptyset$ ($\ddagger$).
  \item $\store'(\basecomps_j) \cap \store'(\basecomps_\ell) =
    \emptyset$, for all $j \in \interv{2}{\ell-1}$: suppose, for a
    contradiction, that there exists $c \in \store'(\basecomps_j) \cap
    \store'(\basecomps_\ell)$, then $c = \store'(x) = \store'(y)$, for
    some $x \in \basecomps_j$ and $y \in \basecomps_\ell$. By
    ($\dagger$), we obtain $x \formeq{\pureform'} y$, contradicting
    the existence of $\abasetuple_0$.
  \end{compactitem}
  Consequently, $\mathcal{D}_\ell \cap \store'(\basecomps_\ell) =
  \emptyset$, for all $\ell\in\interv{2}{h}$. By Lemma
  \ref{lemma:sat-soundness}, there exists a configuration
  $(\comps_\ell, \interacs_\ell, \statemap)$, such that $\comps_\ell
  \cap \mathcal{D}_\ell = \emptyset$ and $(\comps_\ell,
  \interacs_\ell, \statemap) \models^{\store'} \bpred_\ell(x_1,
  \ldots, x_{\arityof{\bpred_\ell}})$, for all
  $\ell\in\interv{2}{h}$. We obtain that $\comps_i \cap \comps_j =
  \emptyset$ and $\interacs_i \cap \interacs_j = \emptyset$, for all
  $1 \leq i < j \leq h$, meaning that the configuration
  $(\comps,\interacs,\statemap) \isdef (\comps_1, \interacs_1,
  \statemap) \comp \ldots \comp (\comps_h, \interacs_h, \statemap)$ is
  defined, which leads to $(\comps,\interacs,\statemap)
  \models^{\store'} \phi$.

  \vspace*{\baselineskip}\noindent For the inductive step $n>1$, by
  Def. \ref{def:dependency}, there exists base tuples
  $\anbasetuple_\ell \in \leastbaseof{\bpred_\ell}$, for all $\ell \in
  \interv{2}{h}$, such that $\apred_1 = \bpred_{i_1}$,
  $\anbasetuple_{i_1-1} = \abasetuple_1$ and:
  \[\abasetuple_0 \in \proj{\left(\basetupleof{\psi*\pureform}{\set{x_1,\ldots,x_{\arityof{\apred}}}}
      \basecomp \Basecomp_{\ell=1}^h \anbasetuple_\ell[x_1/z^\ell_1,
        \ldots,
        x_{\arityof{\bpred_\ell}}/z^\ell_{\arityof{\bpred_\ell}}]\right)}{x_1,\ldots,x_{\arityof{\apred}}}\]
  Then there exists a store $\store'$ that agrees with $\store$ over
  $x_1, \ldots, x_{\arityof{\apred_0}}$ and satisfies ($\dagger$). Let
  $\store'' \isdef \store'[x_1/z^{i_1}_1, \ldots,
    x_{\arityof{\apred_1}}/z^{i_1}_{\arityof{\apred_1}}]$. By the
  inductive hypothesis, since $(\apred_1,\abasetuple_1)
  \depof{\arule_2}{i_2} \ldots \depof{\arule_n}{i_n}
  (\apred_n,\abasetuple_n)$ is a path in $\graphof{\asid}$, there
  exists a configuration $(\comps', \interacs', \statemap)$, such that
  $(\comps', \interacs', \statemap) \models^{\store''}
  \apred_1(z^{i_1}_1, \ldots, z^{i_1}_{\arityof{\apred_1}})$, because
  $(\comps', \interacs', \statemap) \models^{\store'} \phi_1$, for the
  unfolding $\apred_1(x_1, \ldots, x_{\arityof{\apred_1}})
  \unfold{(\arule_2, i_2) \ldots (\arule_n, i_n)}{\asid} \phi_1$. The
  required configuration is defined as $(\comps,\interacs,\statemap)
  \isdef (\comps_1,\interacs_1,\statemap) \comp \ldots \comp
  (\comps_{i_1-1},\interacs_{i_1-1},\statemap) \comp
  (\comps',\interacs',\statemap) \comp
  (\comps_{i_1+1},\interacs_{i_1+1},\statemap) \comp \ldots \comp
  (\comps_h,\interacs_h,\statemap)$, where
  $(\comps_1,\interacs_1,\statemap)$, $\ldots$,
  $(\comps_{i_1-1},\interacs_{i_1-1},\statemap)$ and
  $(\comps_{i_1+1},\interacs_{i_1+1},\statemap)$, $\ldots$,
  $(\comps_h,\interacs_h,\statemap)$ are defined as in the base case,
  by taking $\store''$ instead of $\store'$ and defining, for all
  $\ell \in \interv{2}{h} \setminus \set{i_1}$:
  \[\begin{array}{rcl}
  \mathcal{D}_\ell & \isdef & \comps' \cup \bigcup_{i=1}^{\ell-1} \comps_i \cup \bigcup_{i=\ell+1}^h \store'(\basecomps_i) \\
  \mathcal{J}_\ell & \isdef & \interacs' \cup \bigcup_{i=1}^{\ell-1} \interacs_i \cup \bigcup_{i=\ell+1}^h \store'(\baseinteracs_i) 
  \end{array}\]
  The proof of the fact that $\comps_1$, $\ldots$, $\comps_{i_1-1}$,
  $\comps'$, $\comps_{i_1+1}$, $\ldots$, $\comps_h$ and $\interacs_1$,
  $\ldots$, $\interacs_{i_1-1}$, $\interacs'$, $\interacs_{i_1+1}$,
  $\ldots$, $\interacs_h$ are pairwise disjoint, respectively, follows
  by the same argument as in the base case. \qed
\end{proofE}

\begin{myLemmaE}\label{lemma:unfold-completeness}
  Given an unfolding $\apred_0(x_1, \ldots, x_{\arityof{\apred_0}})
  \unfold{(\arule_1,i_1) \ldots (\arule_n,i_n)}{\asid} \phi$, a
  configuration $(\comps,\interacs,\statemap)$ and a store $\store$,
  such that $(\comps,\interacs,\statemap)\models^\store_\asid \phi$,
  then $\graphof{\asid}$ has a path $(\apred_0,\basetuplen{0})
  \depof{\arule_1}{i_1} \ldots$ \\ $\depof{\arule_n}{i_n}
  (\apred_n,\basetuplen{n})$, for some $\basetuplen{0}, \ldots,
  \basetuplen{n} \in \satbasetuples$, such that $\store(\basecomps_0)
  \subseteq \comps_0$, $\store(\baseinteracs_0) \subseteq \interacs_0$
  and $(\emptyset,\emptyset,\statemap) \models^\store \pureform_0$.
\end{myLemmaE}
\begin{proofE}
  Let $\arule_1$ be the following rule:
  \[\apred_0(x_1, \ldots, x_{\arityof{\apred_0}}) \leftarrow
  \exists y_1 \ldots \exists y_m ~.~ \psi * \pureform *
  \Asterisk_{\ell=2}^h \bpred_\ell(z^\ell_1, \ldots,
  z^\ell_{\arityof{\bpred_\ell}})\] and $\psi*\pureform$ is a
  quantifier- and predicate-free formula and $\pureform$ is, moreover,
  pure. The proof goes by induction on the length $n \geq 1$ of the
  path. For the base case $n=1$, we have $\phi = \exists y_1 \ldots
  \exists y_m ~.~ \psi * \pureform * \Asterisk_{\ell=2}^h
  \bpred_\ell(z^\ell_1, \ldots, z^\ell_{\arityof{\bpred_\ell}})$,
  hence there exists a store $\store'$, that agrees with $\store$ over
  $x_1, \ldots, x_{\arityof{\apred_0}}$, and configurations $(\comps_1,
  \interacs_1, \statemap), \ldots, (\comps_h, \interacs_h,
  \statemap)$, such that: \begin{compactitem}
  \item $(\comps_1, \interacs_1, \statemap) \models^{\store'} \psi *
    \pureform$,
  \item $(\comps_\ell, \interacs_\ell, \statemap)
    \models^{\store'}_{\asid} \bpred_\ell(z^\ell_1, \ldots,
    z^\ell_{\arityof{\bpred_\ell}})$, for all $\ell \in
    \interv{2}{h}$, and
  \item $\aconfig = (\comps_1, \interacs_1, \statemap) \comp \ldots
    \comp (\comps_h, \interacs_h, \statemap)$.
  \end{compactitem}
  We consider the following base tuples: \begin{compactitem}
  \item $\pbasetuplen{1} \isdef \basetupleof{\psi*\pureform}{\set{x_1,
      \ldots, x_{\arityof{\apred_0}}}}$, 
  \item for all $\ell \in \interv{2}{h}$, there exist
    $\pbasetuplen{\ell} \in \leastbaseof{\bpred_\ell}[x_1/z^\ell_1,
    \ldots,
    x_{\arityof{\bpred_\ell}}/z^\ell_{\arityof{\bpred_\ell}}]$, such
    that $\comps_\ell \subseteq \store'(\pbasecomps_\ell)$,
    $\interacs_\ell \subseteq \store'(\pbaseinteracs_\ell)$ and
    $(\emptyset,\emptyset,\statemap) \models^{\store'}
    \ppureform_\ell$, by Lemma \ref{lemma:sat-completeness}.
  \end{compactitem}
  By similar argument to the one from the proof of Lemma
  \ref{lemma:sat-completeness} (base case), we show that the
  composition $\Basecomp_{\ell=1}^h \pbasetuplen{\ell}$ is defined and
  let $\basetuplen{0} \isdef \proj{\left(\Basecomp_{\ell=1}^h
    \pbasetuplen{\ell} \right)}{x_1, \ldots,
    x_{\arityof{\apred}}}$. Moreover, we obtain $\store(\basecomps_0)
  \subseteq \comps_0$, $\store(\baseinteracs_0) \subseteq \interacs_0$
  and $(\emptyset,\emptyset,\statemap) \models^\store \pureform_0$, as
  in the proof of Lemma \ref{lemma:sat-completeness}. Then, by
  Def. \ref{def:dependency}, $\graphof{\asid}$ has an edge $(\apred_0,
  \basetuplen{0}) \depof{\arule_1}{i_1} (\bpred_{i_1-1},
  \pbasetuplen{i_1-1})$.

  \vspace*{\baselineskip}\noindent For the inductive step $n>1$, let
  $\bpred_{i_1-1}(x_1, \ldots, x_{\arityof{\bpred_{i_1-1}}})
  \unfold{(\arule_2,i_2) \ldots (\arule_n,i_n)}{\asid} \phi_1$ be an
  unfolding, such that $\phi$ is obtained from $\exists y_1 \ldots
  \exists y_m ~.~ \psi * \pureform * \Asterisk_{\ell=2}^h
  \bpred_\ell(z^\ell_1, \ldots, z^\ell_{\arityof{\bpred_\ell}})$, by
  replacing $\bpred_{i_1-1}(z^{i_1-1}_1, \ldots,
  z^{i_1-1}_{\arityof{\bpred_{i_1-1}}})$ with $\phi_1[x_1/z^{i_1-1}_1,
    \ldots,
    x_{\arityof{\bpred_{i_1-1}}}/z^{i_1-1}_{\arityof{\bpred_{i_1-1}}}]$. Since
  $(\comps,\interacs,\statemap) \models^\store_\asid \phi$, there
  exists a store $\store'$, that agrees with $\store$ over $x_1,
  \ldots, x_{\arityof{\apred_0}}$, and configurations $(\comps_1,
  \interacs_1, \statemap), \ldots, (\comps_h, \interacs_h,
  \statemap)$, where $\aconfig = (\comps_1, \interacs_1,
  \statemap) \comp \ldots \comp (\comps_h, \interacs_h, \statemap)$
  and the following hold: \begin{compactitem}
  \item $(\comps_1, \interacs_1, \statemap) \models^{\store'} \psi *
    \pureform$,
  \item $(\comps_\ell, \interacs_\ell, \statemap)
    \models^{\store'}_{\asid} \bpred_\ell(z^\ell_1, \ldots,
    z^\ell_{\arityof{\bpred_\ell}})$, for all $\ell \in
    \interv{2}{h} \setminus \set{i_1+1}$,
  \item $(\comps_{i_1-1}, \interacs_{i_1-1}, \statemap)
    \models^{\store'}_\asid \phi_1[x_1/z^{i_1-1}_1, \ldots,
      x_{\arityof{\bpred_{i_1-1}}}/z^{i_1-1}_{\arityof{\bpred_{i_1-1}}}]$,
    hence $(\comps_{i_1-1}, \interacs_{i_1-1}, \statemap)
    \models^{\store'}_\asid \bpred_{i_1-1}(z^{i_1-1}_1, \ldots,
    z^{i_1-1}_{\arityof{\bpred_{i_1-1}}})$. 
  \end{compactitem}
  We consider the following base tuples: \begin{compactitem}
  \item $\pbasetuplen{1} \isdef \basetupleof{\psi*\pureform}{\set{x_1,
      \ldots, x_{\arityof{\apred_0}}}}$, 
  \item for all $\ell \in \interv{2}{h} \setminus \set{i_1+1}$, there
    exist $\pbasetuplen{\ell} \in
    \leastbaseof{\bpred_\ell}[x_1/z^\ell_1, \ldots,
      x_{\arityof{\bpred_\ell}}/z^\ell_{\arityof{\bpred_\ell}}]$, such
    that $\comps_\ell \subseteq \store'(\pbasecomps_\ell)$,
    $\interacs_\ell \subseteq \store'(\pbaseinteracs_\ell)$ and
    $(\emptyset,\emptyset,\statemap) \models^{\store'}
    \ppureform_\ell$, by Lemma \ref{lemma:sat-completeness}.
  \item $\graphof{\asid}$ has a path $(\bpred_{i_1-1},\basetuplen{1})
    \depof{\arule_2}{i_2} \ldots \depof{\arule_n}{i_n}$
    $(\apred_n,\basetuplen{n})$, such that $\comps_{i_1-1} \subseteq
    \store'(\basecomps_1)$, $\interacs_{i_1-1} \subseteq
    \store'(\baseinteracs_1)$ and $(\emptyset,\emptyset,\statemap)
    \models^{\store'} \pureform_1$, by the inductive hypothesis.
  \end{compactitem}
  By an argument similar to the one from Lemma
  \ref{lemma:sat-completeness}, the composition $\pbasetuplen{} \isdef
  \Basecomp_{\ell=1}^{i_1-2} \pbasetuplen{\ell} \basecomp
  \basetuplen{1} \basecomp \Basecomp_{\ell=i_1}^h \pbasetuplen{\ell}$
  is defined and let $\basetuplen{0} \isdef
  \proj{\pbasetuplen{}}{\set{x_1, \ldots,
      x_{\arityof{\apred_0}}}}$. Finally, the conditions
  $\store(\basecomps_0) \subseteq \comps_0$, $\store(\baseinteracs_0)
  \subseteq \interacs_0$ and $(\emptyset,\emptyset,\statemap)
  \models^\store \pureform_0$ follow from a similar argument to the
  one used in Lemma \ref{lemma:sat-completeness}. \qed
\end{proofE}
  
An \emph{elementary cycle} of $\graphof{\asid}$ is a path from some
vertex $(\bpred,\anbasetuple)$ back to itself, such that
$(\bpred,\anbasetuple)$ does not occur on the path, except at its
endpoints. The cycle is, moreover, \emph{reachable} from
$(\apred,\abasetuple)$ if and only if there exists a path
$(\apred,\abasetuple) \depof{\arule_1}{i_1} \ldots
\depof{\arule_n}{i_n} (\bpred,\anbasetuple)$ in $\graphof{\asid}$. We
reduce the complement of the $\bound{\asid}{\apred}$ problem, namely
the existence of an infinite set of models of $\exists x_1 \ldots
\exists x_{\arityof{\apred}} ~.~ \apred(x_1, \ldots,
x_{\arityof{\apred}})$ of unbounded degree, to the existence of a
reachable elementary cycle in $\graphof{\asid'}$, where $\asid'$ is
obtained from $\asid$, as described in the following. 

First, we consider, for each predicate $\bpred \in \defnof{\asid}$, a
predicate $\bpred'$, of arity $\arityof{\bpred}+1$, not in
$\defnof{\asid}$ i.e., the set of predicates for which there exists a
rule in $\asid$. Second, for each rule \(\bpred_0(x_1, \ldots,
x_{\arityof{\bpred_0}}) \leftarrow \exists y_1 \ldots \exists y_m ~.~
\phi * \Asterisk_{\ell=2}^h \bpred_\ell(z^\ell_1, \ldots,
z^\ell_{\arityof{\bpred_\ell}}) \in \asid\), where $\phi$ is a
quantifier- and predicate-free formula and $\interactionvars{\phi}
\subseteq \fv{\phi}$ denotes the subset of variables occurring in
interaction atoms in $\phi$, the SID $\asid'$ has the following rules:
  \begin{eqnarray}
    \bpred'_0(x_1, \ldots, x_{\arityof{\bpred_0}}, x_{\arityof{\bpred_0}+1}) & \leftarrow & 
    \exists y_1 \ldots \exists y_m ~.~ \phi *
    \Asterisk_{\xi \in \interactionvars{\phi}} x_{\arityof{\bpred_0}+1} \not= \xi * \nonumber \\
    & & \hspace{1.5cm} \Asterisk_{\ell=2}^h \bpred'_\ell(z^\ell_1, \ldots,
    z^\ell_{\arityof{\bpred_\ell}}, x_{\arityof{\bpred_0}+1}) \label{eq:interaction-not-bound} \\
    \bpred'_0(x_1, \ldots, x_{\arityof{\bpred_0}}, x_{\arityof{\bpred_0}+1}) & \leftarrow & 
    \exists y_1 \ldots \exists y_m ~.~ \phi * x_{\arityof{\bpred_0}+1} = \xi * \nonumber \\
    & & \hspace{1.5cm} \Asterisk_{\ell=2}^h  \bpred'_\ell(z^\ell_1, \ldots,
    z^\ell_{\arityof{\bpred_\ell}}, x_{\arityof{\bpred_0}+1}) \label{eq:interaction-bound} \\
    \text{for each variable } \xi & \in & \interactionvars{\phi} \text{, that occurs in an interaction
      atom in $\phi$.} \nonumber
  \end{eqnarray}
Intuitively, there exists a family of models (with respect to $\asid$)
of $\exists x_1 \ldots \exists x_{\arityof{\apred}} ~.~ \apred(x_1,
\ldots, x_{\arityof{\apred}})$ of unbounded degree if and only if
these are models of $\exists x_1 \ldots \exists x_{\arityof{\apred}+1}
~.~ \apred'(x_1, \ldots, x_{\arityof{\apred}+1})$ (with respect to
$\asid'$) and the last parameter of each predicate $\bpred' \in
\defnof{\asid'}$ can be mapped, in each of the these models, to a
component that occurs in unboundedly many interactions. The latter
condition is equivalent to the existence of an elementary cycle,
containing a rule of the form (\ref{eq:interaction-bound}), that it,
moreover, reachable from some vertex $(\apred',\abasetuple)$ of
$\graphof{\asid'}$, for some $\abasetuple\in\satbasetuples$. This
reduction is formalized below:

\begin{myLemmaE}\label{lemma:length-unfolding}
  Let $\apred$ be a predicate and $\aconfig$ be a model of $\exists
  x_1 \ldots \exists x_{\arityof{\apred}} ~.~ \apred(x_1, \ldots,
  x_{\arityof{\apred}})$. Then there exists an unfolding $\apred(x_1,
  \ldots, x_{\arityof{\apred}}) \unfold{w}{\asid} \psi$ of length
  $\lenof{w} \geq \frac {\log(\degreeof{\aconfig}) - \log \beta_1}
        {\log \beta_2}$ where $\beta_1$ is the maximal number of
        components and interaction atoms and $\beta_2$ is the maximal
        number of predicate atoms, occurring in a rule of $\asid$.
\end{myLemmaE}
\begin{proofE}
  Let $\aconfig$ be a configuration, $\store$ be a store and
  $\apred$ be a predicate, such that $\aconfig \models_\asid^{\store}
  \apred(x_1, \ldots, x_{\arityof{\apred}})$. We consider the
  derivation tree $T$ induced by the definition of the
  $\models_\asid^\store$ relation.  The nodes of $T$ are labelled by a
  triple $\aconfig' \models_\asid^{\store'} \bpred(x_1, \ldots,
  x_{\arityof{\bpred}})$.  We start from the root labelled by
  $\aconfig \models_\asid^{\store} \apred(x_1, \ldots,
  x_{\arityof{\apred}})$ and define the children of a node
  inductively.

  For each node $\aconfig' \models_\asid^{\store'} \bpred(x_1, \ldots,
  x_{\arityof{\bpred}})$, there exists a rule:
  \[\arule ~:~ \bpred(x_1, \ldots, x_{\arityof{\apred}}) \leftarrow
  \exists y_1 \ldots \exists y_m ~.~ \phi * \bpred_1(z^1_1, \ldots,
  z^1_{\arityof{\bpred_1}}) * \ldots * \bpred_h(z^h_1, \ldots,
  z^h_{\arityof{\bpred_h}})\] and configurations $\aconfig_0, \ldots,
  \aconfig_h$, such that $\aconfig = \aconfig_0 \comp \dots \comp
  \aconfig_h$, $\aconfig_0 \models_\asid^{\store''} \phi$ and
  $\aconfig_i \models_\asid^{\store''} \bpred_i(z^i_1, \ldots,
  z^i_{\arityof{\bpred_i}})$ for every $i \in \interv{1}{h}$, where
  $\phi$ is a predicate-free formula and $\store''$ is a store that
  agrees with $\store'$ over $x_1, \dots, x_{\arityof{\bpred}}$. We
  define $\store_\ell \isdef \store''[x_1/z^\ell, \ldots,
    x_{\arityof{\bpred_\ell}}/z^\ell_{\arityof{\bpred_\ell}}]$, for
  all $\ell \in \interv{1}{h}$. Then the node $\aconfig'
  \models_\asid^{\store'} \bpred(x_1, \ldots, x_{\arityof{\bpred}})$
  has $h$ children in $T$, where the $\ell$-th child is labelled by
  $\aconfig_\ell \models_\asid^{\store_\ell} \bpred_\ell(x_1, \ldots,
  x_{\arityof{\bpred_\ell}})$, for all $\ell \in \interv{1}{h}$.  The
  construction is finite since $\aconfig \models_\asid^{\store}
  \apred(x_1, \ldots, x_{\arityof{\apred}})$ has a finite inductive
  definition.

  We now consider the degree of the configurations which occur in $T$.
  By Def. \ref{def:composition}, we obtain $\degreeof{\aconfig} \leq
  \degreeof{\aconfig_0} + \sum_{i=1}^h \degreeof{\aconfig_i}$. With
  each $\aconfig_i$ associated to a child of this node (except
  $\aconfig_0$), we obtain that $\degreeof{\aconfig_\text{init}}$ does
  not exceed $\beta_1$ times the number of nodes in $T$.  Since $h
  \leq\beta_2$, the height $n$ of $T$ is bound to the degree
  $\degreeof{\aconfig}$ by the inequality $\degreeof{\aconfig} \leq
  \beta_1 \times \sum_{k=0}^n {\beta_2}^k = \beta_1 \times
       {\beta_2}^{n+1} - 1$, leading to:
       \[n+1 \geq \frac {\log \degreeof{\aconfig} - \log \beta_1} {\log \beta_2}\]
       Finally, with $T$ of height $n$, there exists a branch in $T$
       (starting from the root) of length exactly $n+1$.  Yet each
       branch of $T$ corresponds to an unfolding $\apred(x_1, \ldots,
       x_{\arityof{\apred}}) \unfold{w}{\asid} \psi$, with $w$
       obtained by concatenating for every node
       $(\aconfig,\store,\apred)$ of the branch (from root to leaf)
       the couple $(\arule,i)$ consisting of:
       \begin{compactitem}
       \item the rule $\arule \in\asid$ used to unfold $\apred(x_1,
         \ldots, x_{\arityof{\apred}})$, and
       \item the position $i$ of this node among its brothers in $T$
         (take $i=1$ for the root).
       \end{compactitem}
       This unfolding has the length required, which concludes the
       lemma.  \qed
\end{proofE}

\begin{lemmaE}\label{lemma:unbounded}
  There exists an infinite sequence of configurations $\aconfig_1,
  \aconfig_2, \ldots$ such that $\aconfig_i \models_\asid \exists x_1
  \ldots \exists x_{\arityof{\apred}} ~.~ \apred(x_1, \ldots,
  x_{\arityof{\apred}})$ and $\degreeof{\aconfig_i} <
  \degreeof{\aconfig_{i+1}}$, for all $i\geq1$ if and only if
  $\graphof{\asid'}$ has an elementary cycle containing a rule
  (\ref{eq:interaction-bound}), reachable from a node
  $(\apred',\abasetuple)$, for $\abasetuple\in\satbasetuples$.
\end{lemmaE}
\begin{proofE}
  ``$\Rightarrow$'' Let $\store_1, \store_2, \ldots$ be stores such
that $\aconfig_i \models_\asid^{\store_i} \apred(x_1, \ldots,
x_{\arityof{\apred}})$, for all $i \geq 1$. By Lemma
\ref{lemma:length-unfolding}, where exists unfoldings $\apred(x_1,
\ldots, x_{\arityof{\apred}}) \unfold{w_i}{\asid'} \phi_i$ of lengths
$\lenof{w_1} < \lenof{w_2} < \ldots$, such that $\aconfig_i
\models_\asid^{\store_i} \phi_i$, for all $i \geq 1$. For each
configuration $\aconfig_i$, let $d_i \in \universe$ be a component,
such that $\degreeof{\aconfig_i} =
\cardof{\set{(c_1,p_1,\ldots,c_n,p_n) \mid d_i=c_j, j \in
    \interv{1}{n}}}$. By induction on $\lenof{w_i}\geq1$, we build
unfoldings $\apred'(x_1, \ldots, x_{\arityof{\apred}},
x_{\arityof{\apred}+1}) \unfold{w'_i}{\asid'} \phi'_i$ that bind
$x_{\arityof{\apred}+1}$ to all variables bound to $d_i$, using rules
of type (\ref{eq:interaction-bound}).  By Lemma
\ref{lemma:unfold-completeness}, $w'_1, w'_2, \ldots$ are labels of
paths from $\graphof{\asid'}$, that start in $(\apred,\abasetuple_1),
(\apred,\abasetuple_2), \ldots$, respectively. Since
$\graphof{\asid'}$ is finite, we can chose an infinite subsequence of
paths that start in the same node of $\graphof{\asid'}$ and repeat the
same vertex, with a rule of type (\ref{eq:interaction-bound}) in
between.
 
\vspace*{\baselineskip}\noindent ``$\Leftarrow$'' Let
$(\apred',\abasetuple) \depof{\arule'_1}{i_1} \ldots
\depof{\arule'_n}{i_n} (\bpred_n,\abasetuple_n)
\Depof{\arule'_{n+1}}{i_{n+1}} \ldots \Depof{\arule'_{n+p}}{i_{n+p}}
(\bpred_n,\abasetuple_n)$ be a path in $\graphof{\asid'}$, such that
one of the rules $\arule'_{n+1}, \ldots, \arule'_{n+p}$ is of the form
(\ref{eq:interaction-bound}) and let $w'_i \isdef (\arule'_1,i_1)
\ldots (\arule'_n,i_n) [(\arule'_{n+1},i_{n+1}) \ldots
  (\arule'_{n+p},i_{n+p})]^i$, for all $i \geq 1$. By Lemma
\ref{lemma:unfold-soundness}, there exist unfoldings $\apred'(x_1,
\ldots, x_{\arityof{\apred}+1}) \unfold{w'_i}{\asid'} \phi'_i$, stores
$\store_i$ and configurations $\aconfig_i$, such that $\aconfig_i
\models^{\store_i} \phi'_i$. We define: \[\delta_i \isdef
\cardof{\set{(c_1,p_1,\ldots,c_n,p_n) \in \interacs_i \mid
    \store(x_{\arityof{\apred}+1})=c_j, j \in \interv{1}{n}}} \text{,
  for all $i \geq 1$}\] where $\aconfig_i \isdef (\comps_i,
\interacs_i, \statemap_i)$. Since $\aconfig_i \models^{\store_i}
\phi'_i$ and one of the rules $\arule'_{n+1}, \ldots, \arule'_{n+p}$
is of type (\ref{eq:interaction-bound}), the sequence $\delta_1,
\delta_2, \ldots$ is strictly increasing. Moreover, we have $\delta_i
\leq \degreeof{\aconfig_i}$, for all $i \geq 1$, hence there exists a
sequence of integers $1 \leq i_1 < i_2 < \ldots$ such that
$\degreeof{\aconfig_{i_j}} < \degreeof{\aconfig_{i_{j+1}}}$, for all
$j \geq 1$. \qed
\end{proofE}

The complexity result below uses a similar argument on the maximal
size of (hence the number of) base tuples as in Theorem \ref{thm:sat},
leading to similar complexity gaps:

\begin{theoremE}\label{thm:bound}
  $\klbound{\asid}{\apred}{k}{\infty}$ is in \conp,
  $\klbound{\asid}{\apred}{\infty}{\ell}$ is in \exptime\ and
  $\bound{\asid}{\apred}$ is in \twoexptime.
\end{theoremE}
\begin{proofE}
  Lemma \ref{lemma:unbounded} shows the reduction of the
  complement of $\bound{\asid}{\apred}$ to the existence of a
  reachable cycle in the graph $\graphof{\asid'}$, where $\asid'$ is
  constructed from $\asid$ in polynomial time. Moreover, we have
  $\maxarityof{\asid'} = \maxarityof{\asid}+1$ and
  $\maxintersize{\asid'} = \maxintersize{\asid}$. We distinguish
  the three cases below:
  \begin{compactitem} 
  \item $k<\infty$, $\ell=\infty$: in this case, we can define a non-deterministic
    algorithm as follows.  We guess the solution $(\tuple{W_1,\ldots,
      W_K}, W_{K+1}, \tuple{i_1,i_2,\ldots, i_n}$, where: \begin{compactitem}
    \item $\tuple{W_1,\ldots,W_K}$ defines an acyclic witness for a
      satisfiable least solution of $\apred'$ in $\asid'$ constructed
      as in the proof of Thm.~\ref{thm:sat};
    \item $W_{K+1} = (T_{K+1}, r_{K+1}, t_{K+1,1},...,e_{K+1,h_{K+1}})$ is similar to a regular entry
      $W_i$, that is, contains a base tuple $T_{K+1}$, an index 
      $r_{K+1}$ of a rule of $\asid'$ and indices
      $e_{K+1,1}$, $\ldots$, $e_{K+1,h_{K+1}} \in \{1,\ldots, K\}$ such that
      $T_{K+1}$ is computed correctly by applying the rule
      $r_{K+1}$ from base tuples $T_{e_{K+1,1}},\ldots,
      T_{e_{K+1,h_{K+1}}}$ as explained in the proof of Thm.~\ref{thm:sat};
    \item $\tuple{i_1,i_2,\ldots,i_n}$ defines an acyclic path starting at the
      initial node in the directed acyclic graph defined by $W$, that
      is, $1 = i_1 < i_2 < \ldots < i_n \le K$ and moreover $i_{j+1}
      \in \{e_{i_j,1},\ldots, e_{i_j,h_{i_j}}\}$ for all 
      $j\in\{1,2,\ldots,n-1\}$;
    \item the path $\tuple{i_1,i_2,\ldots,i_n}$ can be \emph{closed}
      into a witness reachable cycle from $i_1$ by using $W_{K+1}$ that is,
      whenever (i) rules $r_{i_n}$ and $r_K$ define the same predicate, and moreover
      $T_{i_n} = T_{K+1}$ , (ii) the intersection $X =
      \{e_{K+1,1},\ldots,e_{K+1,h_{K+1}}\} \cap \{i_1, i_2, \ldots,
      i_n\} \not= \emptyset$, (iii) if $i_j = \min X$, that is, the cycle
      starts at $i_j$ then at least one of the rules used along the cycle
      $r_{i_j}, r_{i_{j+1}}, \ldots, r_{i_{n-1}},
      r_{K+1}$ is of the form (\ref{eq:interaction-bound}).
    \end{compactitem}
    The solution is of linear size $\bigO(\size{\asid'})$ by the same
    arguments as in the proof of Thm.~\ref{thm:sat}.  Therefore, it can be
    guessed in polynomial time, and moreover checked in polynomial
    time following the conditions above.  This implies the
    membership of the complement problem in \np, henceforth
    $\klbound{\asid}{\apred}{k}{\infty}$ is in \conp.
  \item $k=\infty$, $\ell<\infty$: in this case, using the algorithm
    from Fig. \ref{fig:base-graph}, the graph $\graphof{\asid'}$ is
    constructed in time $2^{\polynomial(\size{\asid'})}$ as
    previously explained in the proof of Theorem \ref{thm:sat}. Finding a
    reachable cycle with the additional properties required by Lemma
    \ref{lemma:unbounded} can be done in two additional steps,
    respectively, first building the SCCs decomposition of
    $\graphof{\asid'}$ and then checking reachability of SCCs
    containing edges derived from rules of form
    (\ref{eq:interaction-bound}) from SCCs containing vertices
    $(\apred',\abasetuple)$.  Both steps can be done in in linear time
    in the size of $\graphof{\asid'}$ i.e., using Tarjan algorithm for
    SCC decomposition and standard graph traversals.  Therefore, the
    overall time complexity remains
    $2^{\polynomial(\size{\asid'})}$, and as such
    $\klbound{\asid}{\apred}{\infty}{\ell}$ is in \exptime.
  \item $k=\infty$, $\ell=\infty$: following the same argument as in
    the previous point and noticing that the graph $\graphof{\asid'}$
    is constructed in time $2^{2^{\polynomial(\size{\asid'})}}$ we
    conclude that $\bound{\asid}{\apred}$ is in \twoexptime. \qed
  \end{compactitem}
\end{proofE}

\noindent
Moreover, the construction of $\graphof{\asid'}$ allows to prove the
following cut-off result:

\begin{propositionE}\label{prop:bound-cutoff}
  Let $\aconfig$ be a configuration and $\store$ be a store, such that
  $\aconfig \models_\asid^\store \apred(x_1, \ldots,
  x_{\arityof{\apred}})$.  If $\klbound{\asid}{\apred}{k}{\ell}$ then
  \begin{inparaenum}[(1)]
  \item $\degreeof{\aconfig}=\polynomial(\size{\asid})$ if $k<\infty$,
    $\ell=\infty$,
  \item $\degreeof{\aconfig}=2^{\polynomial(\size{\asid})}$ if
    $k=\infty$, $\ell < \infty$ and
  \item $\degreeof{\aconfig}=2^{2^{\polynomial(\size{\asid})}}$ if
    $k=\infty$, $\ell=\infty$.
  \end{inparaenum}
\end{propositionE}
\begin{proofE}
  First, we show that in all cases, the degree is bounded by $2^{B^*}
  \cdot L \cdot I$ where $B^*$ is the maximal length of a satisfiable base
  tuple in $\asid'$, $L$ is the number of predicates in $\asid'$ and $I$ is
  the maximal number of interactions defined in a rule in $\asid'$.  The
  maximal length $B^*$ of a satisfiable base tuples has been considered in
  the proof of Thm.~\ref{thm:sat} to derive an upper bound on the the
  number of distinct satisfiable base tuples for a SID.  Then, $2^{B^*}
  \cdot L$ represents a bound on the number of nodes in the graph
  $\graphof{\asid'}$ as for every predicate there will be at most $2^{B^*}$
  satisfiable base tuples associated to it.  Meantime, this value also
  represents a bound on the longest acyclic path in $\graphof{\asid'}$.  We
  are interested on acyclic paths because cycles in $\graphof{\asid'}$ are
  guaranteed to never connect (use in interactions) the extra variable
  introduced in $\asid'$ (otherwise the system would not be of bounded
  degree).  But then, along the acyclic paths, at most $I$ interactions are
  defined at each step, henceforth, the bound of $2^{B^*} \cdot L \cdot I$
  on the number on total interactions that could involve the extra
  variable.

  Second, let observe that both $L$ and $I$ are the same in $\asid'$
  and in $\asid$ and equal to $\bigO(\size{\asid})$.  Moreover, it was
  shown in the proof of Thm.~\ref{thm:sat} that $B^* = 2\alpha +
  2\alpha^2 + p^{\min(\alpha,\beta)} \alpha^{\min(\alpha,\beta)}$ for
  $\alpha = \maxarityof{\asid'} = \maxarityof{\asid}+1$ and $\beta =
  \maxintersize{\asid'} = \maxintersize{\asid}$, $p = \cardof{\ports}$
  the number of ports.  Henceforth, we distinguished the three cases,
  respectively (i) $B^* = \bigO(1)$ if $k<\infty$, $\ell=\infty$, (ii)
  $B^* = \polynomial(\size{\asid})$ if $k=\infty$, $\ell < \infty$ and
  (iii) $B^* = 2^{\polynomial(\size{\asid})}$ if $k=\infty$,
  $\ell=\infty$. By using the above in the expression $2^{B^*} \cdot
  L \cdot I$ we obtain the values of the bound as stated in the
  Proposition. \qed
\end{proofE}

\section{Entailment}
\label{sec:entailment}

This section is concerned with the entailment problem
$\entl{\asid}{\apred}{\bpred}$, that asks whether $\aconfig
\models^\store_\asid \exists x_{\arityof{\apred}+1} \ldots \exists
x_{\arityof{\bpred}} ~.~ \bpred(x_1,\ldots,x_{\arityof{\bpred}})$, for
every configuration $\aconfig$ and store $\store$, such that $\aconfig
\models^\store_\asid \apred(x_1,\ldots,x_{\arityof{\apred}})$. For
instance, the proof from Fig. \ref{fig:ring} (c) relies on the
following entailments, that occur as the side conditions of the Hoare
logic rule of consequence:
\[\begin{array}{l}
\ring{h}{t}(y) \models_\asid \exists x \exists z . \compactin{y}{\toknotok} * \interac{y}{out}{z}{in} * \chain{h-1}{t}(z,x) * \interac{x}{out}{y}{in} \\
\compactin{z}{\toknotok} * \interac{z}{out}{x}{in} * \chain{h-1}{t}(x,y) * \interac{y}{out}{z}{in} \models_\asid \ring{h}{t}(z)
\end{array}\]
By introducing two fresh predicates $\apred_1$ and $\apred_2$, defined
by the rules:
\begin{align}
  \apred_1(x_1) & \leftarrow \exists y \exists z . \compactin{x_1}{\toknotok} \!*\! \interac{x_1}{out}{z}{in} * \chain{h-1}{t}(z,y) \!*\! \interac{y}{out}{x_1}{in}
  \label{eq:right-rule} \\[-1mm]
  \apred_2(x_1,x_2) & \leftarrow \exists z . \compactin{x_1}{\toknotok} * \interac{x_1}{out}{z}{in} * \chain{h-1}{t}(z,x_2) * \interac{x_2}{out}{x_1}{in}
  \label{eq:left-rule}
\end{align}
the above entailments are equivalent to
$\entl{\asid}{\ring{h}{t}}{\apred_1}$ and
$\entl{\asid}{\apred_2}{\ring{h}{t}}$, respectively, where $\asid$
consists of the rules (\ref{eq:right-rule}) and (\ref{eq:left-rule}),
together with the rules that define the $\ring{h}{t}$ and
$\chain{h}{t}$ predicates (\S\ref{sec:running-example}).

We show that the entailment problem is undecidable, in general
(Thm. \ref{thm:entl-undecidable}), and recover a decidable fragment,
by means of three syntactic conditions, typically met in our
examples. These conditions use the following notion of \emph{profile}:

\begin{definition}\label{def:profile}
  The \emph{profile} of a SID $\asid$ is the pointwise greatest
  function $\profile{\asid} : \preds \rightarrow \pow{\nat}$, mapping each
  predicate $\apred$ into a subset of $\interv{1}{\arityof{\apred}}$,
  such that, for each rule $\apred(x_1, \ldots, x_{\arityof{\apred}})
  \leftarrow \phi$ from $\asid$, each atom $\bpred(y_1, \ldots,
  y_{\arityof{\bpred}})$ from $\phi$ and each $i \in
  \profile{\asid}(\bpred)$, there exists $j \in
  \profile{\asid}(\apred)$, such that $x_j$ and $y_i$ are the same
  variable.
\end{definition}
The profile identifies the parameters of a predicate that are always
replaced by a variable $x_1, \ldots, x_{\arityof{\apred}}$ in each
unfolding of $\apred(x_1,\ldots,x_{\arityof{\apred}})$, according to
the rules in $\asid$; it is computed by a greatest fixpoint iteration,
in time $\polynomial(\size{\asid})$.

\begin{definition}\label{def:pcr}
  A rule $\apred(x_1, \ldots, x_{\arityof{\apred}}) \leftarrow \exists
  y_1 \ldots \exists y_m ~.~ \phi * \Asterisk_{\ell=1}^h
  \bpred_\ell(z^\ell_1, \ldots, z^\ell_{\arityof{\bpred_\ell}})$,
  where $\phi$ is a quantifier- and predicate-free formula, is said to be: \begin{enumerate}
  \item\label{it:progressing} \emph{progressing} if and only if $\phi
    = \compact{x_1} * \psi$, where $\psi$ consists of interaction
    atoms involving $x_1$ and (dis-)equalities, such that
    $\bigcup_{\ell=1}^h \set{z^\ell_1, \ldots,
      z^\ell_{\arityof{\bpred_\ell}}} = \set{x_2, \ldots,
      x_{\arityof{\apred}}} \cup \set{y_1, \ldots, y_m}$,
  \item\label{it:connected} \emph{connected} if and only if, for each
    $\ell\in\interv{1}{h}$ there exists an interaction atom in $\psi$
    that contains both $z^\ell_1$ and a variable from $\set{x_1} \cup
    \set{x_i \mid i \in \profile{\asid}(\apred)}$,
  \item\label{it:restricted} \emph{equationally-restricted
  (e-restricted)} if and only if, for every disequation $x \neq y$
    from $\phi$, we have $\set{x,y} \cap \set{x_i \mid i \in
      \profile{\asid}(\apred)} \neq\emptyset$.
  \end{enumerate}
  A SID $\asid$ is \emph{progressing}, \emph{connected} and
  \emph{e-restricted} if and only if each rule in $\asid$ is
  \emph{progressing}, \emph{connected} and \emph{e-restricted},
  respectively.
\end{definition}
For example, the SID consisting of the rules from
\S\ref{sec:running-example}, together with rules (\ref{eq:right-rule})
and (\ref{eq:left-rule}) is progressing, connected and e-restricted.
\begin{myTextE}
For a configuration $\aconfig = (\comps,\interacs,\statemap)$, let: 
\[\nodesof{\aconfig} \isdef \comps \cup \Set{c_i \mid (c_1, p_1,
  \ldots, c_n, p_n) \in \interacs, i \in \interv{1}{n}}\] be the set
of (possibly absent) components that occur in $\aconfig$.
\end{myTextE}
\begin{myLemmaE}\label{lemma:progressing-sid}
  Given a progressing SID $\asid$ and a predicate
  $\apred\in\defnof{\asid}$, for any configuration $\aconfig =
  (\comps, \interacs, \statemap)$ and store $\store$, such that
  $\aconfig\models^\store_\asid \apred(x_1, \ldots,
  x_{\arityof{\apred}})$, we have $\set{\store(x_1), \ldots,
    \store(x_{\arityof{\apred}})} \subseteq \nodesof{\aconfig} =
  \comps$.
\end{myLemmaE}
\begin{proofE}
  We proceed by fixpoint induction on the definition of
  $\aconfig \models^\store_\asid \apred(x_1, \ldots,
  x_{\arityof{\apred}})$.  By definition, there exists a progressing
  rule
$$\arule ~:~ \apred(x_1, \ldots, x_{\arityof{\apred}}) \leftarrow
  \exists y_1 \ldots \exists y_m ~.~ \compact{x_1} * \psi *
  \Asterisk_{\ell=1}^h \bpred_\ell(z^\ell_1, \ldots,
  z^\ell_{\arityof{\bpred_\ell}})$$ a store $\store'$ and
  configurations $\aconfig_0, \dots, \aconfig_h$ such
  that: \begin{compactitem}
  \item $\aconfig = \aconfig_0 \comp \dots \comp \aconfig_h$,
  \item $\store(x_i) = \store'(x_i)$ for all $i \in
    \interv{1}{\arityof{\apred}}$,
  \item $\aconfig_0 \models_\asid^{\store'} \compact{x_1} * \psi$, and
  \item $\aconfig_\ell \models_\asid^{\store'} \bpred_\ell(z^\ell_1,
    \ldots, z^\ell_{\arityof{\bpred_\ell}})$ for all $\ell \in
    \interv{1}{h}$.
  \end{compactitem}
  For $1\leq \ell\leq h$, let $\store_\ell(x_i) = \store'(z^\ell_i)$
  for $1\leq i\leq \arityof{\bpred_\ell}$.  Now apply the induction
  hypothesis on the derivation of $\aconfig_\ell
  \models_\asid^{\store_\ell} \bpred_\ell(x_1, \ldots,
  x_{\arityof{\bpred_\ell}})$ to obtain that $\set{\store_\ell(x_1),
    \ldots, \store_\ell(x_{\arityof{\bpred_\ell}})} \subseteq
  \nodesof{\aconfig_\ell}$.  Since $\arule$ is progressing, we have:
\begin{equation}
\begin{split}
  &\set{\store(x_1), \ldots, \store(x_{\arityof{\apred}})} 
  \subseteq \set{\store'(x_1), \ldots, \store'(x_{\arityof{\apred}})} \cup \set{\store'(y_1), \ldots, \store'(y_m)} \\
  &= \set{\store'(x_1)} \cup \bigcup_{\ell=1}^h \set{\store'(z^\ell_1), \ldots, \store'(z^\ell_{\arityof{\bpred_\ell}})} 
  = \set{\store'(x_1)} \cup \bigcup_{\ell=1}^h \set{\store_\ell(x_1), \ldots, \store_\ell(x_{\arityof{\bpred_\ell}})} \\
  &\subseteq \nodesof{\aconfig_0} \cup \bigcup_{\ell=1}^h \nodesof{\aconfig_\ell} 
  = \nodesof{\aconfig} \nonumber \text{\qed}
\end{split}
\end{equation}
\end{proofE}

We recall that $\defn{\asid}{\apred}$ is the set of rules from $\asid$
that define $\apred$ and denote by $\defs{\asid}{\apred}$ the least
superset of $\defn{\asid}{\apred}$ containing the rules that define a
predicate from a rule in $\defs{\asid}{\apred}$. The following result
shows that the entailment problem becomes undecidable as soon as the
connectivity condition is even slightly lifted:

\begin{theoremE}\label{thm:entl-undecidable}
  $\entl{\asid}{\apred}{\bpred}$ is undecidable, even when $\asid$ is
  progressing and e-restricted, and only the rules in
  $\defs{\asid}{\apred}$ are connected (the rules in
  $\defs{\asid}{\bpred}$ may be disconnected).
\end{theoremE}
\begin{proofE}
  By a reduction from the known undecidable problem of
  universality of context-free languages \cite{BarHillel61}. A
  context-free grammar $G = \tuple{N,T,S,\Delta}$ consists of a finite
  set $N$ of nonterminals, a finite set $T$ of terminals, a start
  symbol $S \in N$ and a finite set $\Delta$ of productions of the
  form $A \rightarrow w$, where $A \in N$ and $w \in (N \cup
  T)^*$. Given finite strings $u, v \in (N \cup T)^*$, the step
  relation $u \Rightarrow v$ replaces a nonterminal $A$ of $u$ by the
  right-hand side $w$ of a production $A \rightarrow w$ and
  $\Rightarrow^*$ denotes the reflexive and transitive closure of
  $\Rightarrow$. The language of $G$ is the set $\lang{G}$ of finite
  strings $w \in T^*$, such that $s \Rightarrow^* w$. The problem $T^*
  \subseteq \lang{G}$ is known as the universality problem, known to
  be undecidable. Moreover, we assume w.l.o.g.\ that: \begin{itemize}
  \item $T = \set{0,1}$, because every terminal can be encoded as a
    binary string,
  \item $\lang{G}$ does not contain the empty string $\epsilon$,
    because computing a grammar $G'$ such that $\lang{G'} = \lang{G}
    \cap T^+$ is possible and, moreover, we can reduce from the
    modified universality problem problem $T^+ \subseteq \lang{G'}$
    instead of the original $T^* \subseteq \lang{G}$,
  \item $G$ is in Greibach normal form, i.e.\ it contains only
    production rules of the form $\bpred_0 \rightarrow b \bpred_1 \ldots \bpred_n$,
    where $\bpred_0, \ldots \bpred_n \in N$, for some $n \geq 0$ and $b \in T$.
  \end{itemize}
  Let $\ports = \set{p_0,p_1}$ be a set of ports.  For each
  nonterminal $\bpred_0 \in N$, we have a predicate $\bpred_0$ or
  arity two and a rule $\bpred_0(x_1,x_2) \leftarrow \exists y_1
  \ldots \exists y_n ~.~ \compact{x_1} * \interac{x_1}{p_a}{y_1}{p_a}
  * \bpred_1(y_1,y_2) * \ldots * \bpred_n(y_n,x_2)$, for each rule
  $\bpred_0 \rightarrow b \bpred_1 \ldots \bpred_n$ of $G$. Moreover,
  we consider the rules $\apred(x_1,x_2) \leftarrow \exists z ~.~
  \interac{x_1}{p_a}{z}{p_a} * \apred(z,x_2)$ and $\apred(x_1,x_2)
  \leftarrow \interac{x_1}{p_a}{x_2}{p_a}$, for all $a \in
  \set{0,1}$. Let $\asid$ be the SID containing the above rules. It is
  easy to check that the SID is progressing and e-restricted and that,
  moreover, the rules from $\defs{\asid}{\apred}$ are
  connected. Finally, $\apred(x_1,x_2) \modelsid \bpred(x_1,x_2)$ if
  and only if $T^+ \subseteq \lang{G}$.  \qed
\end{proofE}
  
On the positive side, we prove that $\entl{\asid}{\apred}{\bpred}$ is
decidable, if $\asid$ is progressing, connected and e-restricted,
assuming further that $\bound{\asid}{\apred}$ has a positive
answer. In this case, the bound on the degree of the models of
$\apred(x_1, \ldots, x_{\arityof{\apred}})$ is effectively computable,
using the algorithm from Fig. \ref{fig:base-graph} (see
Prop. \ref{prop:bound-cutoff} for a cut-off result) and denote by
$\degreebound$ this bound, throughout this section.

The proof uses a reduction of $\entl{\asid}{\apred}{\bpred}$ to a
similar problem for \seplog, showed to be decidable
\cite{EchenimIosifPeltier21}. We recall the definition of \seplog,
interpreted over heaps $\heap : \universe \finmap \universe^\rank$,
introduced in \S\ref{sec:sl}. \seplog\ rules are denoted as
$\xannot{}{}{\apred}(x_1, \ldots, x_{\#(\xannot{}{}{\apred})})
\leftarrow \phi$, where $\phi$ is a \seplog\ formula, such that
$\fv{\phi} \subseteq \set{x_1, \ldots, x_{\#(\xannot{}{}{\apred})}}$
and \seplog\ SIDs are denoted as $\slsid$. The profile
$\profile{\slsid}$ is defined for \seplog\ same as for
\cl\ (Def. \ref{def:profile}).

\begin{definition}\label{def:pcr-sl}
  A \seplog\ rule $\xannot{}{}{\apred}(x_1, \ldots,
  x_{\#(\xannot{}{}{\apred})}) \leftarrow \phi$ from a SID $\slsid$ is
  said to be: \begin{enumerate}
  \item \emph{progressing} if and only if $\phi = \exists t_1 \ldots
    \exists t_m ~.~ x_1 \mapsto (y_1, \ldots, y_\rank) * \psi$, where
    $\psi$ contains only predicate and equality atoms, 
  \item \emph{connected} if and only if $z_1 \in \set{x_i \mid i \in
    \profile{\slsid}(\xannot{}{}{\apred})} \cup \set{y_1, \ldots,
    y_\rank}$, for every predicate atom
    $\xannot{}{}{\bpred}(z_1,\ldots,z_{\#(\xannot{}{}{\bpred})})$ from
    $\phi$.
    %
    %
  \end{enumerate}
\end{definition}
Note that the definitions of progressing and connected rules are
different for \seplog, compared to \cl\ (Def. \ref{def:pcr}); in the
rest of this section, we rely on the context to distinguish
progressing (connected) \seplog\ rules from progressing (connected)
\cl\ rules. Moreover, e-restricted rules are defined in the same way
for \cl\ and \seplog\ (point \ref{it:restricted} of
Def. \ref{def:pcr}). A tight upper bound on the complexity of the
entailment problem between \seplog\ formul{\ae}, interpreted by
progressing, connected and e-restricted SIDs, is given below:

\begin{theorem}[\cite{EchenimIosifPeltier21}]\label{thm:sl-entailment}
  The \seplog\ entailment problem is in $2^{2^{\polynomial(\width{\slsid} \cdot
        \log\size{\slsid})}}$, for progressing, connected and
  e-restricted SIDs.
\end{theorem}

The reduction of $\entl{\asid}{\apred}{\bpred}$ to
\seplog\ entailments is based on the idea of viewing a configuration
as a logical structure (hypergraph), represented by an indirected
\emph{Gaifman graph}, in which every tuple from a relation (hyperedge)
becomes a clique \cite{Gaifman82}. In a similar vein, we encode a
configuration, of degree at most $\degreebound$, by a heap of degree
$\rank$ (Def. \ref{def:gaifman-heap}), such that $\rank$ is defined using
the following integer function:
\vspace*{-.5\baselineskip}
\[\pos{i}{j}{k} \isdef 1 + \degreebound \cdot \sum_{\ell=1}^{j-1}
\lenof{\intertype_\ell} + i \cdot \lenof{\intertype_j} + k\] where
$\intertypes \isdef \set{\intertype_1, \ldots, \intertype_M}$ is the
set of interaction types and $\states \isdef \set{q_1, \ldots, q_N}$
is the set of states of the behavior
$\beh=(\ports,\states,\arrow{}{})$ (\S\ref{sec:definitions}). Here $i
\in \interv{0}{\degreebound-1}$ denotes an interaction of type $j \in
\interv{1}{M}$ and $k \in \interv{0}{N-1}$ denotes a state. We use $M$
and $N$ throughout the rest of this section, to denote the number of
interaction types and states, respectively.

For a set $\interacs$ of interactions, let $\xituples{\interacs}{j}{c}
\isdef \{\tuple{c_1, \ldots, c_n} \mid (c_1,p_1, \ldots, c_n,p_n) \in
\interacs,~ \intertype_j = \tuple{p_1, \ldots, p_n},~ c \in
\set{c_1,\ldots,c_n}\}$ be the tuples of components from an
interaction of type $\intertype_j$ from $\interacs$, that contain a
given component $c$.

\begin{definition}\label{def:gaifman-heap}
  Given a configuration $\aconfig = (\comps,\interacs,\statemap)$, such
  that $\degreeof{\aconfig} \leq \degreebound$, a \emph{Gaifman heap}
  for $\aconfig$ is a heap $\heap : \universe \finmap
  \universe^\rank$, where $\rank \isdef \pos{0}{M+1}{N}$, $\dom{\heap} =
  \nodesof{\aconfig}$ and, for all $c_0 \in \dom{\heap}$,
  such that $\heap(c_0) = \tuple{c_1,\ldots,c_\rank}$, the following hold: \begin{enumerate}
  \item\label{it21:gaifman-heap} $c_1 = c_0$ if and only if $c_0 \in \comps$,
  \item\label{it22:gaifman-heap} for all $j \in \interv{1}{M}$,
    $\xituples{\interacs}{j}{c} = \set{\vec{c}_1, \ldots, \vec{c}_s}$
    if and only if there exist integers $0 \leq k_1 < \ldots < k_s <
    \degreebound$, such that $\tuple{\heap(c_0)}_{\ipos{k_i}{j}} =
    \vec{c}_i$, for all $i \in \interv{1}{s}$, where $\ipos{i}{j}
    \isdef \interv{\pos{i-1}{j}{0}}{\pos{i}{j}{0}}$ are the entries of
    the $i$-th interaction of type $\intertype_j$ in $\heap(c_0)$,
  \item\label{it23:gaifman-heap} for all $k \in \interv{1}{N}$, we
    have $\tuple{\heap(c_0)}_{\spos{k}} = c_0$ if and only if
    $\statemap(c_0) = q_k$, where the entry $\spos{k} \isdef
    \pos{0}{M+1}{k-1}$ in $\heap(c_0)$ corresponds to the state $q_k
    \in \states$.
\end{enumerate}
We denote by $\gaifman{\aconfig}$ the set of Gaifman heaps for
$\aconfig$.
\end{definition}
Intuitively, if $\heap$ is a Gaifman heap for $\aconfig$ and $c_0 \in
\dom{\heap}$, then the first entry of $\heap(c_0)$ indicates whether
$c_0$ is present (condition \ref{it21:gaifman-heap} of
Def. \ref{def:gaifman-heap}), the next $\degreebound \cdot
\sum_{j=1}^M \lenof{\intertype_j}$ entries are used to encode the
interactions of each type $\intertype_j$ (condition
\ref{it22:gaifman-heap} of Def. \ref{def:gaifman-heap}), whereas the
last $N$ entries are used to represent the state of the component
(condition \ref{it23:gaifman-heap} of Definition
\ref{def:gaifman-heap}). Note that the encoding of configurations by
Gaifman heaps is not unique: two Gaifman heaps for the same
configuration may differ in the order of the tuples from the encoding
of an interaction type and the choice of the unconstrained entries
from $\heap(c_0)$, for each $c_0 \in \dom{\heap}$. On the other hand,
if two configurations have the same Gaifman heap encoding, they must
be the same configuration.

\begin{figure}[t!]
  \vspace*{-\baselineskip}
  \begin{subfigure}[b]{.4\textwidth}
    \scalebox{0.55}{	{\Large
		\begin{tikzpicture}[>=stealth',shorten >=1pt,auto,node distance=2cm]
			\begin{scope}[local bounding box=a, shift={(0,2)}]
				\node[state, fill=lightgray] (t) []     {$\toktoken$};


				\node (Ca) [draw=black, fit= (t)] {};
				\node [xshift=-30pt,yshift=-6pt] at (Ca.south) {$x$};
			\end{scope}

			\begin{scope}[local bounding box=b, shift={(3,2)}]
				\node[state, fill=lightgray] (h)      {$\toknotok$};


				\node (Cb) [draw=black, fit= (h)] {};
				\node [xshift=-30pt,yshift=-6pt] at (Cb.south) {$y$};
			\end{scope}

			\begin{scope}[local bounding box=c, shift={(6,2)}]
				\node[state, fill=lightgray] (h)      {$\toknotok$};


				\node (Cc) [draw=black, fit= (h)] {};
				\node [xshift=-30pt,yshift=-6pt] at (Cc.south) {$z$};
			\end{scope}

			\path [draw=black,fill=black] (Ca.west) circle (0.5mm);
			\path [draw=black,fill=black] (Ca.east) circle (0.5mm);
			\path [draw=black,fill=black] (Cb.west) circle (0.5mm);
			\path [draw=black,fill=black] (Cb.east) circle (0.5mm);
			\path [draw=black,fill=black] (Cc.west) circle (0.5mm);
			\path [draw=black,fill=black] (Cc.east) circle (0.5mm);

			\draw[-] (Ca) edge node[pos=0.25] () {$\tokout$} (Cb);
			\draw[-] (Ca) edge node[pos=0.8] () {$\tokin$} (Cb);
			\draw[-] (Ca) edge node[pos=0.5, below] () {} (Cb);
			\draw[-] (Cb) edge node[pos=0.25] () {$\tokout$} (Cc);
			\draw[-] (Cb) edge node[pos=0.8] () {$\tokin$} (Cc);
			\draw[-] (Cc) edge node[pos=0.5] () {} (Cb);
		\end{tikzpicture}
	}}
    \centerline{\footnotesize(a)}
  \end{subfigure}%
  \begin{subfigure}[b]{.6\textwidth}
    \scalebox{0.4}{{ \LARGE

\begin{tikzpicture}[>=Stealth,shorten >=1pt,auto,node distance=2cm]


	\begin{scope}[local bounding box=a, shift={(0,0)}]
		\matrix (Ma) [matrix of nodes,
			nodes={draw, minimum size=8mm,anchor=center},
			nodes in empty cells, minimum height = 1cm]
		{
		  & & & & & & \\
		};

	\node[yshift=-6pt, xshift=5pt] at (Ma-1-2.north west) () {$1$};
	\node[yshift=-6pt, xshift=5pt] at (Ma-1-3.north west) () {$2$};
	\node[yshift=-6pt, xshift=5pt] at (Ma-1-4.north west) () {$1$};
	\node[yshift=-6pt, xshift=5pt] at (Ma-1-5.north west) () {$2$};


	\node[circle, minimum size=1mm, inner sep=0pt, outer sep=0pt, fill=black]
		at (Ma-1-1.center) () {};
	\node[circle, minimum size=1mm, inner sep=0pt, outer sep=0pt, fill=black]
		at (Ma-1-4.center) () {};
	\node[circle, minimum size=1mm, inner sep=0pt, outer sep=0pt, fill=black]
		at (Ma-1-5.center) () {};
	\node[circle, minimum size=1mm, inner sep=0pt, outer sep=0pt, fill=black]
		at (Ma-1-7.center) () {};

	\draw [decorate, decoration = {brace, raise=5pt, amplitude=5pt}, thick]
		(Ma-1-1.north west) -- (Ma-1-1.north east)
		node[pos=0.5,above=10pt,black] {$\compact{x}$};
	\draw [decorate, decoration = {brace, raise=5pt, amplitude=5pt}, thick]
		(Ma-1-2.north west) -- (Ma-1-5.north east)
		node[pos=0.5,above=10pt,black] {$\tuple{\mathit{out},\mathit{in}}$};
	\node[yshift=10pt] at (Ma-1-6.north) () {$\toknotok$};
	\node[yshift=10pt] at (Ma-1-7.north) () {$\toktoken$};
	\end{scope}

	\begin{scope}[local bounding box=b, shift={(7,0)}]
		\matrix (Mb) [matrix of nodes,
			nodes={draw, minimum size=8mm,anchor=center},
			nodes in empty cells, minimum height = 1cm,
			row 2/.style={nodes={draw=none}},]
		{
		  & & & & & & \\
		};

	\node[yshift=-6pt, xshift=5pt] at (Mb-1-2.north west) () {$1$};
	\node[yshift=-6pt, xshift=5pt] at (Mb-1-3.north west) () {$2$};
	\node[yshift=-6pt, xshift=5pt] at (Mb-1-4.north west) () {$1$};
	\node[yshift=-6pt, xshift=5pt] at (Mb-1-5.north west) () {$2$};

	\node[circle, minimum size=1mm, inner sep=0pt, outer sep=0pt, fill=black]
		at (Mb-1-1.center) () {};
	\node[circle, minimum size=1mm, inner sep=0pt, outer sep=0pt, fill=black]
		at (Mb-1-2.center) () {};
	\node[circle, minimum size=1mm, inner sep=0pt, outer sep=0pt, fill=black]
		at (Mb-1-3.center) () {};
	\node[circle, minimum size=1mm, inner sep=0pt, outer sep=0pt, fill=black]
		at (Mb-1-4.center) () {};
	\node[circle, minimum size=1mm, inner sep=0pt, outer sep=0pt, fill=black]
		at (Mb-1-5.center) () {};
	\node[circle, minimum size=1mm, inner sep=0pt, outer sep=0pt, fill=black]
		at (Mb-1-6.center) () {};

	\draw [decorate, decoration = {brace, raise=5pt, amplitude=5pt}, thick]
		(Mb-1-1.north west) -- (Mb-1-1.north east)
		node[pos=0.5,above=10pt,black]{$\compact{y}$};
	\draw [decorate, decoration = {brace, raise=5pt, amplitude=5pt}, thick]
		(Mb-1-2.north west) -- (Mb-1-5.north east)
		node[pos=0.5,above=10pt,black]{$\tuple{\mathit{out},\mathit{in}}$};
	\node[yshift=10pt] at (Mb-1-6.north) () {$\toknotok$};
	\node[yshift=10pt] at (Mb-1-7.north) () {$\toktoken$};
	\end{scope}

	\begin{scope}[local bounding box=c, shift={(14,0)}]
		\matrix (Mc) [matrix of nodes,
			nodes={draw, minimum size=8mm,anchor=center},
			nodes in empty cells, minimum height = 1cm,
			row 2/.style={nodes={draw=none}},]
		{
		  & & & & & & \\
		};
		
	\node[yshift=-6pt, xshift=5pt] at (Mc-1-2.north west) () {$1$};
	\node[yshift=-6pt, xshift=5pt] at (Mc-1-3.north west) () {$2$};
	\node[yshift=-6pt, xshift=5pt] at (Mc-1-4.north west) () {$1$};
	\node[yshift=-6pt, xshift=5pt] at (Mc-1-5.north west) () {$2$};

	\node[circle, minimum size=1mm, inner sep=0pt, outer sep=0pt, fill=black]
		at (Mc-1-1.center) () {};
	\node[circle, minimum size=1mm, inner sep=0pt, outer sep=0pt, fill=black]
		at (Mc-1-2.center) () {};
	\node[circle, minimum size=1mm, inner sep=0pt, outer sep=0pt, fill=black]
		at (Mc-1-3.center) () {};
	\node[circle, minimum size=1mm, inner sep=0pt, outer sep=0pt, fill=black]
		at (Mc-1-6.center) () {};

	\draw [decorate, decoration = {brace, raise=5pt, amplitude=5pt}, thick]
		(Mc-1-1.north west) -- (Mc-1-1.north east)
		node[pos=0.5,above=10pt,black]{$\compact{z}$};
	\draw [decorate, decoration = {brace, raise=5pt, amplitude=5pt}, thick]
		(Mc-1-2.north west) -- (Mc-1-5.north east)
		node[pos=0.5,above=10pt,black]{$\tuple{\mathit{out},\mathit{in}}$};
	\node[yshift=10pt] at (Mc-1-6.north) () {$\toknotok$};
	\node[yshift=10pt] at (Mc-1-7.north) () {$\toktoken$};
	\end{scope}


	\node[circle, minimum size=1mm, inner sep=0pt, outer sep=0pt, fill=black]
		at (0, -2) (x) {};
	\node[circle, minimum size=1mm, inner sep=0pt, outer sep=0pt, fill=black]
		at (7, -2) (y) {};
	\node[circle, minimum size=1mm, inner sep=0pt, outer sep=0pt, fill=black]
		at (14, -2) (z) {};

	\node[below of=x, yshift=15mm] () {{\huge $x$}};
	\node[below of=y, yshift=15mm] () {{\huge $y$}};
	\node[below of=z, yshift=15mm] () {{\huge $z$}};

	\path[->] (x) edge [bend left, in=60] (Ma-1-1.west);
	\path[->] (Ma-1-1.center) edge [bend left] (x);
	\path[->] (Ma-1-4.center) edge [bend left] (x);
	\path[->] (Ma-1-5.center) edge [bend right] (y);
	\path[->] (Ma-1-7.center) edge [bend left] (x);

	\path[->] (y) edge [bend left, in=60] (Mb-1-1.west);
	\path[->] (Mb-1-1.center) edge [bend left] (y);
	\path[->] (Mb-1-2.center) edge [bend left] (x);
	\path[->] (Mb-1-3.center) edge [bend left] (y);
	\path[->] (Mb-1-4.center) edge [bend left] (y);
	\path[->] (Mb-1-5.center) edge [bend right] (z);
	\path[->] (Mb-1-6.center) edge [bend left] (y);

	\path[->] (z) edge [bend left, in=60] (Mc-1-1.west);
	\path[->] (Mc-1-1.center) edge [bend left] (z);
	\path[->] (Mc-1-2.center) edge [bend left] (y);
	\path[->] (Mc-1-3.center) edge [bend left] (z);
	\path[->] (Mc-1-6.center) edge [bend left] (z);

\end{tikzpicture}

}}
    \centerline{\footnotesize(b)}
  \end{subfigure}%
  \vspace*{-.5\baselineskip}
  \caption{Gaifman Heap for a Chain Configuration}
  \label{fig:gaifman-heap}
  \vspace*{-1.5\baselineskip}
\end{figure}
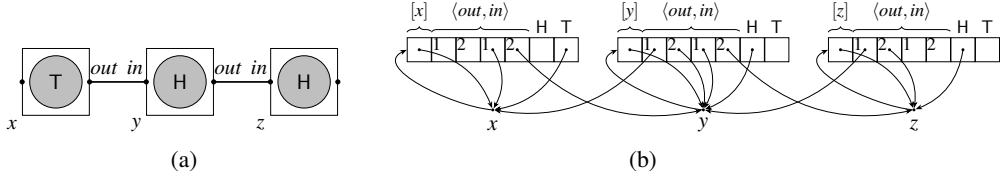

\begin{example}\label{ex:gaifman-heap}
  Fig. \ref{fig:gaifman-heap} (b) shows a Gaifman heap for the
  configuration in Fig. \ref{fig:gaifman-heap} (a), where each
  component belongs to at most $2$ interactions of type
  $\tuple{\mathit{out},\mathit{in}}$. \hfill$\blacksquare$
\end{example}

\begin{myTextE}
We say that a configuration $\aconfig'$ is a \emph{subconfiguration}
of $\aconfig$, denoted $\aconfig' \subconfig \aconfig$ if and only if
$\aconfig = \aconfig' \comp \aconfig''$, for some configuration
$\aconfig''$. The following lemma builds Gaifman heaps for
subconfigurations:

\begin{lemma}\label{lemma:gaifman-subconfig}
  Given configurations $\aconfig$ and $\aconfig'$, such that
  $\aconfig' \subconfig \aconfig$, if $\heap \in \gaifman{\aconfig}$,
  then $\heap' \in \gaifman{\aconfig'}$, where $\dom{\heap'} =
  \dom{\heap} \cap \nodes{\aconfig'}$ and $\heap'(c) = \heap(c)$, for
  all $c\in\dom{\heap'}$.
\end{lemma}
\proof{ We have $\dom{\heap'} = \dom{\heap} \cap \nodesof{\aconfig'} =
  \nodesof{\aconfig} \cap \nodesof{\aconfig'} = \nodesof{\aconfig'}$,
  because $\dom{\heap} = \nodesof{\aconfig} \supseteq
  \nodesof{\aconfig'}$. The points
  (\ref{it21:gaifman-heap}-\ref{it23:gaifman-heap}) of
  Def. \ref{def:gaifman-heap} are by easy inspection. \qed}
\end{myTextE}

We build a \seplog\ SID $\slsid$ that generates the Gaifman heaps of
the models of the predicate atoms occurring in a progressing \cl\ SID
$\asid$. The construction associates to each variable $x$, that occurs
free or bound in a rule from $\asid$, a unique $\rank$-tuple of
variables $\gaifimg{x} \in \vars^\rank$, that represents the image of
the store value $\store(x)$ in a Gaifman heap $\heap$ i.e.,
$\heap(\store(x)) = \store(\gaifimg{x})$. Moreover, we consider, for
each predicate symbol $\apred \in \defnof{\asid}$, an annotated
predicate symbol $\annotate{}{\apred}$ of arity
$\arityof{\annotate{}{\apred}} = (\rank+1)\cdot\arityof{\apred}$,
where $\intermap : \interv{1}{\arityof{\apred}} \times \interv{1}{M}
\rightarrow 2^{\interv{0}{\degreebound-1}}$ is a map associating each
parameter $i \in \interv{1}{\arityof{\apred}}$ and each interaction
type $\intertype_j$, for $j \in \interv{1}{M}$, a set of integers
$\intermap(i,j)$ denoting the positions of the encodings of the
interactions of type $\intertype_j$, involving the value of $x_i$, in
the models of $\annotate{}{\apred}(x_1, \ldots, x_{\arityof{\apred}},
\gaifimg{x_1}, \ldots, \gaifimg{x_{\arityof{\apred}}})$ (point
\ref{it22:gaifman-heap} of Def. \ref{def:gaifman-heap}). Then $\slsid$
contains rules of the form:
\begin{align}
\annotate{}{\apred}(x_1, \ldots, x_{\#(\apred)},\gaifimg{x_1}, \ldots, \gaifimg{x_{\#(\apred)}}) & \leftarrow \label{rule:slsid} \\
\exists y_1 \ldots \exists y_m \exists \gaifimg{y_1} \ldots \exists \gaifimg{y_m} ~.~ \overline{\psi} * \pureform ~* 
& \Asterisk_{\ell=1}^h ~\annotate{\ell}{\bpred}(z^\ell_1, \ldots, z^\ell_{\#(\bpred^\ell)}, 
\gaifimg{z^\ell_1}, \ldots, \gaifimg{z^\ell_{\#(\bpred^\ell)}}) \nonumber
\end{align}
for which $\asid$ has a \emph{stem rule} \(\apred(x_1, \ldots,
x_{\#(\apred)}) \leftarrow \exists y_1 \ldots \exists y_m ~.~ \psi *
\pureform * \Asterisk_{\ell=1}^h \bpred^\ell(z^\ell_1, \ldots,
z^\ell_{\arityof{\bpred^\ell}})\), where $\psi*\pureform$ is a
quantifier- and predicate-free formula and $\pureform$ is the
conjunction of equalities and disequalities from
$\psi*\pureform$. However, not all rules (\ref{rule:slsid}) are
considered in $\slsid$, but only the ones meeting the following
condition:

\begin{definition}\label{def:well-formed}
A rule of the form (\ref{rule:slsid}) is \emph{well-formed} if and
only if, for each $i \in \interv{1}{\arityof{\apred}}$ and each $j \in
\interv{1}{M}$, there exists a set of integers $Y_{i,j} \subseteq
\interv{0}{\degreebound-1}$, such that: \begin{compactitem}
\item $\cardof{Y_{i,j}} = \cardof{\xiatoms{\psi,\pureform}{j}{x_i}}$,
  where $\xiatoms{\psi,\pureform}{j}{x}$ is the set of interaction
  atoms $\interacn{z_1}{p_1}{z_n}{p_n}$ from $\psi$ of type
  $\intertype_j = \tuple{p_1, \ldots, p_n}$, such that $z_s
  \formeq{\pureform} x$, for some $s \in \interv{1}{n}$,
\item $Y_{i,j} \subseteq \intermap(i,j)$ and $\intermap(i,j) \setminus
  Y_{i,j} = \xipos{j}{x_i}$, where $\xipos{j}{x} \isdef
  \bigcup_{\ell=1}^h \bigcup_{k=1}^{\arityof{\bpred^\ell}}
  \set{\intermap^\ell(k,j) \mid x \formeq{\pureform} z^\ell_k}$ is the
  set of positions used to encode the interactions of type
  $\intertype_j$ involving the store value of the parameter $x$, in
  the sub-configuration corresponding to an atom $\bpred_\ell(z^\ell_1,
  \ldots, z^\ell_{\#(\bpred^\ell)})$, for some $\ell\in\interv{1}{h}$.
\end{compactitem}
\end{definition}
We denote by $\slsid$ the set of well-formed rules (\ref{rule:slsid}),
such that, moreover:
\[\begin{array}{l}
\overline{\psi} \isdef x_1 \mapsto \gaifimg{x_1} ~*~ 
\Asterisk_{\!\!x\in\fv{\psi}} ~\compstate{\psi}{x} ~*~ 
\Asterisk_{\!\!i=1}^{\!\!\arityof{\apred}} ~\interparam{\psi}{x_i}
\text{, where:} \\[1mm]
\compstate{\psi}{x} \isdef \Asterisk_{\compact{x} \text{ occurs in } \psi} ~\tuple{\gaifimg{x}}_1 = x ~*~ 
\Asterisk_{\compin{x}{q_k} \text{ occurs in } \psi} ~\tuple{\gaifimg{x}}_{\spos{k}} = x 
\\[1mm]
\interparam{\psi}{x_i} \isdef \Asterisk_{\!\!j=1}^{\!\!M} \Asterisk_{\!\!p=1}^{\!\!r_j} ~\tuple{\gaifimg{x_i}}_{\ipos{j}{k^j_p}} = \vec{x}^j_p
\text{ and } \set{k^j_1, \ldots, k^j_{r_j}} \isdef \intermap(i,j) \setminus \xipos{j}{x_i} 
\end{array}\]
Here for two tuples of variables $\vec{x} = \tuple{x_1, \ldots, x_k}$
and $\vec{y} = \tuple{y_1, \ldots, y_k}$, we denote by
$\vec{x}=\vec{y}$ the formula $\Asterisk_{i=1}^k
x_i=y_i$. Intuitively, the \seplog\ formula $\compstate{\psi}{x}$
realizes the encoding of the component and state atoms from $\psi$, in
the sense of points (\ref{it21:gaifman-heap}) and
(\ref{it23:gaifman-heap}) from Def. \ref{def:gaifman-heap}, whereas
the formula $\interparam{\psi}{x_i}$ realizes the encodings of the
interactions involving a parameter $x_i$ in the stem rule (point
\ref{it22:gaifman-heap} of Def. \ref{def:gaifman-heap}). In
particular, the definition of $\interparam{\psi}{x_i}$ uses the fact
that the rule is well-formed.

\begin{myLemmaE}\label{lemma:gaifman-soundness}
  Let $\asid$ be a progressing SID and $\apred \in \defnof{\asid}$ be
  a predicate, such that $\aconfig \models^\store \apred(x_1, \ldots,
  x_{\arityof{\apred}})$, for some configuration
  $\aconfig=(\comps,\interacs,\statemap)$ and store $\store$. Then,
  for each heap $\heap \in \gaifman{\aconfig}$, there exists a map
  $\intermap : \interv{1}{\arityof{\apred}} \times \interv{1}{M}
  \rightarrow 2^{\interv{0}{\degreebound-1}}$ and a store $\slstore$,
  such that the following hold: \begin{enumerate}
  \item\label{it1:gaifman-soundness} $\slstore(x_i) = \store(x_i) \in
    \dom{\heap}$ and $\slstore(\gaifimg{x_i}) = \heap(\store(x_i))$,
    $\forall i \in \interv{1}{\arityof{\apred}}$,
  \item\label{it2:gaifman-soundness}
    $\xituples{\interacs}{j}{\slstore(x_i)} =
    \Set{\tuple{\slstore(\gaifimg{x_i})}_{\ipos{j}{k}} \mid k \in
      \intermap(i,j)}$, $\forall i \in \interv{1}{\arityof{\apred}}~
    \forall j \in \interv{1}{M}$,
  \item\label{it3:gaifman-soundness} $\heap \slmodelsid^{\slstore}
    \annotate{}{\apred}(x_1, \ldots, x_{\arityof{\apred}},
    \gaifimg{x_1}, \ldots, \gaifimg{x_{\#(\apred)}})$.
  \end{enumerate}
\end{myLemmaE}
\begin{proofE}
  By induction on the definition of $\aconfig \models^\store
  \apred(x_1, \ldots, x_{\arityof{\apred}})$, assume that $\aconfig
  \models^\store\exists y_1 \ldots \exists y_m ~.~ \psi * \pureform *
  \Asterisk_{\ell=1}^h \bpred^\ell(z^\ell_1, \ldots,
  z^\ell_{\#(\bpred^\ell)})$, where $\apred(x_1, \ldots,
  x_{\#(\apred)}) \leftarrow \exists y_1 \ldots \exists y_m ~.~ \psi *
  \pureform * \Asterisk_{\ell=1}^h \bpred^\ell(z^\ell_1, \ldots,
  z^\ell_{\#(\bpred^\ell)})$ is a rule from $\asid$, such that
  $\psi*\pureform$ is quantifier- and predicate-free and $\pureform$
  is the conjunction of equalities and disequalities from
  $\psi*\pureform$. Then there exists a store $\store'$, that agrees
  with $\store$ over $x_1, \ldots, x_{\arityof{\apred}}$, and
  configurations $\aconfig_0 = (\comps_0,\interacs_0,\statemap),
  \ldots, \aconfig_h = (\comps_h,\interacs_h,\statemap)$, such
  that: \begin{compactitem}
  \item $\aconfig_0 \models^{\store'}_\asid \psi*\pureform$, 
  \item $\aconfig_\ell \models^{\store'}_\asid \bpred_\ell(z^\ell_1,
    \ldots, z^\ell_{\arityof{\bpred_\ell}})$, for all $\ell \in
    \interv{1}{h}$, and
  \item $\aconfig = \aconfig_0 \comp \ldots \comp \aconfig_h$.
  \end{compactitem}
  We define the heaps $\heap_0, \ldots, \heap_h$, as
  follows: \begin{compactitem}
  \item for each $\ell\in\interv{1}{h}$, let $\dom{\heap_\ell} =
    \nodesof{\aconfig_\ell}$, $\heap_\ell(c) = \heap(c)$, for all
    $c\in\dom{\heap_\ell}$,
  \item $\heap_0 \isdef \heap \setminus (\bigcup_{\ell=1}^h \heap_\ell)$. 
  \end{compactitem}
  By Lemma \ref{lemma:gaifman-subconfig}, we obtain that $\heap_\ell
  \in \gaifman{\aconfig_\ell}$, for all $\ell\in\interv{1}{h}$. We
  define $\heap \isdef \heap_0 \cup \ldots \cup \heap_h$ and prove
  that this is indeed a heap, by showing $\dom{\heap_i} \cap
  \dom{\heap_j} = \emptyset$, for all $0 \leq i < j \leq h$. If $i =
  0$, we have $\dom{\heap_i} \cap \dom{\heap_j} = \emptyset$, by the
  definition of $\heap_i$. Else, suppose, for a contradiction, that $c
  \in \dom{\heap_i} \cap \dom{\heap_j}$, for some $1 \leq i < j \leq
  h$. Then $c \in \nodesof{\aconfig_i} \cap \nodes{\aconfig_j}$. Since
  $\aconfig_\ell \models \exists x_1 \ldots \exists
  x_{\arityof{\bpred_\ell}} ~.~ \bpred(x_1, \ldots,
  x_{\arityof{\bpred_\ell}})$, by Lemma \ref{lemma:progressing-sid},
  we obtain $c \in \comps_i \cap \comps_j$, which contradicts the fact
  that $\aconfig_i \comp \aconfig_j$ is defined. Next, we apply the
  inductive hypothesis to find stores $\slstore_\ell$ and maps
  $\intermap^\ell$ such that, for $\ell \in \interv{1}{h}$, we
  have: \begin{compactitem}
  \item $\slstore(z^\ell_i) = \store'(z^\ell_i) \in \dom{\heap_\ell}$
    and $\heap_\ell(\store'(z^\ell_i)) =
    \slstore(\gaifimg{z^\ell_i})$, $\forall i \in
    \interv{1}{\arityof{\bpred_\ell}}$,
  \item $\xituples{\interacs_\ell}{j}{\slstore(z^\ell_i)} =
    \Set{\tuple{\slstore_\ell(\gaifimg{z^\ell_i})}_{\ipos{j}{k}}
      \mid k \in \intermap^\ell(i,j)}$, $\forall i \in
    \interv{1}{\#(\bpred_\ell)}~ \forall j \in \interv{1}{M}$, and
  \item $\heap_\ell \slmodelsid^{\slstore_\ell}
    \annotate{}{\bpred_\ell}(z^\ell_1, \ldots,
    z^\ell_{\arityof{\bpred_\ell}}, \gaifimg{z^\ell_1}, \ldots,
    \gaifimg{z^\ell_{\#(\bpred_\ell)}})$.
  \end{compactitem}
  First, for each $i \in \interv{1}{\arityof{\apred}}$ and each $j \in
  \interv{1}{M}$, we define $\intermap(i,j) \isdef \set{k_1, \ldots,
    k_{s_i}} \cup \bigcup_{\ell=1}^h
  \bigcup_{k=1}^{\arityof{\bpred_\ell}} \Set{\intermap^\ell(k,j) \mid
    x_i \formeq{\pureform} z^\ell_k}$, where: \begin{compactitem}
  \item $\xituples{\interacs_0}{j}{\store(x_i)} = \set{\vec{c}_1,
    \ldots, \vec{c}_{s_{i,j}}}$, and 
  \item $0 \leq k^{i,j}_1 < \ldots < k^{i,j}_{s_{i,j}} < \degreebound$
    are integers, such that
    $\tuple{\heap(\store(x_i))}_{\ipos{j}{k^{i,j}_\ell}} =
    \vec{c}_\ell$, $\forall \ell \in \interv{1}{s_{i,j}}$; the
    existence of these integers is stated by point
    (\ref{it22:gaifman-heap}) of Def. \ref{def:gaifman-heap}, relative
    to $\store(x_i)$.
  \end{compactitem}
  Second, we define the store $\slstore$ as follows: \begin{compactitem}
  \item $\slstore(x_1) = \store(x_1)$ and $\slstore(\gaifimg{x_1})
    \isdef \heap(\store(x_1))$,
  \item $\slstore(z^\ell_i) \isdef \store'(z^\ell_i)$, $\forall
    \ell\in\interv{1}{h}~ \forall
    i\in\interv{1}{\arityof{\bpred_\ell}}$,
  \item $\slstore(\gaifimg{z^\ell_i}) \isdef
    \heap(\store'(z^\ell_i))$, $\forall \ell\in\interv{1}{h}~ \forall
    i\in\interv{1}{\arityof{\bpred_\ell}}$,
  \item $\slstore$ is arbitrary everywhere else.
  \end{compactitem}
  The points (\ref{it1:gaifman-soundness}) and
  (\ref{it2:gaifman-soundness}) of the statement follow from the
  definitions of $\slstore$ and $\iota$, respectively. To prove point
  (\ref{it3:gaifman-soundness}), suppose, for a contradiction, that $k
  \in \set{k^{i,j}_1, \ldots, k^{i,j}_{s_{i,j}}} \cap
  \intermap^\ell(t,j) \neq \emptyset$, for some $i \in
  \interv{1}{\arityof{\apred}}$, $j \in \interv{1}{M}$,
  $t\in\interv{1}{\arityof{\bpred_\ell}}$ and $\ell\in\interv{1}{h}$,
  such that $x_i \formeq{\pureform} z^\ell_t$.  Then there exists a
  tuple of components $\vec{c} \in
  \xituples{\interacs_0}{j}{\store(x_i)}$, such that $\vec{c} =
  \tuple{\slstore_\ell(\gaifimg{z^\ell_i})}_{\ipos{j}{t}} \in
  \xituples{\interacs_\ell}{j}{\slstore(z^\ell_i)}$. Hence
  $\interacs_0 \cap \interacs_\ell \neq \emptyset$, which contradicts
  the fact that the composition $\aconfig_0 \comp \aconfig_\ell$ is
  defined. Hence, the rule:
  \begin{align*}
    \annotate{}{\apred}(x_1, \ldots, x_{\#(\apred)},\gaifimg{x_1}, \ldots, \gaifimg{x_{\#(\apred)}}) & \leftarrow \\
    \exists y_1 \ldots \exists y_m \exists \gaifimg{y_1} \ldots \exists \gaifimg{y_m} ~.~ \overline{\psi} * \pureform ~* 
    & \Asterisk_{\ell=1}^h ~\annotate{\ell}{\bpred}(z^\ell_1, \ldots, z^\ell_{\#(\bpred^\ell)}, 
    \gaifimg{z^\ell_1}, \ldots, \gaifimg{z^\ell_{\#(\bpred^\ell)}}) \
  \end{align*}
  is well-formed and thus belongs to $\slsid$. To obtain $\heap
  \slmodelsid^{\slstore} \annotate{}{\apred}(x_1, \ldots,
  x_{\arityof{\apred}}, \gaifimg{x_1}, \ldots,
  \gaifimg{x_{\#(\apred)}})$, by the definition of:
  \[\overline{\psi} \isdef x_1 \mapsto \gaifimg{x_1} * 
  \Asterisk_{\!\!x\in\fv{\psi}} ~\compstate{\psi}{x} *
  \Asterisk_{\!\!i=1}^{\!\!\arityof{\apred}} ~\interparam{\psi}{x_i}\]
  it is sufficient to prove the following points: \begin{compactitem}
  \item $\heap_0 \slmodels^{\slstore} x_1 \mapsto \gaifimg{x_1}$: by
    the definition $\slstore$, we have $\slstore(\gaifimg{x_1}) =
    \heap(\store(x_1))$, hence it is sufficient to prove that
    $\dom{\heap_0} = \set{\store(x_1)}$. ``$\subseteq$'' Let $c \in
    \dom{\heap_0}$ be a component. By the definition of $\heap_0 =
    \heap \setminus \bigcup_{\ell=1}^h \heap_\ell$, we have
    $\dom{\heap_0} = \dom{\heap} \setminus \bigcup_{\ell=1}^h
    \dom{\heap_\ell} = \nodesof{\aconfig} \setminus \bigcup_{\ell=1}^h
    \nodesof{\aconfig_\ell} = \nodesof{\aconfig_0}$, because $\heap
    \in \gaifman{\aconfig}$ and $\heap_\ell \in
    \gaifman{\aconfig_\ell}$, for all $\ell \in \interv{1}{h}$. Since
    $\aconfig_0 \models^{\store'}_{\asid} \psi$, we have $c =
    \store'(x)$, for some $x \in \fv{\psi}$. Suppose, for a
    contradiction, that $x$ and $x_1$ are not the same variable, then
    $x \in \set{z^\ell_1, \ldots, z^\ell_{\arityof{\bpred_\ell}}}$,
    for some $\ell \in \interv{1}{h}$, because $\asid$ is progressing
    (Def. \ref{def:pcr}). By Lemma \ref{lemma:progressing-sid}, we
    obtain $c \in \nodes{\aconfig_\ell}$, contradiction. Then $c =
    \store'(x_1) = \store(x_1)$. ``$\supseteq$'' Because $\asid$ is
    progressing, $\psi = \compact{x_1} * \varphi$, hence $\store(x_1)
    = \store'(x_1) \in \nodesof{\aconfig_0} = \dom{\heap_0}$, because
    $\aconfig_0 \models^{\store'} \psi$. 
  \item $\emptyset \slmodels^{\slstore} \compstate{\psi}{x}$, for each
    $x\in\fv{\psi}$: by the definition of $\slstore$, $\heap \in
    \gaifman{\aconfig}$ and points \ref{it21:gaifman-heap} and
    \ref{it23:gaifman-heap} of Def. \ref{def:gaifman-heap}.
  \item $\emptyset \slmodels^{\slstore} \interparam{\psi}{x_i}$, for
    each $i \in \interv{1}{\arityof{\apred}}$: by the definition of
    $\slstore$, $\heap \in \gaifman{\aconfig}$, definition of
    $\intermap(i,j)$, for all $i \in \interv{1}{\arityof{\apred}}$ and
    $j\in\interv{1}{M}$ and point \ref{it22:gaifman-heap} of
    Def. \ref{def:gaifman-heap}.
  \item $\emptyset \slmodels^{\slstore} \pureform$: because
    $(\emptyset,\emptyset,\statemap) \models^{\store'} \pureform$ and
    $\slstore$ agrees with $\store'$ over $\fv{\pureform}$. \qed
  \end{compactitem}
\end{proofE}
  
\begin{myLemmaE}\label{lemma:gaifman-completeness}
  Let $\asid$ be a progressing SID and $\apred \in \defnof{\asid}$ be
  a predicate, such that $\heap \slmodelsid^{\slstore}
  \annotate{}{\apred}(x_1, \ldots, x_{\#(\apred)}, \gaifimg{x_1},
  \ldots, \gaifimg{x_{\#(\apred)}})$, for a map $\intermap :
  \interv{1}{\#(\apred)} \times \interv{1}{m} \rightarrow
  2^{\interv{0}{\degreebound-1}}$ and a store $\slstore$. Then, the
  following hold: \begin{enumerate}
  \item\label{it1:gaifman-completeness} $\slstore(x_i) \in
    \dom{\heap}$ and $\heap(\slstore(x_i)) = \slstore(\gaifimg{x_i})$,
    for all $i \in \interv{1}{\arityof{\apred}}$,
  \item\label{it2:gaifman-completeness} there exists a configuration
    $\aconfig$, such that $\heap \in \gaifman{\aconfig}$ and $\aconfig
    \models^{\slstore}_\asid \apred(x_1, \ldots,
    x_{\arityof{\apred}})$.
  \end{enumerate}
\end{myLemmaE}
\begin{proofE}
  By fixpoint induction on the definition of $\heap
  \slmodelsid^{\slstore} \annotate{}{\apred}(x_1, \ldots,
  x_{\#(\apred)}, \gaifimg{x_1}, \ldots,
  \gaifimg{x_{\#(\apred)}})$. Consider the following well-formed rule
  from $\slsid$:
  \begin{align*}
    \annotate{}{\apred}(x_1, \ldots, x_{\#(\apred)},\gaifimg{x_1}, \ldots, \gaifimg{x_{\#(\apred)}}) & \leftarrow \\
    \exists y_1 \ldots \exists y_m \exists \gaifimg{y_1} \ldots \exists \gaifimg{y_m} ~.~ \overline{\psi} * \pureform ~* 
    & \Asterisk_{\ell=1}^h ~\annotate{\ell}{\bpred}(z^\ell_1, \ldots, z^\ell_{\#(\bpred^\ell)}, 
    \gaifimg{z^\ell_1}, \ldots, \gaifimg{z^\ell_{\#(\bpred^\ell)}}) 
  \end{align*}
  such that \(\heap \slmodelsid^{\slstore'} \overline{\psi} *
  \pureform ~* \Asterisk_{\ell=1}^h ~\annotate{\ell}{\bpred}(z^\ell_1,
  \ldots, z^\ell_{\#(\bpred^\ell)}, \gaifimg{z^\ell_1}, \ldots,
  \gaifimg{z^\ell_{\#(\bpred^\ell)}})\), where $\slstore'$ is a store
  that agrees with $\store$ over $x_1, \ldots, x_{\#(\apred)}$ and
  $\gaifimg{x_1}, \ldots, \gaifimg{x_{\#(\apred)}}$.

  \vspace*{\baselineskip}\noindent(\ref{it1:gaifman-completeness}) If $i = 1$ then $x_1
  \mapsto \gaifimg{x_1}$ is a subformula of $\overline{\psi}$, thus
  $\slstore(x_1) = \slstore'(x_1) \in \dom{\heap}$ and
  $\heap(\slstore(x_1)) = \slstore'(\gaifimg{x_1}) =
  \slstore(\gaifimg{x_1})$. Otherwise, because $\asid$ is progressing,
  $x_i \in \set{z^\ell_1, \ldots, z^\ell_{\arityof{\bpred_\ell}}}$,
  for some $\ell \in \interv{1}{h}$ and point
  (\ref{it1:gaifman-completeness}) follows from the inductive
  hypothesis.

  \vspace*{\baselineskip}\noindent(\ref{it2:gaifman-completeness})
  There exist heaps $\heap_0, \ldots, \heap_h$, such that the
  following hold: \begin{compactitem}
  \item $\dom{\heap_i} \cap \dom{\heap_j} = \emptyset$, for all $0
    \leq i < j \leq h$ and $\heap = \heap_0 \cup \ldots \cup \heap_h$,
  \item $\heap_0 \slmodels^{\slstore'} \overline{\psi} * \pureform$,
  \item $\heap_\ell \slmodelsid^{\slstore'}
    \annotate{\ell}{\bpred}(y^\ell_1, \ldots,
    y^\ell_{\#(\bpred^\ell)}, \gaifimg{y^\ell_1}, \ldots,
    \gaifimg{y^\ell_{\#(\bpred^\ell)}})$, for all $\ell \in
    \interv{1}{h}$.
  \end{compactitem}
  By the inductive hypothesis, there exist configurations $\aconfig_1
  = (\comps_1, \interacs_1, \statemap_1), \ldots, \aconfig_h =
  (\comps_h, \interacs_h, \statemap_h)$, such that $\heap_\ell \in
  \gaifman{\aconfig_\ell}$ and $\aconfig_\ell \models^{\slstore}_\asid
  \bpred_\ell(z^\ell_1, \ldots, z^\ell_{\#(\bpred_\ell)})$, for all
  $\ell \in \interv{1}{h}$. We define the configuration $\aconfig_0 =
  (\comps_0, \interacs_0, \statemap_0)$, as follows: \begin{compactitem}
  \item $\comps_0 \isdef \set{\slstore(x_1)}$,
  \item $\interacs_0 \isdef \set{(\slstore'(z_1), p_1, \ldots,
    \store'(z_n), p_n) \mid \interacn{z_1}{p_1}{z_n}{p_n} \text{
      occurs in } \psi}$,
  \item $\statemap_0(\slstore'(z)) \isdef q_k$ if and only if
    $\compin{z}{q_k}$ occurs in $\psi$, otherwise $\statemap_0$ is
    arbitrary.
  \end{compactitem}
  Moreover, we define the state map $\statemap$ as $\statemap(c)
  \isdef \statemap_\ell(c)$ if $c \in \dom{\heap_\ell}$, for all
  $\ell\in\interv{0}{h}$ and $\statemap(c)$ is arbitrary, for $c
  \not\in \bigcup_{\ell=0}^h \dom{\heap_\ell}$. Since $\dom{\heap_0},
  \ldots, \dom{\heap_h}$ are pairwise disjoint, $\statemap$ is
  properly defined. First, we prove that the composition $(\comps_0,
  \interacs_0, \statemap) \comp \ldots \comp (\comps_h, \interacs_h,
  \statemap)$ is defined, namely that, for all $0 \leq i < j \leq h$,
  we have:\begin{compactitem}
  \item $\comps_i \cap \comps_j = \emptyset$: If $i = 0$ then either
    $\comps_0 = \emptyset$, in which case we are done, or $\comps_0 =
    \set{\slstore(x_1)} = \set{\slstore'(x_1)}$. By the definition of
    $\overline{\psi} = x_1 \mapsto \gaifimg{x_1} * \varphi$ and
    $\heap_0 \slmodels^{\slstore'} \overline{\psi}$, we obtain
    $\slstore'(x_1) \in \dom{\heap_0}$. Since $\dom{\heap_0} \cap
    \dom{\heap_j} = \emptyset$, we obtain $\slstore(x_1) \not\in
    \dom{\heap_j}$. Since $\heap_j \in \gaifman{\aconfig_j}$, we
    obtain $\comps_j \subseteq \nodesof{\aconfig_j} = \dom{\heap_j}$,
    thus $\comps_i \cap \comp_j = \emptyset$. Else $i > 0$ and, since
    $\heap_i \in \gaifman{\aconfig_i}$ and $\heap_j \in
    \gaifman{\aconfig_j}$, we obtain $\comps_i \subseteq
    \nodesof{\aconfig_i} = \dom{\heap_i}$ and $\comps_j \subseteq
    \nodesof{\aconfig_j} = \dom{\heap_j}$. But $\dom{\heap_i} \cap
    \dom{\heap_j} = \emptyset$, leading to $\comps_i \cap \comps_j =
    \emptyset$. 
  \item $\interacs_i \cap \interacs_j = \emptyset$: If $i = 0$, by the
    definition of $\interacs_0$, each interaction from $\interacs_0$
    is of the form $(\slstore'(z_1), p_1, \ldots, \slstore'(z_n),
    p_n)$, such that $\interacn{z_1}{p_1}{z_n}{p_n}$ is an interaction
    atom occuring in $\psi$. Since, moreover, $\asid$ is progressing,
    we have $x_1 \in \set{z_1, \ldots, z_n}$, hence $\slstore(x_1) \in
    \set{\slstore'(z_1), \ldots, \slstore'(z_n)}$. Let $(c_1, p_1,
    \ldots, c_n, p_n) \in \interacs_j$ be an interaction. Since
    $\heap_j \in \gaifman{\aconfig_j}$, we obtain $\set{c_1, \ldots,
      c_n} \subseteq \nodesof{\aconfig_j} = \dom{\heap_j}$, hence
    $\set{c_1, \ldots, c_n} \cap \dom{\heap_0} = \set{c_1, \ldots,
      c_n} \cap \set{\slstore(x_1)} = \emptyset$, leading to
    $\interacs_i \cap \interacs_j = \emptyset$, because the choices of
    $(\slstore'(z_1), p_1, \ldots, \slstore'(z_n), p_n) \in
    \interacs_i$ and $(c_1, p_1, \ldots, c_n, p_n) \in \interacs_j$
    are arbitrary. Else, $i > 0$ and let $(c^i_1, p_1, \ldots, c^i_n,
    p_n) \in \interacs_i$, $(c^j_1, p_1, \ldots, c^j_n, p_n) \in
    \interacs_j$ be two interactions of the same type. Since $\heap_i
    \in \gaifman{\aconfig_i}$ and $\heap_j \in \gaifman{\aconfig_j}$,
    we have $\set{c^i_1, \ldots, c^i_n} \subseteq \nodesof{\aconfig_i}
    = \dom{\heap_i}$ and $\set{c^j_1, \ldots, c^j_n} \subseteq
    \nodesof{\aconfig_j} = \dom{\heap_j}$, respectively. Since
    $\dom{\heap_i} \cap \dom{\heap_j} = \emptyset$, we obtain
    $\set{c^i_1, \ldots, c^i_n} \cap \set{c^j_1, \ldots, c^j_n} =
    \emptyset$. Since the choices of $(c^i_1, p_1, \ldots, c^i_n, p_n)
    \in \interacs_i$ and $(c^j_1, p_1, \ldots, c^j_n, p_n) \in
    \interacs_j$ are arbitrary, we obtain $\interacs_i \cap
    \interacs_j = \emptyset$.
  \end{compactitem}
  Consequently, we define $\aconfig \isdef (\comps_0, \interacs_0,
  \statemap) \comp \ldots \comp (\comps_h, \interacs_h, \statemap)$
  and conclude by proving the following points: \begin{compactitem}
  \item $\heap = \gaifman{\aconfig}$: we prove that $\dom{\heap} =
    \bigcup_{\ell=0}^h \dom{\heap_\ell} = \nodesof{\aconfig}$, as
    required by Def. \ref{def:gaifman-heap}. The conditions
    (\ref{it21:gaifman-heap}-\ref{it23:gaifman-heap}) for
    $c_0=\slstore(x_1)$ are by the definition of $\overline{\psi}$;
    for $c_0 \in \dom{\heap_\ell}$ these conditions follow from
    $\heap_\ell \in \gaifman{\aconfig_\ell}$. ``$\subseteq$'' Let $c
    \in \dom{\heap}$ be a component. If $c \in \dom{\heap_0} =
    \set{\slstore{x_1}}$, then $c \in \comps_0 \subseteq
    \nodesof{\aconfig_0} \subseteq \nodesof{\aconfig}$. Else $c \in
    \dom{\heap_\ell}$, for some $\ell \in \interv{1}{h}$, then $c \in
    \nodesof{\aconfig_\ell} \subseteq \nodesof{\aconfig}$, because
    $\heap_\ell \in \gaifman{\aconfig_\ell}$. ``$\supseteq$'' Let $c
    \in \nodesof{\aconfig} = \bigcup_{\ell=0}^h
    \nodesof{\aconfig_\ell}$ be a component. If $c \in
    \nodesof{\aconfig_0}$ then either $c \in \comps_0$ or $c$ occurs
    in some interaction from $\interacs_0$. If $c \in \comps_0$ then
    $c = \slstore(x_1) \in \dom{\heap_0} \subseteq \dom{\heap}$, by
    the definition of $\comps_0$. Else there exists an interaction
    $(c_1, p_1, \ldots, c_n, p_n) \in \interacs_0$, such that $c \in
    \set{c_1, \ldots, c_n}$. In this case $c = \slstore'(z)$, for some
    variable $z$ that occurs in an interaction atom from $\psi$. Since
    $\asid$ is progressing, $z \in \set{z^\ell_1, \ldots,
      z^\ell_{\arityof{\bpred_\ell}}}$, for some $\ell \in
    \interv{1}{h}$. Because $\aconfig_\ell \models^{\slstore'}_\asid
    \bpred_\ell(z^\ell_1, \ldots, z^\ell_{\arityof{\bpred_\ell}})$, we
    obtain $c = \slstore'(z) \in \nodesof{\aconfig_\ell}$, by Lemma
    \ref{lemma:progressing-sid}, and $c \in \dom{\heap_\ell} \subseteq
    \dom{\heap}$, because $\heap_\ell \in \gaifman{\aconfig_\ell}$. If
    $c \in \nodesof{\aconfig_\ell}$, for some $\ell \in
    \interv{1}{h}$, we have $c \in \dom{\heap_\ell} \subseteq
    \dom{\heap}$, because $\heap_\ell \in \gaifman{\aconfig_\ell}$.
  \item $\aconfig \models^{\slstore}_\asid \apred(x_1, \ldots,
    x_{\arityof{\apred}})$: Let the stem of the above rule from
    $\slsid$ be: \[\apred(x_1, \ldots, x_{\arityof{\apred}})
    \leftarrow \exists y_1 \ldots \exists y_m ~.~ \psi * \pureform *
    \Asterisk_{\ell=1}^h \bpred_\ell(z^\ell_1, \ldots,
    z^\ell_{\arityof{\bpred_\ell}})\] Since $(\comps_\ell,
    \interacs_\ell, \statemap_\ell) \models^{\slstore'}_\asid
    \bpred_\ell(z^\ell_1, \ldots, z^\ell_{\arityof{\bpred_\ell}})$ and
    $\statemap$ agrees with $\statemap_\ell$ over
    $\nodesof{\aconfig_\ell}$, it follows that $(\comps_\ell,
    \interacs_\ell, \statemap) \models^{\slstore'}_\asid
    \bpred_\ell(z^\ell_1, \ldots, z^\ell_{\arityof{\bpred_\ell}})$,
    for all $\ell \in \interv{1}{h}$. Moreover, $(\comps_0,
    \interacs_0, \statemap) \models^{\slstore'} \psi$ by definition,
    and $(\emptyset, \emptyset, \statemap) \models^{\slstore'}
    \pureform$, because $\emptyset \slmodels^{\slstore'}
    \pureform$. Altogether, we obtain $\aconfig
    \models^{\slstore'}_\asid \psi * \pureform * \Asterisk_{\ell=1}^h
    \bpred_\ell(z^\ell_1, \ldots, z^\ell_{\arityof{\bpred_\ell}})$,
    leading to $\aconfig \models^{\slstore}_\asid \apred(x_1, \ldots,
    x_{\arityof{\apred}})$. \qed
  \end{compactitem}
\end{proofE}
We state below the main result of this section on the complexity of
the entailment problem. The upper bounds follow from a many-one
reduction of $\entl{\asid}{\apred}{\bpred}$ to the \seplog\ entailment
$\xannot{}{\iota}{\apred}(x_1,\ldots,x_{\arityof{\apred}},
\gaifimg{x_1}, \ldots,\gaifimg{x_{\arityof{\apred}}}) \slmodelsid
\exists x_{\arityof{\bpred}+1} \ldots \exists x_{\arityof{\bpred}}
\exists \gaifimg{x_{\arityof{\bpred}+1}} \ldots \exists
\gaifimg{x_{\arityof{\bpred}}} ~.~ \\
\xannot{}{\intermap'}{\bpred}(x_1,\ldots,x_{\arityof{\bpred}},
\gaifimg{x_1},\ldots,\gaifimg{x_{\arityof{\bpred}}})$, in combination
with the upper bound provided by Theorem \ref{thm:sl-entailment}, for
\seplog\ entailments. If $k < \infty$, the complexity is tight for
\cl, whereas gaps occur for $k=\infty, \ell<\infty$ and $k=\infty,
\ell=\infty$, due to the cut-off on the degree bound
(Prop. \ref{prop:bound-cutoff}), which impacts the size of $\slsid$
and time needed to generate it from $\asid$.

\begin{theoremE}\label{thm:entailment}
  If $\asid$ is progressing, connected and e-restricted and, moreover,
  $\bound{\asid}{\apred}$ has a positive answer,
  $\klentl{\asid}{\apred}{\bpred}{k}{\ell}$ is in \twoexptime,
  $\klentl{\asid}{\apred}{\bpred}{\infty}{\ell}$ is in
  \threeexptime\ $\cap$ \twoexptime-hard, and
  $\entl{\asid}{\apred}{\bpred}$ is in \fourexptime\ $\cap$
  \twoexptime-hard.
\end{theoremE}
\begin{proofE}
  The proof consists of three parts. (1) We reduce
  $\entl{\asid}{\apred}{\bpred}$ to an equivalent \seplog\ entailment
  problem, for a progressing, connected and e-restricted SID. (2) This
  reduction provides upper bounds for
  $\klentl{\asid}{\apred}{\bpred}{k}{\ell}$, in the cases
  $k,\ell<\infty$, $k=\infty,\ell<\infty$ and $k=\ell=\infty$,
  respectively. (3) We give a lower bound for
  $\entl{\asid}{\apred}{\bpred}$, by reduction from
  \seplog\ entailment.

  \vspace*{\baselineskip}\noindent(1) We prove that, for each map
  $\iota : \interv{1}{\arityof{\apred}} \times \interv{1}{M}
  \rightarrow 2^{\interv{0}{\degreebound-1}}$ there exists a map
  $\iota' : \interv{1}{\arityof{\bpred}} \times \interv{1}{M}
  \rightarrow 2^{\interv{0}{\degreebound-1}}$, such that:
  \begin{align*}
    \apred(x_1,\ldots,x_{\arityof{\apred}}) & \models_\asid & \exists x_{\arityof{\bpred}+1} \ldots \exists x_{\arityof{\bpred}} ~.~ \bpred(x_1,\ldots,x_{\arityof{\bpred}}) ~\iff \\
    \xannot{}{\iota}{\apred}(x_1,\ldots,x_{\arityof{\apred}},\gaifimg{x_1},\ldots,\gaifimg{x_{\arityof{\apred}}}) & \slmodelsid & 
    \exists x_{\arityof{\bpred}+1} \ldots \exists x_{\arityof{\bpred}} \exists \gaifimg{x_{\arityof{\bpred}+1}} \ldots \exists \gaifimg{x_{\arityof{\bpred}}} ~.~ \\
    && \xannot{}{\intermap'}{\bpred}(x_1,\ldots,x_{\arityof{\bpred}},\gaifimg{x_1},\ldots,\gaifimg{x_{\arityof{\bpred}}})
  \end{align*}
  \noindent``$\Rightarrow$'' Let $\heap$ be a heap and $\slstore$ be a
  store, such that $\heap \slmodelsid^{\slstore}
  \xannot{}{\iota}{\apred}(x_1,\ldots,x_{\arityof{\apred}},\gaifimg{x_1},
  \ldots, \gaifimg{x_{\arityof{\apred}}})$. By Lemma
  \ref{lemma:gaifman-completeness}, we have $\heap(\slstore(x_i)) =
  \slstore(\gaifimg{x_i})$, for all $i \in
  \interv{1}{\arityof{\apred}}$ and, moreover, there exists a
  configuration $\aconfig$, such that $\heap \in \gaifman{\aconfig}$
  and $\aconfig \models^{\slstore}_\asid \apred(x_1, \ldots,
  x_{\arityof{\apred}})$. By the hypothesis, we obtain $\aconfig
  \models^{\slstore}_\asid \exists x_{\arityof{\apred}+1} \ldots
  \exists x_{\arityof{\bpred}} ~.~ \bpred(x_1, \ldots,
  x_{\arityof{\bpred}})$, hence there exists a store $\slstore'$, that
  agrees with $\slstore$ over $x_1, \ldots, x_{\arityof{\apred}}$,
  such that $\aconfig \models^{\slstore'}_\asid \bpred(x_1, \ldots,
  x_{\arityof{\bpred}})$. By Lemma \ref{lemma:gaifman-soundness},
  there exists a store $\slstore''$ that agrees with $\slstore'$ over
  $x_1, \ldots, x_{\arityof{\bpred}}$, such that
  $\heap(\slstore''(x_i)) = \slstore''(\gaifimg{x_i})$, for all $i \in
  \interv{1}{\arityof{\bpred}}$ and $\heap \slmodelsid^{\slstore''}
  \xannot{}{\iota'}{\bpred}(x_1, \ldots, x_{\arityof{\bpred}},
  \gaifimg{x_1}, \ldots, \gaifimg{x_{\arityof{\bpred}}}))$, for some
  map $\iota' : \interv{1}{\arityof{\bpred}} \times \interv{1}{M}
  \rightarrow 2^{\interv{0}{\degreebound-1}}$, because $\heap \in
  \gaifman{\aconfig}$. Hence, $\slstore''$ agrees with $\slstore$ over
  $x_1, \ldots, x_{\arityof{\apred}}, \gaifimg{x_1}, \ldots,
  \gaifimg{\arityof{\apred}}$, thus we obtain \(\heap
  \slmodelsid^{\slstore} \exists x_{\arityof{\apred}+1} \ldots \exists
  x_{\arityof{\bpred}} ~.~ \xannot{}{\iota'}{\bpred}(x_1, \ldots,
  x_{\arityof{\bpred}}, \gaifimg{x_1}, \ldots,
  \gaifimg{x_{\arityof{\bpred}}})\).

  \vspace*{\baselineskip}\noindent''$\Leftarrow$'' Let $\aconfig$ be a
  configuration, $\store$ be a store such that $\aconfig
  \models_\asid^\store \apred(x_1, \ldots, x_{\arityof{\apred}})$ and
  $\heap \in \gaifman{\aconfig}$ be a heap. Cleary, such a heap
  exists, for any given configuration, by
  Def. \ref{def:gaifman-heap}. By Lemma \ref{lemma:gaifman-soundness},
  there exists a map $\iota : \interv{1}{\arityof{\apred}} \times
  \interv{1}{M} \rightarrow 2^{\interv{0}{\degreebound-1}}$ and a
  store $\slstore$, that agrees with $\store$ over $x_1, \ldots,
  x_{\arityof{\apred}}$, such that $\slstore(\gaifimg{x_i}) =
  \heap(\store(x_i))$, for all $i \in \interv{1}{\arityof{\apred}}$
  and $\heap \slmodelsid^{\slstore}
  \xannot{}{\iota}{\apred}(x_1,\ldots,x_{\arityof{\apred}},\gaifimg{x_1},
  \ldots, \gaifimg{x_{\arityof{\apred}}})$. By the hypothesis, we have
  $\heap \slmodelsid^{\slstore'}
  \xannot{}{\intermap'}{\bpred}(x_1,\ldots,x_{\arityof{\bpred}},\gaifimg{x_1},\ldots,\gaifimg{x_{\arityof{\bpred}}})$,
  for some map $\iota' : \interv{1}{\arityof{\bpred}} \times
  \interv{1}{M} \rightarrow 2^{\interv{0}{\degreebound-1}}$ and a
  store $\slstore'$ that agrees with $\slstore$ over
  $x_1,\ldots,x_{\arityof{\apred}}$ and $\gaifimg{x_1}, \ldots,
  \gaifimg{x_{\arityof{\apred}}}$. By Lemma
  \ref{lemma:gaifman-completeness}, we have $\slstore'(\gaifimg{x_i})
  = \heap(\slstore(x_i)) = \slstore(\gaifimg{x_i})$, for all $i \in
  \interv{1}{\arityof{\apred}}$ and there exists a configuration
  $\aconfig'$, such that $\heap \in \gaifman{\aconfig'}$ and
  $\aconfig' \models^{\slstore'}_\asid \bpred(x_1, \ldots,
  x_{\arityof{\bpred}})$. Since $\heap \in \gaifman{\aconfig} \cap
  \gaifman{\aconfig'}$, by Def. \ref{def:gaifman-heap}, we obtain
  $\aconfig=\aconfig'$, hence $\aconfig \models^{\slstore'}_\asid
  \bpred(x_1, \ldots, x_{\arityof{\bpred}})$. Since, moreover
  $\slstore'$, $\slstore$ and $\store$ all agree over $x_1, \ldots,
  x_{\arityof{\apred}}$, we obtain $\aconfig \models^\store_\asid
  \exists x_{\arityof{\apred}+1} \ldots \exists x_{\arityof{\bpred}}
  ~.~ \bpred(x_1, \ldots, x_{\arityof{\bpred}})$.

  \vspace*{\baselineskip}\noindent Since $\asid$ is progressing and
  connected, $\slsid$ is progressing and connected as well. Moreover,
  $\slsid$ is e-restricted, because $\asid$ is e-restricted and the
  construction of $\slsid$ only introduces equalities, not
  disequalities.

  \vspace*{\baselineskip}\noindent(2) The upper bound relies on the
  result of \cite[Theorem 32]{EchenimIosifPeltier21b}, that gives a
  $2^{2^{\poly{\width{\slsid} \cdot \log\size{\slsid}}}}$ upper bound
  for \seplog\ entailments. Note that the number of variables in each
  rule from $\slsid$ is the number of variables in its stem rule
  multiplied by $\rank+1$, hence \(\maxwidthof{\slsid} \leq
  \maxwidthof{\asid} \cdot (\rank+1) = \maxwidthof{\asid} \cdot
  \degreebound \cdot \maxintersize{\asid} \cdot
  2^{\bigO(\maxintersize{\asid})}\), because $\rank = \pos{0}{M+1}{N}
  = 1 + \degreebound \cdot \sum_{\ell=1}^M \lenof{\intertype_\ell} + N
  = \degreebound \cdot \maxintersize{\asid} \cdot
  2^{\bigO(\maxintersize{\asid})}$. The time needed to build $\slsid$
  and its size are bounded as follows:
  \[\begin{array}{rcl}
  \sizeof{\slsid} & \leq & \cardof{\slsid} \cdot \maxwidthof{\slsid}\text{, since there are $2^{\degreebound \cdot M \cdot \maxarityof{\asid}}$ maps
    $\iota : \interv{1}{\arityof{\apred}} \times \interv{1}{M} \rightarrow 2^{\interv{0}{\degreebound-1}}$}  \\
  & \leq & 2^{\degreebound \cdot M \cdot \maxarityof{\asid}} \cdot \cardof{\asid} \cdot \maxwidthof{\slsid} \\
  & = & 2^{\degreebound \cdot 2^{\maxintersize{\asid}} \cdot \maxarityof{\asid}} \cdot \cardof{\asid} \cdot \degreebound \cdot \maxintersize{\asid} \cdot
  2^{\bigO(\maxintersize{\asid})} \\
  & = & \sizeof{\asid} \cdot 2^{\polynomial{(\degreebound \cdot 2^{\maxintersize{\asid}} \cdot \maxarityof{\asid})}}
  \end{array}\]
  By Prop. \ref{prop:bound-cutoff}, we consider the following cases: \begin{compactitem}
  \item if $k,\ell < \infty$ then
    $\degreebound=\polynomial{(\sizeof{\asid})}$, thus
    $\sizeof{\slsid} = 2^{2^{\bigO(\sizeof{\sizeof{\asid}})}}$
  \item if $k=\infty$ and $\ell < \infty$ then $\degreebound =
    2^{\polynomial(\size{\asid})}$, thus $\sizeof{\slsid} =
    2^{2^{2^{\bigO(\sizeof{\sizeof{\asid}})}}}$
  \item if $k=\infty$ and $\ell=\infty$ then $\degreebound =
    2^{2^{\polynomial(\size{\asid})}}$, thus $\sizeof{\slsid} =
    2^{2^{2^{2^{\bigO(\sizeof{\sizeof{\asid}})}}}}$
  \end{compactitem}

  \vspace*{\baselineskip}\noindent(3) The \twoexptime-hard lower bound
  for $\klentl{\asid}{\apred}{\bpred}{\infty}{\ell}$ and
  $\entl{\asid}{\apred}{\bpred}$ is obtained by reduction from the
  \seplog\ entailment problem $\xannot{}{}{\apred}(x_1, \ldots, x_k)
  \slmodelsid \xannot{}{}{\bpred}(x_1, \ldots, x_k)$, where $\slsid$
  is a progressing and connected SID, with no disequalities
  \cite[Theorem 18]{EchenimIosifPeltier20}. Note that the maximum
  arity of $\asid$ cannot be bounded to a constant, in order to obtain
  \twoexptime-hardness of the \seplog\ entailment problem, hence the
  lower bound does not apply to
  $\klentl{\asid}{\apred}{\bpred}{k}{\ell}$. The idea of the reduction
  is to encode each \seplog\ atomic proposition of the form $x \mapsto
  (y_1, \ldots, y_\rank)$ by the formula $\compact{x} *
  \interacn{x}{p_0}{y_\rank}{p_\rank}$. Then each model $\heap$ of a
  \seplog\ predicate atom $\xannot{}{}{\apred}(x_1, \ldots,
  x_{\#(\xannot{}{}{\apred})})$ is represented by a configuration
  $\aconfig = (\comps, \interacs, \statemap)$, such that $\comps =
  \dom{\heap}$ and $\interacs = \set{(c_0,p_0,\ldots,c_\rank,p_\rank)
    \mid \heap(c_0) = \tuple{c_1, \ldots, c_\rank}}$. Since $\slsid$
  is progressing and connected, the \cl\ SID $\asid$, obtained from
  the reduction, is progressing and connected. Since, moreover, the
  reduction does not introduce disequalities, $\asid$ is trivially
  e-restricted. Because the reduction takes polynomial time, we obtain
  a \twoexptime-hard lower bound.  \qed
\end{proofE}

\section{Conclusions and Future Work}

We study the satisfiability and entailment problems in a logic used to
write proofs of correctness for dynamically reconfigurable distributed
systems. The logic views the components and interactions from the
network as resources and reasons also about the local states of the
components. We reuse existing techniques for Separation Logic
\cite{Reynolds02}, showing that our configuration logic is more
expressive than \seplog, fact which is confirmed by a number of
complexity gaps. Closing up these gaps and finding tight complexity
classes in the more general cases is considered for future work. In
particular, we aim at lifting the boundedness assumption on the degree
of the configurations that must be considered to check the validity of
entailments.

\bibliographystyle{abbrv}
\bibliography{refs}

\end{document}